\documentclass[aps,prb,twocolumn,longbibliography,superscriptaddress]{revtex4-1}
\usepackage{graphicx, bm, amsmath, amsfonts,amssymb}
\usepackage{subcaption}
\usepackage[varg]{txfonts}
\usepackage{dsfont}
\usepackage{cancel}
\usepackage{epsfig}
\usepackage{epstopdf}
\usepackage{dcolumn}
\usepackage{grffile}
\usepackage{verbatim}
\usepackage{mathrsfs}
\usepackage{amssymb}
\usepackage[colorlinks=true,linkcolor=blue,citecolor=blue, urlcolor=blue]{hyperref}

\def \C{\mathbb{C}}
\def \Z{\mathbb{Z}}

\def \tr{\operatorname{tr}}
\def \sgn{\operatorname{sgn}}

\def\v#1{\textbf{\emph{#1}}}
\def\v#1{\textbf{\emph{#1}}}
\def\dd{\mathrm{d}}

\newcommand{\expect}[1]{\langle{#1}\rangle}

\graphicspath{
    {./figures/}
}

\begin{document}


\title{Spin-charge separation and unconventional superconductivity in \textit{t}-\textit{J} model on honeycomb lattice}

\author{Jian-Jian Miao}
\thanks{These authors contribute equally.}
\affiliation{Department of Physics, The Chinese University of Hong Kong, Sha Tin, New Territories, Hong Kong, China}

\author{Zheng-Yuan Yue}
\thanks{These authors contribute equally.}
\affiliation{Department of Physics, The Chinese University of Hong Kong, Sha Tin, New Territories, Hong Kong, China}

\author{Hao Zhang}
\affiliation{School of Physics and Astronomy, University of Minnesota, Minneapolis, MN 55455, USA}

\author{Wei-Qiang Chen}
\email{chenwq@sustech.edu.cn}
\affiliation{Department of Physics and Shenzhen Institute for Quantum Science and Engineering, Southern University of Science and Technology, Shenzhen 518055, China}
\affiliation{Shenzhen Key Laboratory of Advanced Quantum Functional Materials and Devices, Southern University of Science and Technology, Shenzhen 518055, China}

\author{Zheng-Cheng Gu}
\email{zcgu@phy.cuhk.edu.hk}
\affiliation{Department of Physics, The Chinese University of Hong Kong, Sha Tin, New Territories, Hong Kong, China}

\date{\today}

\begin{abstract}
The physical nature of doped Mott-insulator has been intensively studied for more than three decades. It is well known that the single band Hubbard model or $t$-$J$ model on the bipartite lattice is the simplest model to describe a doped Mott insulator. Unfortunately, the key mechanism of superconductivity in these toy models is still under debate.  
In this paper, we propose a new mechanism for the $d+id$-wave superconductivity (SC) that occurs in the small-doping region of the honeycomb lattice $t$-$J$ model based on the Grassmann tensor product state numerical simulation and spin-charge separation formulation.
Moreover, in the presence of anti-ferromagnetic order, a continuum effective field theory for holons is developed near half-filling.  It reveals the competition between repulsive and attractive holon interactions induced by spinon fluctuations and gauge fluctuations, respectively. 
At a large value of $t/J$, the repulsive interaction dominates, leading to the non-Fermi liquid like behavior; while in a moderate range of $t/J$, the attractive interaction dominates, leading to the SC order. 
Possible experimental detection of spin-charge separation phenomena is also discussed. 

\end{abstract}

\maketitle


\section{Introduction}

Since the discovery of cuprates \cite{Bednorz}, the mechanism of high-$T_c$ superconductivity (SC) remains one of the long-standing hardcore problems in condensed matter physics. The $t$-$J$ model is a well-accepted microscopic model which potentially captures the essential physics of $\mathrm{CuO_2}$ layers in high-$T_c$ cuprates, and can be derived from the large-$U$ limit of three-band Hubbard model for the $\mathrm{CuO_2}$ layers \cite{Zhang_Rice}. The $t$-$J$ model Hamiltonian reads:
\begin{align}
    H = -t {\sum_{\langle ij \rangle,\sigma}} (
        \hat{c}_{i\sigma}^{\dagger}\hat{c}_{j\sigma} + h.c.
    ) 
    + J {\sum_{\langle ij \rangle}} \left(
        \mathbf{S}_i \cdot \mathbf{S}_j
        - \frac{1}{4} n_i n_j
    \right),
    \label{eq:tJ-ham-exact}
\end{align}
where 
$n_i = \sum_\sigma c_{i\sigma}^{\dagger}c_{i\sigma} \equiv \sum_\sigma n_{i\sigma}$ is the electron density operator,
$\hat{c}_{i\sigma} = (1 - n_{i,-\sigma})c_{i\sigma}$ is the electron annihilation operator defined in no-double-occupancy subspace, and
\(
\mathbf{S}_i = (1/2) \sum_{\alpha\beta}
c^\dagger_{i\alpha} 
\boldsymbol{\sigma}_{\alpha\beta} c_{i\beta}
\)
is the spin-1/2 operator. 
The $t$-term permits the motion of holes while the $J$-term is the superexchange interaction. 

Even though having been studied for decades, the global phase diagram of the $t$-$J$ model on a square lattice is still controversial. Recently, the study of the $t$-$J$ model on a honeycomb lattice attracts a lot of interest since these two types of lattices share similar features -- both of them are bipartite lattices that stabilize the anti-ferromagnetic (AFM) order at half-filling. 
Remarkably, Grassmann tensor product state methods discovered 
the emergence of uniform $d+id$-wave SC order at very small doping, while the stripe and the $d$-wave SC orders coexist at relatively large doping in the honeycomb lattice $t$-$J$ model \cite{Gu_honeycomb,Yang}. In the weak coupling limit, the uniform $d+id$ superconductivity may also occur in the doped graphene systems revealed by various numerical methods, such as renormalized mean-field theory \cite{Schaffer-PRB}, quantum Monte Carlo \cite{Pathak,Ma,Ying,Jiang,Wang}, renormalization group \cite{Wang,Kiesel,Nandkishore}, and dynamical cluster approximation \cite{Xu}. Other pairing symmetries, such as $s$ and $p+ip$ waves, are also discovered \cite{Fay,Uchoa,Gu_pwave}. Apparently, the phase diagram of the $t$-$J$ model on a honeycomb lattice can provide deep insights into the key mechanism of high-$T_c$ superconductivity on a square lattice, which is closely related to realistic experimental materials.

Three decades ago, P. W. Anderson proposed the resonating valence bond (RVB) picture to account for the high-$T_c$ superconductivity \cite{Anderson}, and the corresponding spin-charge separation scenario is systematically studied in terms of slave-boson formalism for the $t$-$J$ model \cite{Lee_RMP}. Considering the AFM order as a characteristic feature in the small doping region, the slave-fermion formalism is a more appealing approach for small doping; the condensation of Schwinger bosons leads to a long-range magnetic order \cite{Auerbach-book}.
Nevertheless, previous slave-fermion studies of the square lattice $t$-$J$ model suggested the spiral order at finite doping instead of the AFM order \cite{Jayaprakash}, which contradicts numerical and experimental results. However, as there are two sites per unit cell on honeycomb lattice, mean-field solutions with AFM order is still possible at finite doping. We believe that such kind of translation invariant mean-field theory might be crucial for the emergence of uniform $d+id$-wave SC order.  

In this paper, we first use the state-of-the-art Grassmann tensor product state numerical method to obtain the global phase diagram of $t$-$J$ model on honeycomb lattice at finite doping. 
We adopt the slave-fermion mean-field approach and it gives rise to an AFM order on a honeycomb lattice that is consistent with our numerical results. At small doping, we find coexisting short-range AFM and FM correlations, and holon mobility. The mutual dependence of ferromagnetic (FM) correlation and holon mobility is also reproduced, which is consistent with the physics of the Nagaoka state in the infinite-$U$ Hubbard model near half-filling \cite{Nagaoka}. We further employ the functional field integral formalism to derive the effective interacting holon theory. The effective holon Hamiltonian contains repulsive interaction generated by exchanging the spinons and attractive interaction via gauge fluctuations. For a moderate $t/J$, the attractive interaction overcomes the repulsive one and leads to the SC orders. 
We stress that the attractive interaction 
is only possible in the systems with spin-charge separation. We hope that such a mechanism can shed light on the emergence of SC orders in the $t$-$J$ model on the square lattice as well and help us to understand the origin of high temperature superconductivity in cuprates.

The rest of the paper is organized as follows: Sec.~\ref{sec:tensor} presents the Grassmann tensor product state numerical results, including the phase diagram at finite doping of the $t$-$J$ model on a honeycomb lattice. The slave-fermion mean-field theory of the \(t\)-\(J\) model on honeycomb lattice is presented in Sec.~\ref{sec:MFT-and-gauge}. In Sec.~\ref{sec:mechanism-sc}, we derive the low-energy effective interacting holon theory by including the spinon-holon coupling and gauge fluctuations. 
Finally, we summarize the results and discuss the potential experimental detection of spin-charge separation phenomena in Sec.~\ref{discussion}.

\begin{figure}[t]
\begin{center}
\includegraphics[width=6cm]{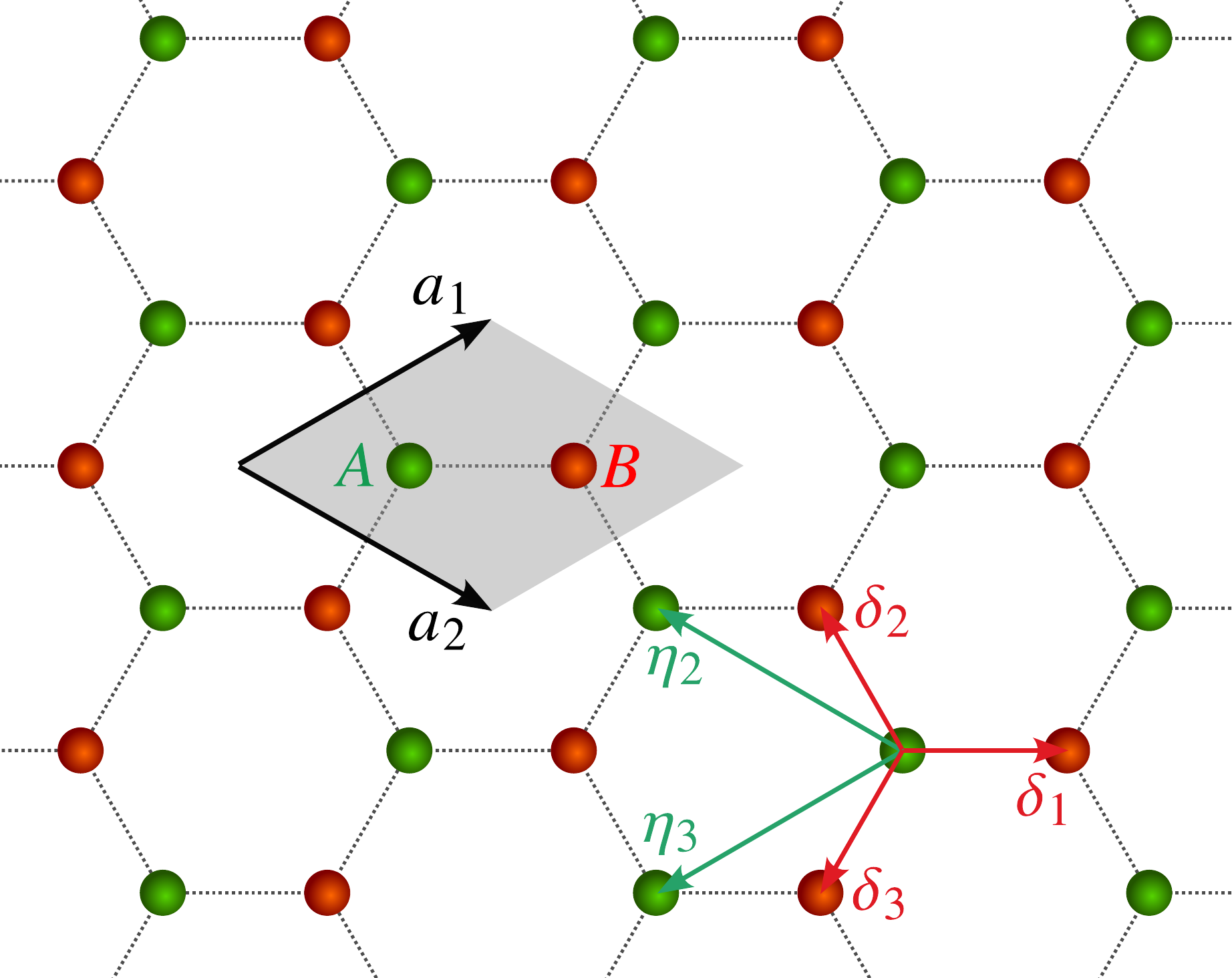}
\end{center}
\caption{The honeycomb lattice. The two triangular sub-lattices (with basis vectors \(\{a_n\}_{n=1,2}\)) are named \(A, B\) and colored in green, red respectively. A unit cell is shown by the shaded area. \(\{\delta_n\}_{n=1,2,3}\) are nearest neighbor vectors (with \(\delta_1 = a\hat{x}\), \(a\) being the nearest neighbor distance). \(\{\eta_n\}_{n=1,2,3}\) are vectors to the unit cells containing the nearest neighbors, which are related to \(\delta_n\) by \(\eta_n = \delta_n - a\hat{x}\) (in particular \(\eta_1 = 0\)).}
\label{fig:honeycomb}
\end{figure}

\begin{figure}[h]
\begin{center}
\includegraphics[width=8cm]{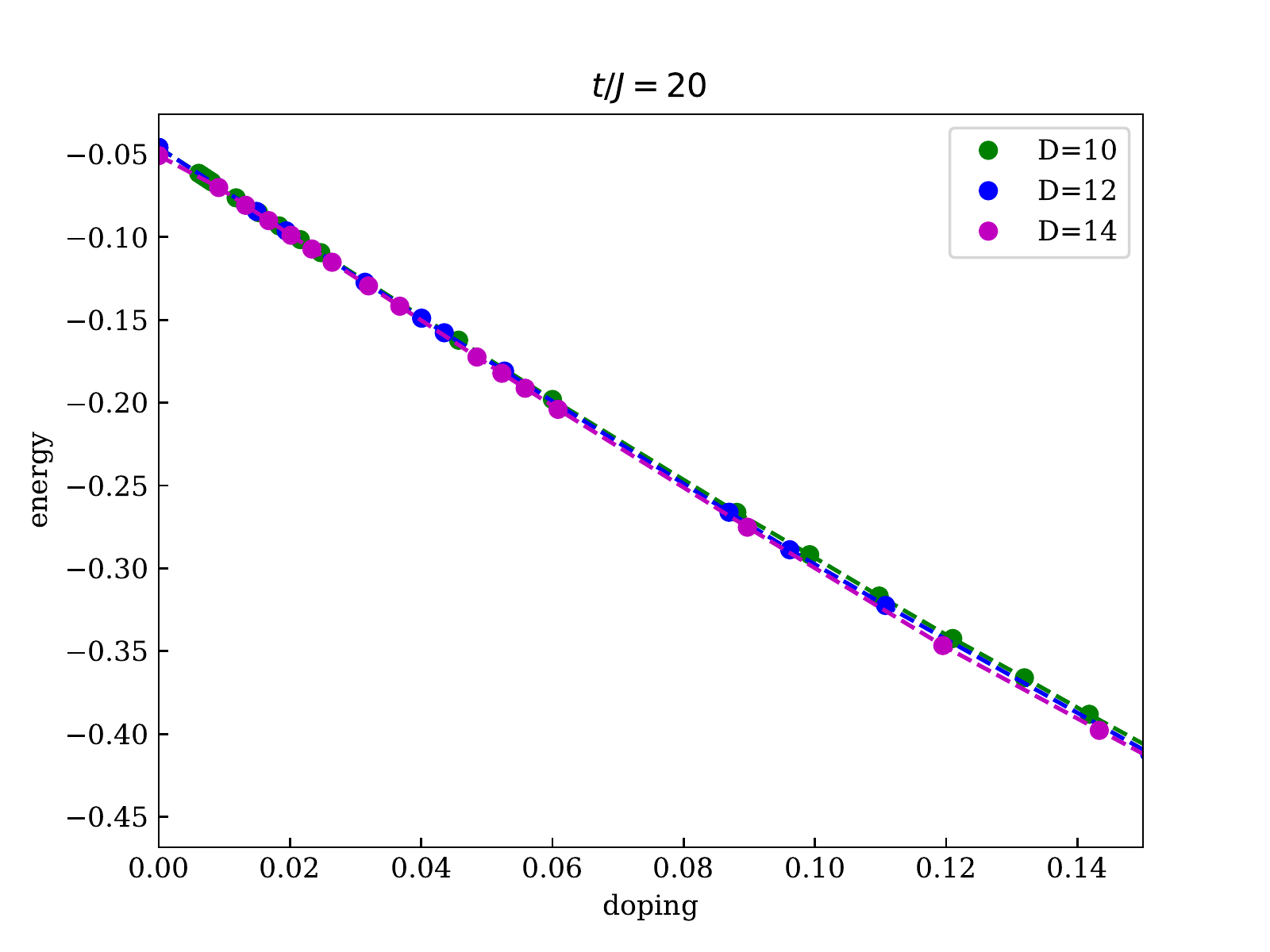}
\end{center}
\caption{Ground state energy versus doping at $t/J=20$.}
\label{fig:energy20}
\end{figure}

\begin{figure}[h]
    \def\svgwidth{2.0\columnwidth}
    \includegraphics[width=8cm]{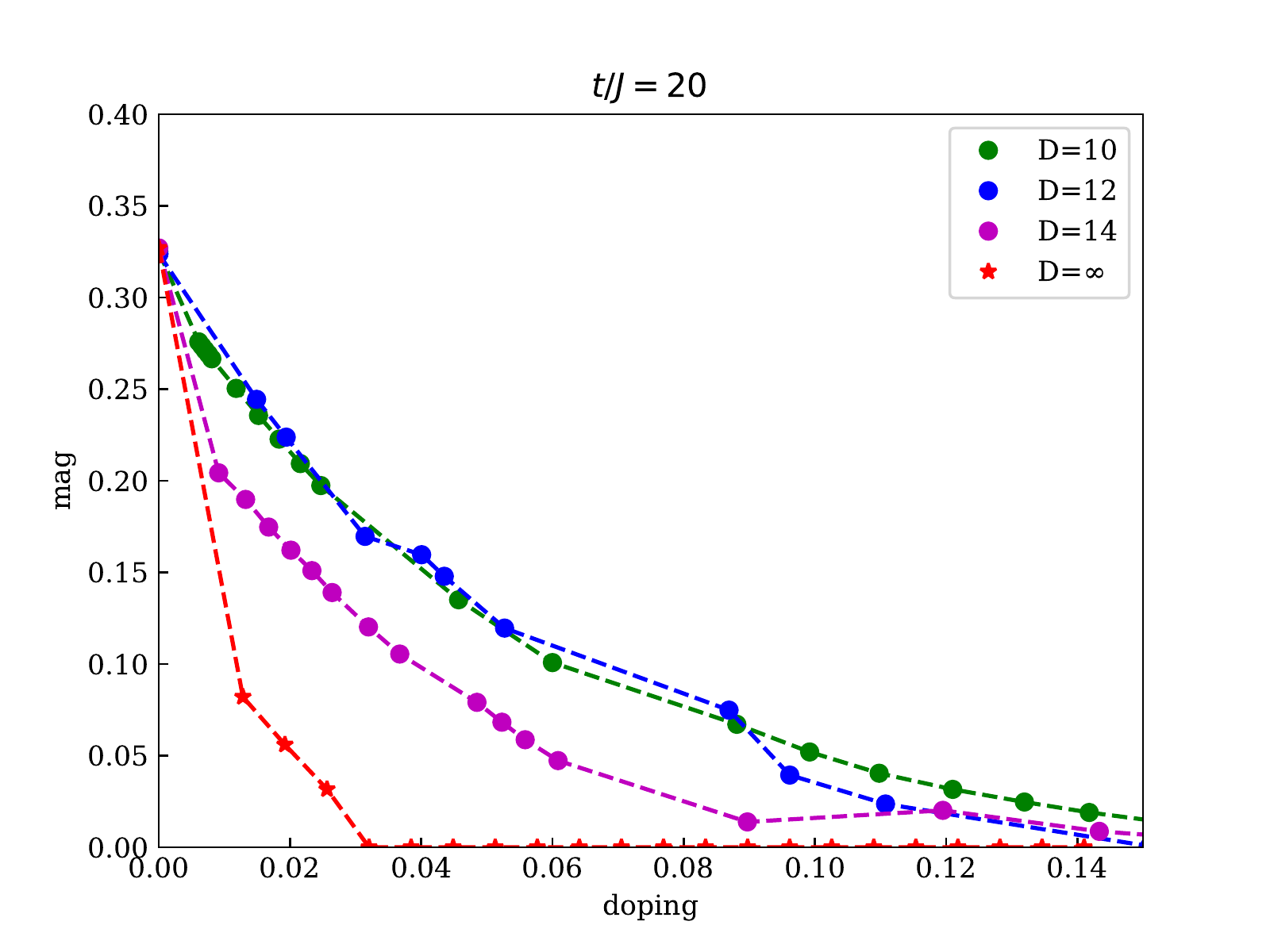}
    \includegraphics[width=8cm]{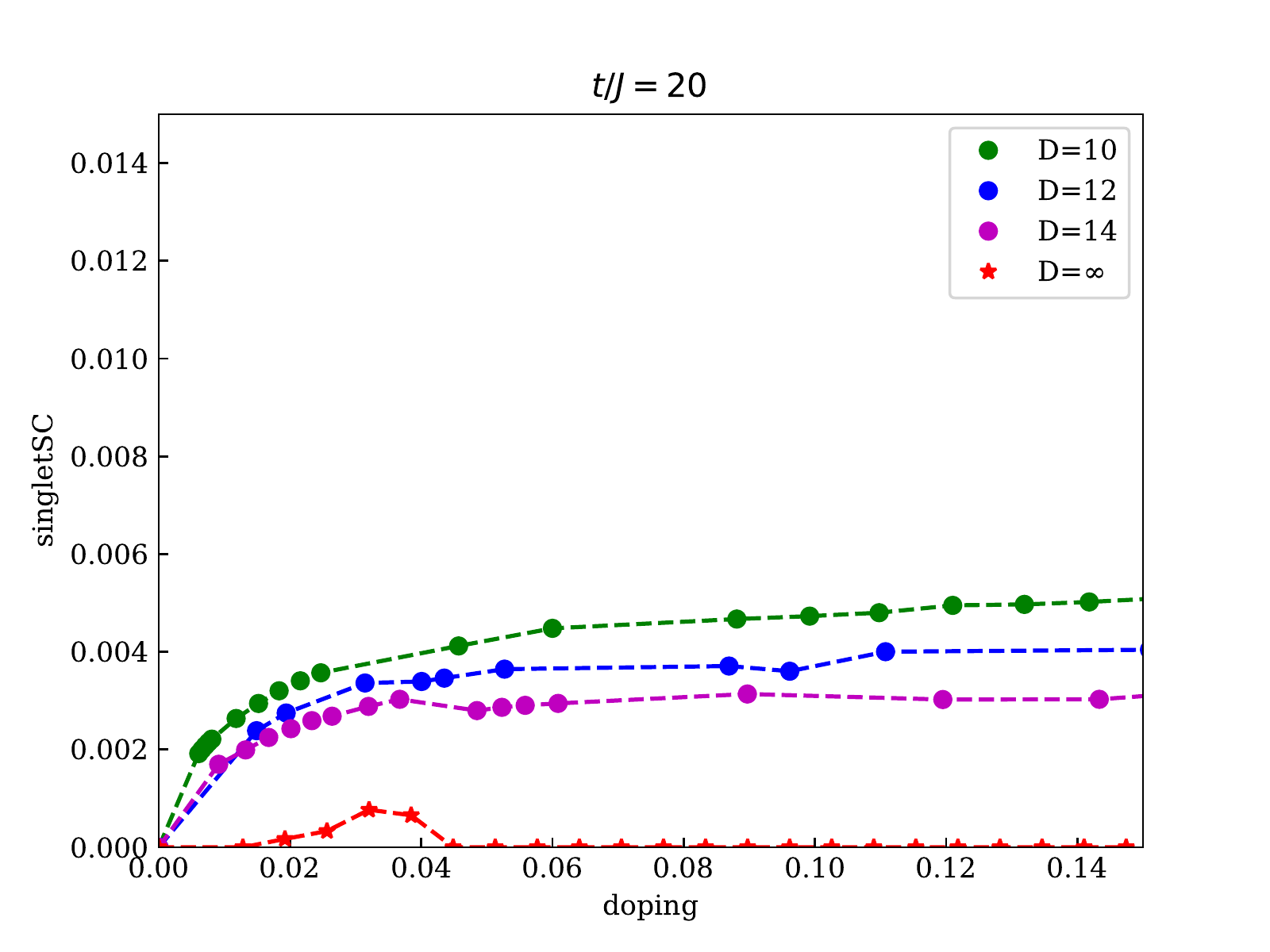}
    \includegraphics[width=8cm]{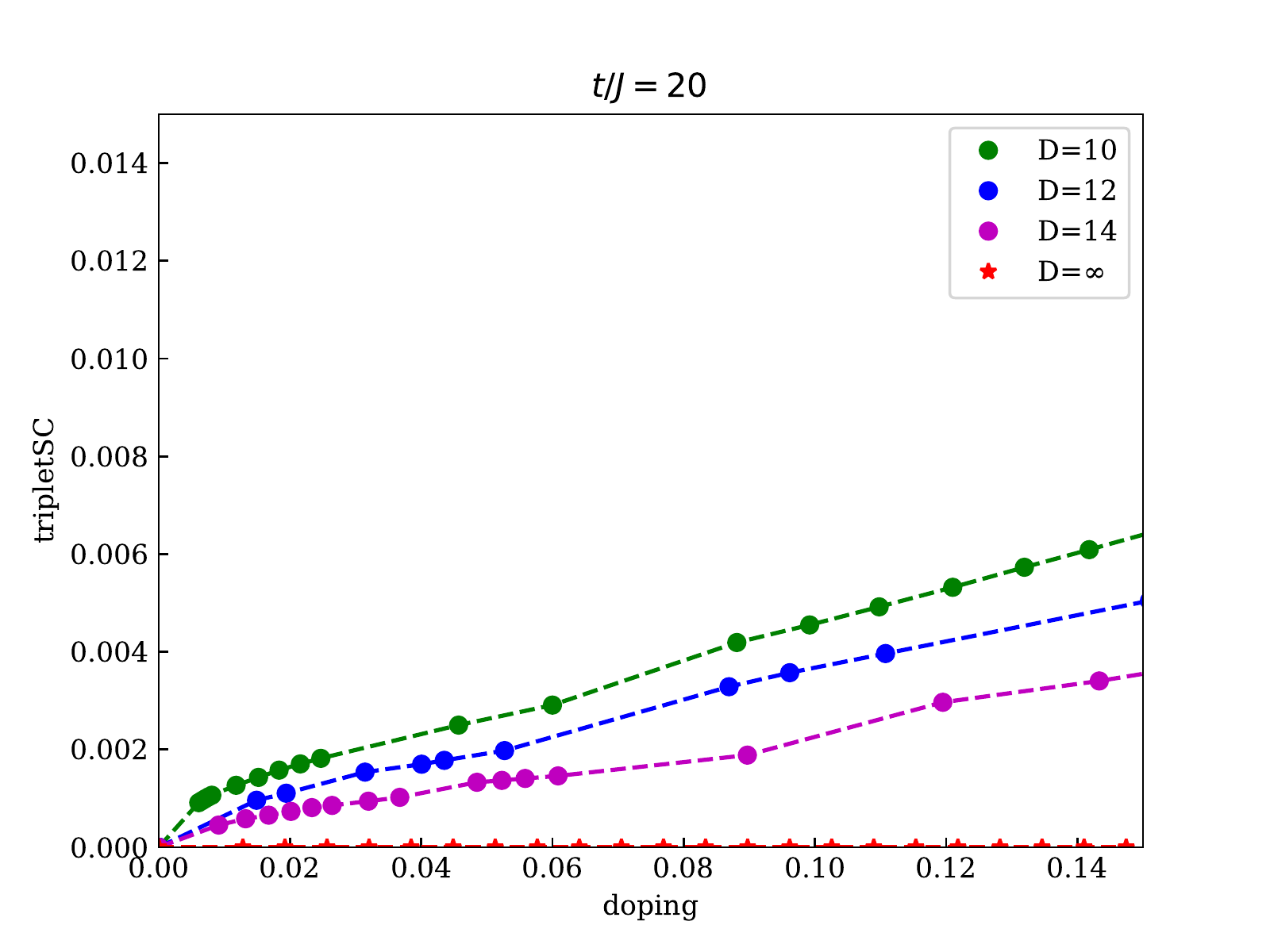}
    \caption{Staggered magnetization, amplitudes of singlet and triplet SC order parameters versus doping at $t/J=20$.}
    \label{fig:grassmann-results-main}
\end{figure}

\section{Grassmann Tensor Product State Numerical results}\label{sec:tensor}

We first use the state-of-the-art Grassmann tensor product state method to investigate the ground state phase diagram of the $t$-$J$ model on a honeycomb lattice. The ground state wave function is obtained via the so-called simple update (SU) scheme, and we also choose different bond dimensions $D=10,12,14$ to examine the $D$ dependence of the ground state energy and physical measurement such as the staggered magnetization $m = \sqrt{\langle S^x_i \rangle^2+\langle S^y_i \rangle^2+\langle S^z_i \rangle^2}$
(with \(\expect{\mathbf{S}_{i\in A}} = - \expect{\mathbf{S}_{i\in B}}\); \(A, B\) are the two sub-lattices of the honeycomb lattice), singlet superconductivity (SC) order parameters \(
\Delta^n_s = \langle  
c_{i\uparrow}c_{i+\delta_n \downarrow}
- c_{i\downarrow}c_{i+\delta_n \uparrow}
\rangle / \sqrt{2}
\)
and triplet SC order parameters \(
\boldsymbol{\Delta}^n_t 
= \langle \sum_{\alpha,\beta} 
c_{i\alpha} (i\sigma^y\boldsymbol{\sigma})_{\alpha\beta}
c_{i+\delta_n, \beta}\rangle / \sqrt{2}
\) (\(\{\delta_n\}_{n=1,2,3}\) are the 3 nearest neighbor vectors of a site \(i \in A\), as shown in Fig.~\ref{fig:honeycomb}).

The calculated singlet SC order parameters show \(d+id\) pairing, characterized by: 
\(
    \Delta^2_s / \Delta^1_s
    \simeq \Delta^3_s / \Delta^2_s
    \simeq \Delta^1_s / \Delta^3_s
    \simeq e^{2\pi i / 3}
\). 
Similar to a previous study of $t/J=3$ case\cite{Yang}, we find the ground state energy shows very good convergence for different choices of $D$, e.g., for the case with $t/J=20$ in Fig.~\ref{fig:energy20} . Nevertheless, very different from the $t/J=3$ case, we find that both magnetization and SC order parameters have a rather strong $D$ dependence, especially for larger doping. The calculated magnetization, magnitudes of singlet and triplet SC order parameters (averaged over the 3 nearest neighbors) at $t/J=20$ are shown in Fig.~\ref{fig:grassmann-results-main}.  After proper extrapolation to the infinite $D$ limit (see more details Appendix \ref{app:tensor}), we estimate that the stagger magnetization will vanish for $\delta>0.03$ and the singlet SC order parameter will vanish for $\delta>0.05$. The triplet SC order parameter has a rather strong $D$ dependence even at very small doping, and our simple extrapolation suggests it should be very close to zero at any doping.

\begin{figure}[t]
\begin{center}
\includegraphics[width=9cm]{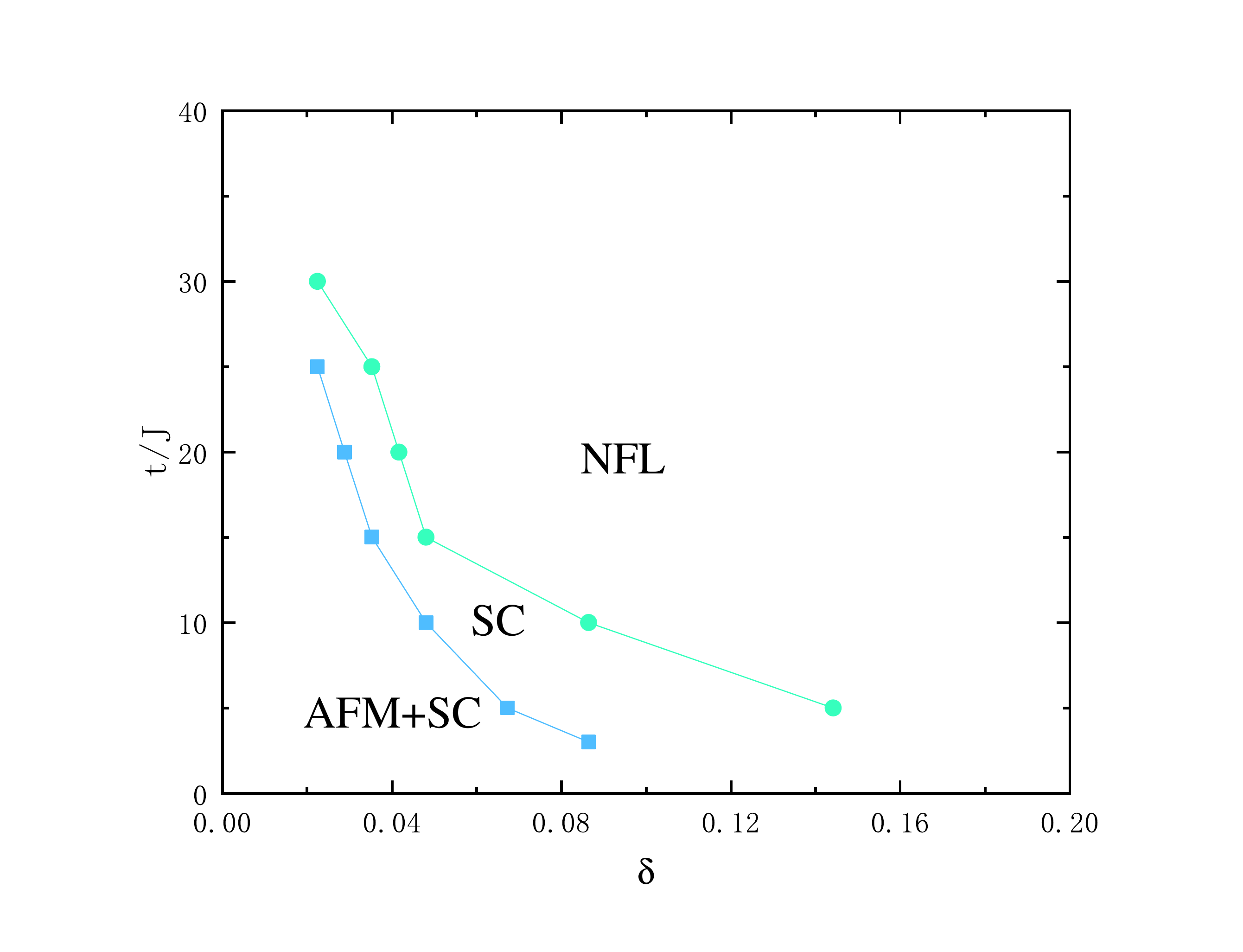}
\end{center}
\caption{The phase diagram of $t$-$J$ model on honeycomb lattice at finite doping. AFM, SC, and NFL denote antiferromagnetic order, superconducting order and non-Fermi liquid phase respectively.}
\label{fig:phase}
\end{figure}

By carefully analyzing the data at other $t/J$ ratios, we obtain the approximate phase diagram at finite doping as shown in Fig.~\ref{fig:phase}. Exactly at half-filling, the Hamiltonian of $t$-$J$ model reduces to the AFM Heisenberg model. Upon doping, the AFM order is suppressed and finally vanishes, while the $d+id$ SC orders emerge. The singlet SC order dominates and exists persistently to relatively large doping (for each fixed $t/J$). 
Interestingly, for large enough $t/J$ at each fixed doping, both AFM and SC orders eventually vanish. We conjecture that the system should enter a potentially non-Fermi liquid phase, and we shall explore the physical nature of such an exotic phase in the next two sections.
The raw data determining the phase diagram are given in Appendix~\ref{app:tensor}.

\section{Slave-fermion approach}\label{sec:MFT-and-gauge}

\subsection{Mean-field theory}\label{sec:MFT}

\begin{figure*}[t]
    \def\svgwidth{2.0\columnwidth}
    \includegraphics[height=7cm]{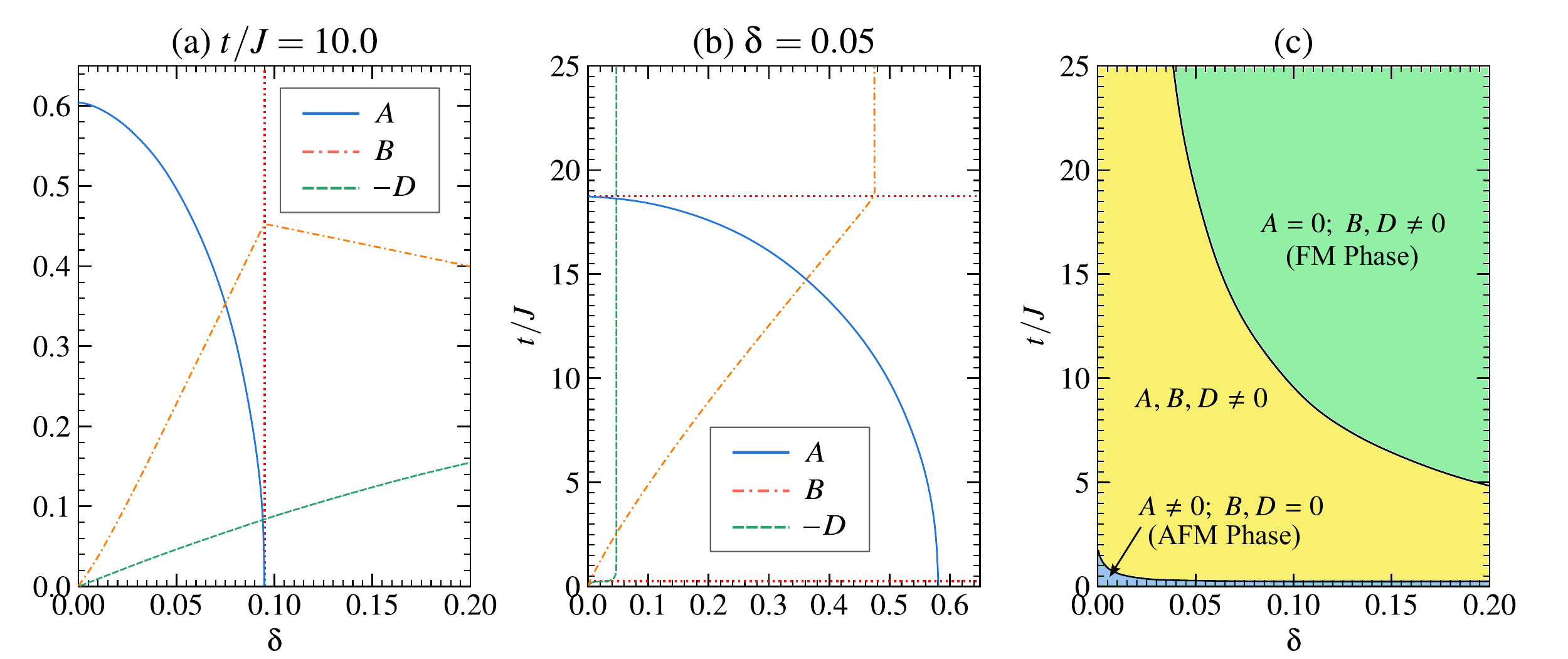}
    \caption{
        Solution to the mean-field equations for doping \(0 \le \delta \le 0.2\) on a system of \(64 \times 64\) unit cells at low temperature \(\beta = 800\). (a) The values of \(A, B\) and \(-D\) at fixed \(t/J=10\). A second-order phase transition (indicated by a vertical dotted line) happens at \(\delta = 0.09\). (b) The values of \(A, B\) and \(-D\) at fixed doping \(\delta=0.05\). Phase transitions (indicated by horizontal dotted lines) happens at \(t/J \approx 0.3\) (first-order) and \(18.7\) (second-order). (c) The mean-field phase diagram.
    }
    \label{fig:mean-field}
\end{figure*}

To gain more physical understanding for the result from numerical simulations, we first adopt the slave-fermion mean-field theory to analytically study the $t$-$J$ model on the honeycomb lattice. 
In the slave-fermion representation, the electron operator can be decomposed into fermionic holons and bosonic spinons in the no-double-occupancy subspace as:
\begin{equation}
    c^{\dagger}_{i\sigma} = b^{\dagger}_{i\sigma} h_i,
    \label{eq:slave-fermion}
\end{equation}
where $b_{i\sigma}$ are Schwinger-boson (spinon) operators and $h_i$ are slave-fermion (holon) operators, with no-double-occupancy constraint at each site \(i\):
\begin{equation}
    {\textstyle \sum_\sigma}
    b_{i\sigma}^{\dagger} b_{i\sigma} 
    + h^\dagger_i h_i
    = 1.
\end{equation}
In the no-double-occupancy subspace, the spin and electron density operators on each site can be expressed in spinons only:
\begin{equation}
    \mathbf{S}_i
    = \frac{1}{2} \sum_{\alpha,\beta} 
    b^\dagger_{i\alpha} 
    \boldsymbol{\sigma}_{\alpha\beta} b_{i\beta}
    , \quad
    n_i = \hat{n}^b_i
    \equiv \sum_\sigma 
    b^\dagger_{i\sigma} b_{i\sigma}.
\end{equation}
Then we can re-express the \(t\)-\(J\) model Eq.~(\ref{eq:tJ-ham-exact}) as:
\begin{align}
    H  = t \sum_{\expect{ij}, \sigma} (
        h^\dagger_j h_i 
        b^\dagger_{j\sigma} b_{i\sigma}
        + h.c.
    ) 
    - \frac{J}{2} \sum_{\expect{ij}} 
    \sum_{\sigma, \sigma'} \sigma \sigma' 
    b^\dagger_{j,-\sigma} b^\dagger_{i\sigma} 
    b_{i\sigma'} b_{j,-\sigma'} .
    \label{eq:tJ-slave-fermion}
\end{align}
Through out the whole paper, we label spin-up as \(+1\), and spin-down as \(-1\). In the summation over bonds, we choose $i$ to be on sub-lattice \(A\). To obtain the mean-field theory, the no-double-occupancy constraint is only imposed \emph{on average}:
\begin{equation}
    \expect{h^\dagger_i h_i} =\delta, \quad
    {\textstyle \sum_\sigma}
    \expect{b^{\dagger}_{i\sigma} b_{i\sigma}}
    = 1 - \delta
    \quad (0 \le \delta \le 1),
    \label{eq:doping}
\end{equation}
which is done by introducing chemical potentials $\lambda_i$ for spinons and $\mu_i$ for holons at each site. We further define the spinon pairing (RVB) operators and spinon hopping operators on each bond \(\expect{ij}\) as: 
\begin{equation}
    \hat{A}_{ij} = \frac{1}{2} \sum_\sigma
    \sigma b_{i\sigma} b_{j,-\sigma}, 
    \quad
    \hat{B}_{ij} = \frac{1}{2} \sum_\sigma
    b_{i\sigma}^{\dagger}b_{j\sigma}.
    \label{eq:AB-operator}
\end{equation}
Both are invariant under global $SU(2)$ spin rotations. The mean-field ansatz reads:
\begin{equation}
    A_{ij} = \expect{\hat{A}_{ij}}, \quad
    B_{ij} = \expect{\hat{B}_{ij}}, \quad
    D_{ij} = \expect{h^\dagger_i h_j}.
    \label{eq:abd}
\end{equation}
Here \(\expect{\,\cdot\,}\) refers to the mean-field average, and \(j\) is the nearest neighbor of \(i\). All these parameters are in general complex numbers. We choose \(i \in A\) to avoid redundancy, since \(A_{ji} = -A_{ij}\), \( B_{ji} = B^*_{ij}\), and \(D_{ji} = D^*_{ij}\). Physically, $A$ and $B$ represent the short-range AFM and FM correlations respectively, and $D$ represents the mobility of holons \cite{Jayaprakash}. In addition, we further require that:
\begin{equation}
\begin{aligned}
    \expect{b_{i\sigma} b_{j\sigma'}}
    &= \sigma A_{ij} 
    \delta_{\sigma,-\sigma'},
    \\
    \expect{b^\dagger_{i\sigma} b_{j\sigma'}}
    &= B_{ij} \delta_{\sigma \sigma'},
    \\
    \expect{b^\dagger_{i\sigma} b_{i\sigma'}}
    &= (1/2)(1-\delta) \delta_{\sigma \sigma'}.
\end{aligned}
\label{eq:abd2}
\end{equation}
The mean-field Hamiltonian becomes (see details in Appendix \ref{app:mf-decouple}):
\begingroup
\allowdisplaybreaks
\begin{align}
    H_{\text{MF}}
    &= H_h + H_b + H_0,
    \label{eq:mf-ham}
    \\
    H_h &= 2t \sum_{\expect{ij}} (
        h^\dagger_i h_j B^*_{ij} + h.c.
    ) - \sum_i \mu_i h^\dagger_i h_i,
    \label{eq:mf-ham-holon}
    \\
    H_b &= 2t \sum_{\expect{ij}} (
        D_{ij} \hat{B}^\dagger_{ij} + h.c.
    ) + \sum_{i,\sigma} \lambda_i b^\dagger_{i\sigma} b_{i\sigma}
    \nonumber \\ &\quad
    + J \sum_{\expect{ij}} (
        B^*_{ij} \hat{B}_{ij} 
        - 2 A^*_{ij} \hat{A}_{ij}
        + h.c.
    ),
    \label{eq:mf-ham-spinon}
    \\
    H_0 &= - 2t \sum_{\expect{ij}} (
        D_{ij} B^*_{ij} + h.c.
    ) 
    - J \sum_{\expect{ij}} (
        |B_{ij}|^2 - 2|A_{ij}|^2
    )
    \nonumber \\ &\quad
    + \sum_i [
        \mu_i \delta
        - \lambda_i(1-\delta)
    ]
    - \frac{3}{4} N J (1 - \delta)^2,
    \label{eq:mf-ham-number}
\end{align}
\endgroup
where \(N\) is the number of \emph{unit cells}. To keep the translation symmetry and $C_3$ rotation symmetry on the honeycomb lattice, we choose the \emph{real and uniform} ansatz:
\begin{gather}
A_{ij} = A, \ \ 
B_{ij} = B, \ \ 
D_{ij} = D, \ \ 
\mu_i = \mu, \ \ 
\lambda_i = \lambda.
\label{eq:mf-params}
\end{gather} 
Then we perform Fourier transform
\begin{align*}
    h_{i \in s} &= \frac{1}{\sqrt{N}} 
    \sum_k h^s_k e^{ik \cdot R_i},
    \\
    b_{i\in s, \sigma} &= \frac{1}{\sqrt{N}} 
    \sum_k b^s_{k\sigma} e^{ik \cdot R_i}.
\end{align*}
where \(s = A, B\) labels the sub-lattice, and \(R_i\) is the position of the unit cell containing site \(i\). Defining
\begin{gather}
    h_k = \begin{bmatrix}
        h^A_k \\[0.3em] h^B_k
    \end{bmatrix}, \quad
    b_{k\sigma} = \begin{bmatrix}
        b^A_{k\sigma} \\[0.3em] b^B_{k\sigma}
    \end{bmatrix},
    \\
    r = t B, \quad
    p = t D + J B / 2, \quad
    q = J A,
\end{gather}
the mean-field Hamiltonian in momentum space is:
\begin{align}
    H_h &= \sum_k h^\dagger_k H^h_k h_k,
    \label{eq:mf-holon}
    \\
    H_b &= \sum_k \begin{bmatrix}
        b^\dagger_{k\uparrow} & b_{-k\downarrow}
    \end{bmatrix} 
    H^b_k \begin{bmatrix}
        b_{k\uparrow} \\ b^\dagger_{-k\downarrow}
    \end{bmatrix}
    - 2 N \lambda,
    \label{eq:mf-spinon}
\end{align}
where the Hermitian matrices \(H^h_k, H^b_k\) are defined as: 
\begin{align}
    H^h_k &= \begin{bmatrix}
        -\mu & \xi^f_k \\
        \xi^{f*}_k & -\mu
    \end{bmatrix},
    \label{eq:mf-holon-mat}
    \\
    H^b_k &= \begin{bmatrix}
        \lambda & \xi^b_k & & -\Delta_k\\
        \xi^{b*}_k & \lambda & \Delta^*_k & \\
        & \Delta_k & \lambda & \xi^b_k \\
        -\Delta^*_k & & \xi^{b*}_k & \lambda
    \end{bmatrix},
    \label{eq:mf-spinon-mat}
    \\
    \xi^f_k &= 2r \gamma_k,
    \quad
    \xi^b_k = p \gamma_k, 
    \quad
    \Delta_k = q \gamma_k.
\end{align}
Here $\gamma_{k}=\sum_{\eta}e^{ik\cdot\eta}$ and $\eta$ sums over vectors to the 3 unit cells containing nearest neighbor sites (see Fig.~\ref{fig:honeycomb}). 
To diagonalize Eq.~(\ref{eq:mf-holon}), we define quasi-holons \(f^s_k\) (fermions) by a unitary transformation:
\begin{equation}
    \begin{bmatrix}
        h^A_k \\[0.1em] h^B_k
    \end{bmatrix} = U_k \begin{bmatrix}
        f^A_k \\[0.1em] f^B_k
    \end{bmatrix}
    , \quad
    U_k \equiv \frac{1}{\sqrt{2}} \begin{bmatrix}
        s_r & e^{i\theta_k} \\ e^{-i\theta_k} & -s_r
    \end{bmatrix},
\end{equation}
where \(\theta_k\) is the phase of \(\gamma_k\) (i.e. \(\gamma_k = |\gamma_k| e^{i\theta_k}\)), and \(s_r \equiv \sgn(r)\), with the definition \(\sgn(0) = 1\). To diagonalize Eq.~(\ref{eq:mf-spinon}), we define quasi-spinons \(\beta^s_{k\sigma}\) (bosons) by a Bogoliubov transformation
\begin{equation}
\begin{gathered}
    \begin{bmatrix}
        b_{k\uparrow} \\ b^\dagger_{-k\downarrow}
    \end{bmatrix} = W_k \begin{bmatrix}
        \beta_{k\uparrow} \\ \beta^\dagger_{-k\downarrow}
    \end{bmatrix},
    \\
    W_k = \frac{1}{\sqrt{2}} 
    {\small \begin{bmatrix}
    u_k s_q e^{i\theta_k} &
    u_k s_q e^{i\theta_k} &
    v_k s_q e^{i\theta_k} & 
    v_k s_q e^{i\theta_k}
    \\      
    u_k s_p s_q & -u_k s_p s_q &
    -v_k s_p s_q & v_k s_p s_q
    \\
    -v_k s_p e^{i\theta_k} &
    v_k s_p e^{i\theta_k} &
    u_k s_p e^{i\theta_k} &
    -u_k s_p e^{i\theta_k}
    \\
    v_k & v_k & u_k & u_k
    \end{bmatrix}},
\end{gathered}
\end{equation}
where \(s_p = \sgn(p)\), \(s_q = \sgn(q)\) and
\begin{gather}
    u_k = \bigg(
        \frac{\lambda + \chi_k}{2\chi_k}
    \bigg)^{1/2}, \quad
    v_k = \bigg(
        \frac{\lambda - \chi_k}{2\chi_k}
    \bigg)^{1/2},
    \\[0.1em]
    \chi_k = \sqrt{\lambda^2 - |\Delta_k|^2}.
\end{gather}
The diagonalized mean-field Hamiltonian reads:
\begin{align}
    H_\text{MF} 
    &= \sum_{k,s} 
    E^f_{ks} f^{s\dagger}_k f^s_k
    + \sum_{k,s,\sigma} E^b_{ks} 
    \beta^{s\dagger}_{k\sigma} \beta^s_{k\sigma}
    \nonumber \\ &\quad
    + H_0 - 2 N \lambda
    + \sum_{k,s} E^b_{ks},
    \label{eq:mf-ham-diag}
\end{align}
with quasi-holon \(f^s_k\) and quasi-spinon \(\beta^s_{k\sigma}\) dispersion:
\begin{align}
E_{k\pm}^{f} & =\pm |\xi^f_k| - \mu,
\label{eq:quasi-holon-disp}
\\
E_{k\pm}^{b} & =\pm |\xi^b_k| + \chi_k.
\label{eq:quasi-spinon-disp}
\end{align}
The self-consistency equations are derived by minimizing the mean-field free energy with respect to \(A\), \(B\), \(D\), \(\mu\) and \(\lambda\):
\begin{align}
    \delta &= \frac{1}{2N} \sum_k [
        n_f(E^f_{k+}) + n_f(E^f_{k-})
    ],
    \label{eq:mfeq-delta}
    \\
    1 - \delta
    &= \frac{1}{N} \sum_k \Big\{
        \frac{\lambda}{\chi_k} [
            1 + n_b(E^b_{k+}) 
            + n_b(E^b_{k-})
        ] - 1
    \Big\},
    \label{eq:mfeq-lambda}
    \\
    A &= \frac{q}{2\alpha N} \sum_k 
    \frac{|\gamma_k|^2}{\chi_k} [
        1 + n_b(E^b_{k+}) + n_b(E^b_{k-})
    ],
    \label{eq:mfeq-a}
    \\
    B &= \frac{\sgn(p)}{2\alpha N} 
    \sum_k |\gamma_k| [
        n_b(E^b_{k+}) - n_b(E^b_{k-}) 
    ],
    \label{eq:mfeq-b}
    \\
    D &= \frac{\sgn(r)}{2\alpha N} 
    \sum_k |\gamma_k| [
        n_f(E^f_{k+}) - n_f(E^f_{k-})
    ].
    \label{eq:mfeq-d}
\end{align}
Here \(\alpha = 3\) is the coordination number; $n_{b}$ and $n_{f}$ are usual boson and fermion distributions defined as:
\begin{equation}
    n_b(x) = \frac{1}{e^{\beta x}-1}, 
    \quad
    n_f(x) = \frac{1}{e^{\beta x}+1}.
\end{equation}

Fig.~\ref{fig:mean-field} shows the solution of the self-consistency equations on a honeycomb lattice of \(64 \times 64\) unit cells with periodic boundary condition, calculated at \(\beta = 800\), which is large enough to approximate the zero temperature behavior. Similar to previous slave-fermion mean-field theory on the square lattice \cite{Jayaprakash, Yoshioka_hubbard}, we obtain 3 phases in the small-doping region. When $t/J$ is small or at zero doping, there is an AFM phase in which \(A > 0\) and \(B,D=0\). As \(t/J\) increases above a certain value at finite doping, the system enters a phase with \(A,B,D \ne 0\) via a first-order transition, where short-range FM and AFM correlations coexist. Note that the minimum of the spinon dispersion \(E^b_{k-}\) is always at \(k = 0\); thus at zero temperature (when spinon condensation occurs), the spiral magnetic order is avoided, in contrast to the mean-field theory on the square lattice \cite{Jayaprakash}. The behavior of \(A,B,D\) in this phase is shown in Fig.~\ref{fig:mean-field} (a) and (b). In the upper-right part of the phase diagram there is an FM phase with \(A=0\) and \(B, D \ne 0\), which is separated from the \(A,B,D \ne 0\) phase by a second-order transition. This phase appears because of a large holon mobility, similar to the physics of Nagaoka states in the Hubbard model with infinite \(U\) and near half-filling. However, the FM phase is not observed from numerical simulations at large but finite $t/J$. We believe that the long-range FM order in this phase is an artifact of the mean-field theory, 
and the FM order is expected to be eliminated by gauge fluctuations induced by no-double-occupancy constraints. Below we shall discuss more details for the gauge structure for mean-field Hamiltonian. 

\subsection{Gauge structure of the mean-field Hamiltonian}\label{sec:MFT-gauge}

Since the mean field theory does not impose the no-double-occupancy constraint on each site exactly, we must consider the gauge fluctuations to restore such a nontrivial constraint for the low energy subspace. 
In the slave-fermion representation Eq.~(\ref{eq:slave-fermion}), the physical electron operator $c_{i\sigma}$ is invariant under the local $U(1)$ gauge transformation, i.e., for any site \(i \in A, B\):
\begin{equation}
h_i \to h_i e^{-i\theta_i}, \quad
b_{i\sigma} \to b_{i\sigma} e^{-i\theta_i}.
\label{eq:u1-trans}
\end{equation}
After projection into the no-double-occupancy physical space, each of the three phases from the mean field theory is characterized by its invariant gauge group (IGG) \cite{Wen-psg, Wang-psg}, which consists of local gauge transformations of the form Eq.~(\ref{eq:u1-trans}) that leave the ansatz unchanged. 
The FM phase ($A=0$ and $B,D\neq0$) has the usual $U(1)$ gauge structure Eq.~(\ref{eq:u1-trans}), with IGG \(= U(1)\). However, the AFM phase ($A\neq0$ and $B,D=0$) has a \emph{staggered} $U(1)$ gauge structure
\begin{equation}
    h_{i \in s} \rightarrow 
    h_{i} e^{-is\theta_{i}},
    \quad
    b_{i\in s, \sigma} \rightarrow 
    b_{i\sigma} e^{-is\theta_{i}}.
\end{equation}
where we interpret \(s = +1\) on sub-lattice \(A\), and \(s = -1\) on \(B\). This opposite sign on two sub-lattices indicate opposite gauge charges on $A$ and $B$ sub-lattices.
In the next section, we will show that such a staggered $U(1)$ gauge structure eventually leads to an effective attraction between holons on different sub-lattices, and is crucial for the emergence of superconductivity. 

Finally, the $A,B,D\neq0$ phase has a \(\Z_2\) gauge structure
\begin{equation}
h_i \to s_i h_i, \quad
b_{i\sigma} \to s_i b_{i\sigma}.
\end{equation}
where \(s_i = \pm 1\) is a sign factor depending on the site \(i\), and the corresponding IGG is \(\Z_2\). In this phase, besides the holon and spinon quasi-particle excitations, there is another excitation corresponding to the $\Z_2$ flux of $\Z_2$ gauge fields, dubbed as vison \cite{Senthil_PRB}. 
The $\Z_2$ gauge fluctuation corresponds to phase fluctuations of order parameters. It is coupled to the holon and spinon in a $\Z_2$ gauge invariant way and has a finite gap(if we assume the spinon does not condense), which does not confine the holon and spinon; hence it does not change the qualitative results obtained in this section. 
The existence of the vison excitation is an indication of fractionalization in topological order \cite{Senthil_PRL}, and the corresponding topological ground state degeneracy can be detected by DMRG calculations on cylinder or torus. 


\section{Mechanism of superconductivity}\label{sec:mechanism-sc}
To understand the key mechanism of SC order in the $t$-$J$ model, we need to investigate the effective interactions among holons. We begin with the AFM phase with $A\neq0$ and $B,D=0$, and restrict the discussion to the small doping region (\(\delta \ll 1\)). After we establish a low energy effective field theory to understand the interactions among holons in this phase, we shall try to understand the whole phase diagram later.

\subsection{Repulsive holon interaction at large \textit{t} / \textit{J}}\label{sec:eff-holon-spinon}

 The first step beyond the mean-field theory is to add the spinon-holon interaction from Eq.~(\ref{eq:tJ-slave-fermion}), and the total Hamiltonian reads:
\begin{align}
    H
    &= H_h + H_b + H_{\text{int}} ,
    \\
    H_{\text{int}}
    &= t \sum_{\expect{ij},\sigma} (
        h^\dagger_j h_i 
        b_{i\sigma}^{\dagger} b_{j\sigma}
        +h.c.
    ).
    \label{eq:spinon-holon-int-ham}
\end{align}
To study the effective interaction between holons induced by the spinons, we shall use the coherent state path integral formalism. The spinon (or holon) operators are replaced by the corresponding complex (or Grassmann) numbers, with the following notations:
\begin{equation*}
    (b_{i\sigma}, b_{i\sigma}^\dagger) 
    \to (b_{i\sigma}, \bar{b}_{i\sigma})
    , \quad
    (h_i, h^\dagger_i) \to (h_i, \bar{h}_i).
\end{equation*} 
The partition function of fermion-boson coupled theory in the functional field integral representation is (number terms are dropped)
\begin{align}
Z &= \int D\bar{h}\,Dh\,D\bar{b}\,Db\,
e^{-S[\bar{h},h;\bar{b},b]},
\\
S &= S_h + S_b + S_\text{int},
\label{eq:tJ-mf_action}
\\
S_h
&= \int_{0}^{\beta} d\tau \Big[
\sum_i \bar{h}_i \partial_{\tau} h_i + H_h(\tau)
\Big],
\label{eq:holon-action}
\\
S_b
&= \int_{0}^{\beta} d\tau \Big[
\sum_{i,\sigma} \bar{b}_{i\sigma} \partial_{\tau} b_{i\sigma}
+ H_b(\tau)
\Big],
\label{eq:spinon-action}
\\
S_\text{int}
&= \int_{0}^{\beta} d\tau \, H_\text{int}(\tau).
\label{eq:spinon-holon-int-action}
\end{align}
Now $h$ and $b$ are Grassmann and complex variables, denoting the holon and spinon degrees of freedom in the coherent state representation. We then integrate out spinons to derive the effective holon Hamiltonian:
\begin{align}
    H^{\text{eff}}_h
    &= H_h + H^{\text{eff}}_\text{int},
    \\
    H^{\text{eff}}_\text{int}
    &= \frac{1}{N}
    {\sum_{\{k\}}}' V_{k_1 k_2 k'_1 k'_2}
    h^{B\dagger}_{k_1} h^{A\dagger}_{k_2} 
    h^A_{k'_1} h^B_{k'_2},
    \label{eq:eff-holon-int-spinon}
\end{align}
where ${\sum_{\{k\}}}'$ is summation over all momentum with conservation (\(k_1+k_2 = k'_1+k'_2\)). In the small doping limit, \(B, D \approx 0\), and the interaction strength is (see Appendix \ref{app:eff-holon})
\begin{align}
    & V_{k_1 k_2 k'_1 k'_2}
    = -\frac{t^2}{2 N} \sum_p \bigg\{ 
        \frac{1}{
            \chi_p \chi_{p'} (\chi_p + \chi_{p'})
        } 
        \nonumber \\ & \times
        \Big[
            J_b^2 (
                \gamma^*_p \gamma^*_{p'}
                \gamma^*_{k'_1-p} \gamma_{k_2+p'}
                + \gamma_p \gamma_{p'}
                \gamma^*_{k'_1+p} \gamma_{k_2-p}
            ) \nonumber \\ &\quad
            + (\chi_p \chi_{p'} - \lambda^2)(
                \gamma^*_{k'_1+p} \gamma_{k_2+p'}
                + \gamma^*_{k'_1-p} \gamma_{k_2-p}
            )
        \Big]
    \bigg\},
    \label{eq:eff-holon-int-v}
\end{align}
where $p'=p+k_1-k'_1$, and \(J_b = J A\). We have already take the static limit approximation to extract the frequency-independent effective interaction, which is mediated by exchanging two spinons with momentum $p$ and $p'$. As the minimum of the spinon dispersion is at $k = 0$, the dominant contributions to the interaction coefficient in the low energy come from $p \sim p'\sim 0$, and we replace \(\gamma_{k'_1-p} \to \gamma_{k'_1}\) and \(\gamma_{k_2+p'} \to \gamma_{k_2}\) in Eq.~(\ref{eq:eff-holon-int-v}), leading to
\begin{align}
    & V_{k_1 k_2 k'_1 k'_2}
    \approx -t^2 \gamma_{-k'_1} \gamma_{k_2} f(k_1 - k'_1),
    \\[0.5em]
    & f(k) \equiv \frac{1}{N} \sum_p 
    \left. \frac{
        J_b^2 \operatorname{Re}(\gamma_p \gamma_{p'})
        + (\chi_p \chi_{p'} - \lambda^2)
    }{
        \chi_p \chi_{p'} (\chi_p + \chi_{p'})
    } \right|_{p'=p+k}.
    \label{eq:eff-holon-int-v-approx}
\end{align}
In the small doping limit, the holon momenta \(k_1, k'_1\) are both expected to be small. Let us then calculate:
\begin{align}
    f(0) &= \frac{1}{N} \sum_p 
    \frac{1}{2 \chi_p^3} [
        J_b^2 \operatorname{Re}(\gamma_p^2)
        + (\chi_p^2 - \lambda^2)
    ] \nonumber \\
    &= -\frac{1}{N} \sum_p \frac{J_b^2}{\chi_p^3}
    (\operatorname{Im} \gamma_p)^2 
    < 0,
    \label{eq:fk-estim}
\end{align}
which is \emph{negative}. Assuming a weak momentum dependence of \(f(k)\), we can approximate \(f(k)\) by a \emph{negative} constant \(C/J_b\) (with \(C < 0\)), an inverse Fourier transform of Eq.~(\ref{eq:eff-holon-int-spinon}) leads to a simple interaction of holons in real space:
\begin{align}
    H^{\text{eff}}_{\text{int}}
    &= - \frac{t^2 C}{J_b N} {\sum_{\{k\}}}'
    \gamma_{-k'_1} \gamma_{k_2}
    h^{B\dagger}_{k_1} h^{A\dagger}_{k_2} 
    h^A_{k'_1} h^B_{k'_2}
    \nonumber \\
    &= - \frac{t^2 C}{J_b} 
    \sum_{j \in B} \sum_{\delta,\delta'}
    h^{\dagger}_j h^{\dagger}_{j-\delta'} 
    h_{j-\delta} h_j ,
\end{align}
where \(\delta, \delta'\) sum over the nearest neighbor vectors \(\{\delta_i\}_{i=1}^3\) (see Fig.~\ref{fig:honeycomb}).
In particular, \(H^{\text{eff}}_{\text{int}}\) contains the following nearest-neighbor interaction terms (when $\delta = \delta'$)
\begin{equation}
    - \frac{t^2 C}{J_b} 
    \sum_{j \in B} \sum_\delta
    h^{\dagger}_j h^{\dagger}_{j-\delta} 
    h_{j-\delta} h_j
    = -\frac{t^2C}{J_b} \sum_{\expect{ij}} 
    h^\dagger_i h_i h^\dagger_j h_j .
\end{equation}
$C < 0$ implies that the nearest neighbor interaction between holons induced by exchanging spinons is \emph{repulsive}.

\subsection{\textit{U}(1) gauge fluctuations in the \textit{t}-\textit{J} Model}\label{u(1)-gauge-field}

Next we impose the no-double-occupancy constraint on each site exactly. 
This is done by a delta-function in the partition function:
\begin{align}
    &\prod_j \delta \Big(
        \bar{h}_j h_j
        + \sum_\sigma \bar{b}_{j\sigma} b_{j\sigma} - 1
    \Big) 
    \nonumber \\
    &= \int D\lambda \, \exp \bigg[{
        -i \sum_j \lambda_j \Big(
            \bar{h}_j h_j
            + \sum_\sigma \bar{b}_{j\sigma} b_{j\sigma} - 1
        \Big)
    }\bigg],
\end{align}
where \(\{\lambda_j\}\) are real Lagrange multipliers, and \(D\lambda \equiv \prod_j d\lambda_j\). With the Hamiltonian Eq.~(\ref{eq:tJ-slave-fermion}), the coherent state path integral representation of the partition function is:
\begin{align*}
    & Z = \int D\bar{b} \, Db \, 
    D\bar{h} \, Dh \, D\lambda \, 
    e^{
        - \int_0^\beta d\tau \, L
    },
    \\
    & L = \sum_j \Big[
        \bar{h}_j 
        (\partial_\tau + i\lambda_j) h_j
        + \sum_\sigma  \bar{b}_{j\sigma} 
        (\partial_\tau + i\lambda_j) b_{j\sigma}
        - i \lambda_j
    \Big]
    \\ &
    + t\sum_{\expect{ij},\sigma} (
        \bar{h}_j h_i \bar{b}_{i\sigma} b_{j\sigma} 
        + h.c.
    )
    - \frac{J}{2} \sum_{\substack{
        \expect{ij} \\ \sigma,\sigma'
    }}
    \sigma \sigma'
    \Bar{b}_{j,-\sigma} \Bar{b}_{i\sigma}
    b_{i\sigma'} b_{j,-\sigma'}.
\end{align*}
The \(J\)-term is then decoupled by a Hubbard-Stratonovich transformation in the AFM channel (the FM channel is unimportant in the small-doping region). Introducing auxiliary complex fields \(\Delta_{ij}\) (\(i \in A\), \(j \in B\)) and their conjugate \(\bar{\Delta}_{ij}\) on each bond, we arrive at
\begin{align}
    Z &= \int D\bar{b} \, Db \, D\bar{h} \, Dh \, 
    D\bar{\Delta} \, D\Delta \, D\lambda \, 
    e^{- \int_0^\beta d\tau \, L},
    \\
    L &= \sum_j \Big[
        \bar{h}_j 
        (\partial_\tau + i\lambda_j) h_j
        + \sum_\sigma  \bar{b}_{j\sigma} 
        (\partial_\tau + i\lambda_j) b_{j\sigma}
    \Big]
    \nonumber \\ &\quad
    - i \sum_j \lambda_j
    + t\sum_{\expect{ij},\sigma} (
        \bar{h}_j h_i \bar{b}_{i\sigma} b_{j\sigma} 
        + h.c. 
    )
    \nonumber \\ &\quad 
    + \frac{J}{2} \sum_{\expect{ij}} \Big[
        |\Delta_{ij}|^2 
        - \sum_\sigma \sigma (
            \Bar{\Delta}_{ij} 
            b_{i\sigma} b_{j,-\sigma} + h.c.
        )
    \Big],
    \label{eq:tJ-lag-general}
\end{align}
where \(D\Delta \equiv \prod_{\expect{ij}} d\Delta_{ij}\) (with \(i \in A\)), and \(|\Delta_{ij}|^2 \equiv \bar{\Delta}_{ij} \Delta_{ij}\). The Lagrangian Eq.~(\ref{eq:tJ-lag-general}) is invariant (up to a total \(\tau\)-derivative term) under the staggered $U(1)$ gauge transformation: 
\begin{equation}
\begin{aligned}
    b_{i \in s, \sigma} &\to 
    b_{i\sigma} e^{-is\theta_i},
    \\
    h_{i\in s} &\to 
    h_i e^{-is\theta_i},
\end{aligned} \quad \begin{aligned}
    \Delta_{ij}
    &\to \Delta_{ij} e^{
        -i(\theta_i - \theta_j)
    } ,
    \\
    \lambda_{i \in s}
    &\to \lambda_i + s \, \partial_\tau \theta_i.
\end{aligned}
\label{eq:gaugetrans-spinon}
\end{equation}
The transformation can be reformulated using $U(1)$ gauge fields: define the vector potential \(\mathbf{A}\) (on bonds) and the scalar potential \(A_\tau\) (on sites)
\begin{equation}
    \theta_i - \theta_j
    = x_{ij} \cdot \mathbf{A}(i,j)
    , \quad
    A_\tau(i) = \partial_\tau \theta_i.
\end{equation}
Here \(x_{ij}\) is the vector from site \(j\) to site \(i\). The gauge transformation of \(\Delta, \lambda\) are then
\begin{equation}
\begin{aligned}
    \Delta_{ij}
    &\to \Delta_{ij} e^{
        - i x_{ij} \cdot \mathbf{A}(i,j)
    },
    \\
    \lambda_{i \in s}
    &\to \lambda_i + s A_\tau(i).
\end{aligned}
\end{equation}
In the partition function, we keep configurations of \(\Delta_{ij}\), \(\bar{\Delta}_{ij}\), \(\lambda_i\) at the saddle point with fluctuations of the gauge field:
\begin{equation}
\begin{aligned}
    \Delta_{i,i+\delta}
    &\to \Delta e^{
        i \delta \cdot \mathbf{A}(i,i+\delta)
    }
    & (i \in A),
    \\
    \bar{\Delta}_{i,i+\delta}
    &\to \Delta e^{
        -i \delta \cdot \mathbf{A}(i,i+\delta)
    }
    & (i \in A),
    \\[0.2em]
    \lambda_{i \in s}
    &\to -i \lambda + s A_\tau(i),
\end{aligned}
\label{eq:saddle-gauge}
\end{equation}
where \(\Delta = 2A\) and \(\lambda\) are real solutions from the mean-field theory. The fluctuation of the norm of \(\Delta_{ij}\) is ignored. Then
\begin{align}
    Z &= \int D\bar{b}\, Db\, D\bar{h}\, Dh\, DA\,
    e^{- \int_0^\beta d\tau \, L},
    \\
    L &= L_h + L_\text{heis} + L_t + L_0,
\end{align}
with each part in the Lagrangian given by
\begin{align}
    & \ \ \ 
    L_h = \sum_s \sum_{i\in s} \Big[
        \bar{h}_i (\partial_\tau + i s A_\tau(i)) h_i
        + \lambda \bar{h}_i h_i
    \Big],
    \label{eq:l-holon}
    \\
    & L_\text{heis} 
    = \sum_{s} \sum_{i \in s} \Big[
        \bar{b}_{i\sigma} 
        (\partial_\tau + i s A_\tau(i)) b_{i\sigma}
        + \lambda \bar{b}_{i\sigma} b_{i\sigma}
    \Big]
    \nonumber \\ &
    - \frac{J \Delta}{2} 
    \sum_{i\in A} \sum_{\delta,\sigma} 
    \sigma \Big[
        e^{i \delta \cdot \mathbf{A}(i,i+\delta)}
        \bar{b}_{i\sigma} \bar{b}_{i+\delta,-\sigma}
        + h.c.
    \Big],
    \label{eq:l-heis}
    \\
    & \ \ \ \ 
    L_t = t \sum_{i \in A} \sum_{\delta,\sigma}
    \bar{h}_{i+\delta} h_i 
    \bar{b}_{i\sigma} b_{i+\delta,\sigma}
    + h.c.,
    \label{eq:l-spinon-holon}
    \\
    & \ \ \ \ 
    L_0 = - \sum_s \sum_{i \in s} 
    (\lambda + i s A_\tau(i))
    + \frac{\alpha J N}{2} \Delta^2.
    \label{eq:l0}
\end{align}
Here $s$ sums over the two sub-lattices $A, B$. We also used the fact that the nearest neighbors of a site \(i \in A\) are at positions \(\{i + \delta_l\}_{l=1}^3\) on the honeycomb lattice. 

We then take the continuum limit by setting the nearest neighbor distance \(a \to 0\) and system size \(\to \infty\). A summation over sites in one sub-lattice is replaced by an integral over space:
\begin{equation}
    \sum_{i\in A} \text{ or } \sum_{j \in B} 
    \to \int \frac{d^2x}{2\Omega}, \quad
    \Omega = \frac{3\sqrt{3}}{4} a^2.
\end{equation}
Here \(a\) is the nearest neighbor distance, and \(2\Omega\) is the area of a unit cell. The continuum fields are
\begin{equation}
\begin{aligned}
    h_{i \in s} &\to h^s(x), &
    b_{i \in s, \sigma} &\to b^s_\sigma(x),
    \\
    A_\tau(i) &\to A_\tau(x), &
    \mathbf{A}(i,i+\delta) &\to
    \mathbf{A}(x+\tfrac{\delta}{2}).
\end{aligned}
\end{equation}
where \(x\) is the position of the site \(i\). For convenience, we redefine spinon-holon coupling \(\alpha t \to t\), and introduce \(Q \equiv \alpha J \Delta / 8\). The continuum Lagrangian is (see details in Appendix \ref{app:continuum})
\begin{align}
    & L_h = \int\frac{d^2x}{2\Omega} 
    \sum_s \Big[
        \bar{h}^s (\partial_\tau + i s A_\tau) h^s
        + \lambda \bar{h}^s h^s
    \Big],
    \label{eq:l-holon-con}
    \\
    & L_\text{heis}
    = \int\frac{d^2x}{2\Omega}
    \sum_{s,\sigma} \Big[
        \bar{b}^s_\sigma 
        (\partial_\tau + i s A_\tau) b^s_\sigma
        + \lambda \bar{b}^s_\sigma b^s_\sigma 
    \Big]
    \nonumber \\ &\ 
    + Q a^2 \int\frac{d^2x}{2\Omega} 
    \sum_\sigma \sigma \Big[
        (\nabla - i\mathbf{A}) \bar{b}^A_\sigma 
        \cdot
        (\nabla + i\mathbf{A}) \bar{b}^B_{-\sigma} 
        + h.c.
    \Big]
    \nonumber \\ &\ 
    - 4Q \int\frac{d^2x}{2\Omega} 
    \sum_\sigma \sigma \Big[
        \bar{b}^A_\sigma  \bar{b}^B_{-\sigma}
        + h.c.
    \Big],
    \label{eq:l-heis-con}
    \\
    & L_t = 
    t \int\frac{d^2x}{2\Omega} 
    \sum_\sigma \Big[
        \bar{h}^B h^A \bar{b}^A_\sigma b^B_\sigma
        + h.c.
    \Big],
    \label{eq:l-t-con}
    \\
    & L_0 = \int\frac{d^2x}{2\Omega} \Big[
        - 2\lambda
        + \frac{\alpha J}{2} \Delta^2
    \Big].
    \label{eq:l-0-con}
\end{align}
Define \(A_\mu = (A_\tau, \mathbf{A})\) and \(\partial_\mu = (\partial_\tau, \nabla)\); the gauge transformation in continuum limit is
\begin{equation}
\begin{aligned}
    b^s_\sigma(x) &\to b^s_\sigma(x) e^{-i s \theta(x)},
    \\
    h^s(x) &\to h^s(x) e^{-i s \theta(x)},
    \\
    A_\mu(x) &\to A_\mu(x) + \partial_\mu \theta(x).
\end{aligned}
\label{eq:gaugetrans-continuum}
\end{equation}

\subsection{Effective theory of holons}\label{effective-theory-of-holons}

Following Ref.~\citenum{read-sachdev}, we combine the Schwinger bosons into a low energy field \(z\) and a high energy field \(\pi\):
\begin{equation}
z_\sigma = 
(b^A_\sigma + \sigma \bar{b}^B_{-\sigma}) / 2, 
\quad
\pi_\sigma = 
(b^A_\sigma - \sigma \bar{b}^B_{-\sigma}) / 2.
\label{eq:z-pi-field}
\end{equation}
From Eq.~(\ref{eq:gaugetrans-continuum}), the gauge transformation of \(z,\pi\) fields are
\begin{equation}
    z_\sigma \to z_\sigma e^{-i\theta}, \quad
    \pi_\sigma \to \pi_\sigma e^{-i\theta}.
\end{equation}
We then express the Lagrangian \(L\) in terms of \(z, \pi\) fields. The transformed \(L_\text{heis}\) (Eq.~(\ref{eq:l-heis-con})) is
\begin{align}
    & L_\text{heis} 
    = \int\frac{d^2x}{\Omega} 
    \sum_\sigma \bar{z}_\sigma [
        (\lambda - 4Q)
        - Q a^2 (\nabla + i\mathbf{A})^2
    ] z_\sigma
    \nonumber \\ &
    + \int\frac{d^2x}{\Omega} 
    \sum_\sigma [
        \bar{\pi}_\sigma (\partial_\tau + iA_\tau) z_\sigma
        - \pi_\sigma (\partial_\tau - iA_\tau) \bar{z}_\sigma
    ]
    \nonumber \\ &
    + \int\frac{d^2x}{\Omega} 
    \sum_\sigma \bar{\pi}_\sigma [
        (\lambda + 4Q)
        + Q a^2 (\nabla + i\mathbf{A})^2
    ] \pi_\sigma,
\end{align}
and the spinon-holon interaction \(L_t\) (Eq.~(\ref{eq:l-t-con})) becomes
\begin{equation}
L_t = (-t) \int\frac{d^2x}{\Omega} 
\sum_\sigma \sigma (
    \bar{\pi}_\sigma  \bar{h}^B h^A \bar{z}_{-\sigma} 
    + h.c.
).
\end{equation}

To derive the effective theory of holons, we first integrate over the high-energy field \(\pi\). 
The low-energy effective Lagrangian of \(h,z,A\) fields is
\begin{align}
    & L_{hzA}
    = L_0 + L_h
    \nonumber \\ &
    + \int\frac{d^2x}{\Omega} \sum_\sigma \Big\{
        (\lambda - 4Q) |z_\sigma|^2
        + Q a^2 |(\nabla + i\mathbf{A}) z_\sigma|^2
        \nonumber \\ & \quad 
        + \frac{1}{\lambda + 4Q} \big[
            |(\partial_\tau + iA_\tau) z_\sigma|^2
            - t^2 \bar{h}^A h^B \bar{h}^B h^A 
            z_\sigma  \bar{z}_\sigma  
            \nonumber \\ &\qquad
            + t \sigma (
                \bar{h}^A h^B
                z_{-\sigma} \partial_\tau z_\sigma
                - \bar{h}^B h^A 
                \bar{z}_{-\sigma} \partial_\tau \bar{z}_\sigma
            )
        \big]
    \Big\}.
\end{align}
After scaling $\tau \to \tau/c$ (therefore $A_\tau \to c A_\tau$) and defining
\begin{equation}
\begin{aligned}
    \Gamma &= \sqrt{\lambda^2 - 16Q^2},
    &
    m &= \Gamma/c,
    \\
    c &= \sqrt{Q(\lambda+4Q)}a, 
    &
    g^{-1} &= Q a^2 / (c \Omega),
\end{aligned}
\end{equation}
we obtain the effective action
\begin{align}
    S_{hzA} &= S_h + S_{zA} + S_t
    , \quad
    S_t = S_t^1 + S_t^2,
    \\
    S_h &= \frac{1}{\Omega} \int d^3x \sum_s 
    \Big[
        \bar{h}^s (\partial_\tau + i s A_\tau) h^s
        + \frac{\lambda}{c} \bar{h}^s h^s
    \Big],
    \\
    S_{zA} &= \frac{1}{g} \int d^3x 
    \sum_\sigma \Big[
        m^2 |z_\sigma|^2
        + |(\partial_\mu + iA_\mu) z_\sigma|^2
    \Big],
    \label{eq:S-zA}
    \\
    S_t &= \frac{t}{cg} 
    \int d^3x \sum_\sigma \sigma (
        \bar{h}^A h^B 
        z_{-\sigma} \partial_\tau z_\sigma
        - \bar{h}^B h^A 
        \bar{z}_{-\sigma} \partial_\tau \bar{z}_\sigma
    )
    \nonumber \\ &\quad
    +\frac{t^2}{c^2 g} \int d^3x \,
    \bar{h}^A h^A \bar{h}^B h^B 
    \sum_\sigma \bar{z}_\sigma z_\sigma.
    \label{eq:S-t}
\end{align}
Here \(\int d^3x \equiv \int_0^{c\beta} d\tau \int d^2x\), and constant terms from \(L_0\) are dropped.
Note that at half-filling, the holons are eliminated, and the effective action reduces to Eq.~(\ref{eq:S-zA}) only, which is the \(\mathrm{\C P^1}\) model \cite{read-sachdev}. Let us re-express \(z\) fields as
\begin{equation}
\begin{aligned}
z_\sigma(\tau,x) 
&= \sqrt{\rho_\sigma(\tau,x)} 
\exp[i\phi_\sigma(\tau,x)],
\\
\bar{z}_\sigma(\tau,x) 
&= \sqrt{\rho_\sigma(\tau,x)} 
\exp[-i\phi_\sigma(\tau,x)].
\end{aligned}
\end{equation}
At zero temperature, the spinon (and thus the \(z\)-field) condenses. At sufficiently low temperature, we can replace both \(\rho_\uparrow\) and \(\rho_\downarrow\) by a number \(\rho_0/2\), where \(\rho_0\) is of the same order as the density \(n_0\) of condensed spinons; the fluctuations of the \(\rho_\sigma\) fields will be ignored. Then
\begin{align}
    & \sum_\sigma
    |(\partial_\mu + iA_\mu) z_\sigma|^2
    \nonumber \\
    &= \sum_\sigma \frac{1}{4\rho_\sigma} \Big[
        (\partial_\mu \rho_\sigma)^2
        + 4 \rho_\sigma^2
        (A_\mu + \partial_\mu \phi_\sigma)^2
    \Big]
    \nonumber \\ &
    \approx \sum_\sigma
    \rho_\sigma A_\mu^2
    \approx \rho_0 A_\mu^2.
    \label{eq:A-absorb-phi}
\end{align}
In the last line, we dropped the derivative \(\partial_\mu \rho_\sigma\). By a gauge transformation, the derivative of \(\phi\) is absorbed into \(A_\mu\), and \(A_\mu\) acquires a mass \(\sqrt{\rho_0}\) via the Anderson-Higgs mechanism. 
For analysis near the mean-field saddle point, we can assume a weak \(\tau\)-dependence of \(z\) and drop the \(\tau\)-derivatives of \(z\) in Eq.~(\ref{eq:S-t}). Then
\begin{align}
    S_{zA} &\approx 
    \frac{\rho_0}{g} 
    \int d^3x \, (m^2 + A_\mu^2),
    \\
    S_t &\approx \frac{t^2 \rho_0}{c^2 g} 
    \int d^3x \, \bar{h}^A h^A \bar{h}^B h^B,
    \label{eq:action-st}
\end{align}
yielding an effecive action independent of \(\phi\). The integration over \(\phi\) simply gives an (infinite) constant factor, and is dropped thereafter (this removes the gauge redundancy). We reorganize terms in \(S_{hzA}\) according to the order of \(A\):
\begin{align}
    S_{hzA} &= S_0 + S_1 + S_2,
    \\
    S_0 &= S_t 
    + \int d^3x \, m^2 g^{-1} \rho_0
    \nonumber \\ & \quad
    + \Omega^{-1} \int d^3x \sum_s \Big(
        \bar{h}^s \partial_\tau h^s
        + \frac{\lambda}{c} n^h_s
    \Big),
    \\
    S_1 &= i \Omega^{-1} \int d^3x \, 
    A_\tau (n^h_A - n^h_B),
    \\
    S_2 &= g^{-1} 
    \int d^3x \, \rho_0 A_\mu^2.
\end{align}
Here \(n^h_s \equiv \bar{h}^s h^s\) (\(s = A,B\)). Finally we integrate over \(A\):
\begin{align*}
    & Z_{hzA} 
    \cong \int D\bar{h} \, Dh \, DA \, e^{-S_{hzA}}
    \\
    &= \int D\bar{h} \, Dh \, e^{-S_0} \underbrace{
        \int DA \, e^{-(S_1+S_2)}
    }_{\equiv Z'}
    \\
    & Z'
    \equiv \underbrace{
        \int D\mathbf{A} \exp \bigg\{
            \int d^3x \, 
            (- g^{-1} \rho_0 \mathbf{A}^2)
        \bigg\}
    }_{\text{a constant factor}}
    \\ & \times 
    \int DA_\tau \exp \bigg\{
        \int d^3x \, 
        \Big[{
            - \frac{1}{g} \rho_0 A_\tau^2
            - \frac{i}{\Omega} (n^h_A - n^h_B) A_\tau
        }\Big]
    \bigg\}
    \\
    &\propto \exp \bigg\{
        \int d^3x \, 
        \frac{g}{4\rho_0} \Big[
            \frac{i}{\Omega}(n^h_A - n^h_B)
        \Big]^2
    \bigg\}
    \\
    &= \exp \bigg\{
        - \int d^3x \, 
        \frac{g}{4\rho_0 \Omega^2} 
        (n^h_A - n^h_B)^2
    \bigg\},
\end{align*}
Then the effective action of the holons is:
\begin{equation}
    S_\text{eff} = S_0 + \int d^3x \, 
    \frac{g}{4\rho_0 \Omega^2}  (n^h_A - n^h_B)^2.
\end{equation}
We identify terms representing the interaction between \(A\)-holons and \(B\)-holons (recall we redefined \(\alpha t \to t\) earlier, now we restore the original \(t\)):
\begin{equation}
    H_\text{int}^{AB}
    = \bigg[
        \frac{\alpha^2 t^2 \rho_0}{c^2 g}
        - \frac{g}{2 \rho_0 \Omega^2}
    \bigg] \int d^2x \, n^h_A n^h_B.
    \label{eq:competing-interaction}
\end{equation}
The first term (inherited from Eq.~(\ref{eq:action-st})) is repulsion, produced by the exchange of two spinons between holons, in agreement with the previous calculation (in Sec.~\ref{sec:eff-holon-spinon}) on a discrete lattice; the second term is attraction, produced by the gauge fluctuation.

The most important implication of Eq.~(\ref{eq:competing-interaction}) is that there exists a critical \(t/J\) below which the attraction due to gauge fluctuation should overcome the repulsion due to spinon exchange, leading to a superconducting order. For a large value of \(t/J\), the repulsion dominates, hence the absence of SC order. This critical value of \(t/J\) is estimated from
\begin{equation}
    \frac{\alpha^2 t^2 \rho_0}{c^2 g}
    \lesssim \frac{g}{2 \rho_0 \Omega^2}.
\end{equation}
Note that \(\lambda \simeq 4Q\) at low temperature and small doping. Thus 
\begin{equation}
    \frac{t}{J} \lesssim \frac{\Delta}{\sqrt{2} \rho_0}.
\end{equation}
From the mean-field calculation at half-filling and zero temperature (see Appendix \ref{app:spinon-condense}), we obtain \(\Delta \approx 1.2\), and \(\rho_0 \sim n_0 \approx 0.5\), leading to $t/J \lesssim 1.7$. However, from the renormalization group perspective (by integrating over high-energy holons) \cite{Nagaosa-book}, the effective attractive  between holons will renormalize to a bigger value at low energy. Thus the actual range of \(t/J\) that allows superconductivity is much larger than our naive estimation. The net attractive interaction leads to (\(p+ip\))-wave pairing of the spinless holons \cite{Cheng}. Together with the singlet-pairing of the spinons, 
we obtain the (\(d+id\))-wave pairing of the electrons. 

For the $A,B,D\neq 0$ region, the order parameter $B$ can be regarded as an additional Higgs field which further breaks down the staggered $U(1)$ gauge field into a $\Z_2$ gauge field, and we believe that such an additional gauge field will not affect the attractive interaction in the presence of AFM long-range order. In fact, a nonzero B is also crucial for the holon mobility, e.g., for the introduction of holon kinetic term, and superconductivity is only possible in this phase. Nevertheless, at very small doping, our estimation for the effective holon interactions in previous sections is still valid since the correction induced by B is always proportional to holon concentration $\delta$.
Physically, as long as we start from the short-range RVB phase $A\neq 0$, the staggered $U(1)$ gauge fluctuations will always induce attractive interactions between holons belonging to different sub-lattices, and this naturally explains why \(d+id\) SC order still survive even in the absence of AFM long-range order in this phase. We stress that the spin-charge separation scenario plays an essential role for the emergence of SC order here, especially for the region without AFM order, and there is no other conventional picture can explain such a novel $d+id$ Sc oder emerged in a doped-Mott insulator!   

Finally, for the region $A=0$ and $B,D\neq 0$, the usual uniform $U(1)$ gauge fluctuations will kill the FM long-range order at finite $t/J$ and induce repulsive interactions among holons. Such a deconfined phase of $U(1)$ gauge field does not support superconductivity and would be a natural candidate for the non-Fermi liquid phase emerged in the large $t/J$ limit. 

\section{Conclusion and discussion}\label{discussion}

In conclusion, we investigate the global phase diagram of the $t$-$J$ model on a honeycomb lattice at small doping region using the Grassmann tensor network numerical method. Furthermore, the slave-fermion mean-field theory is employed to account for the global phase diagram. The effective interacting holon theory is rigorously derived in the presence of the AFM background. The novel phenomena of spin-charge separation and attractive interaction among holons naturally explain the emergence of $d+id$ SC orders. Remarkably, the predicted phase boundary for $d+id$ SC orders via self-consistent mean-field theory approach and effective field theory analysis is intrinsically close to the results from Grassmann tensor product numerical simulations. We stress that the spin-charge separation mechanism occurs only in strongly correlated systems 
and is essentially different from the weak coupling mechanism in BCS theory. 

We also would like to point out that the attractive interaction between holons belonging to different sub-lattices is induced by gauge fluctuations, and the \emph{staggered} $U(1)$ gauge structure is crucial to the attractive interaction, since in this case the two sub-lattices $A$ and $B$ carry opposite gauge charges. Although our derivation is in the region for $A\neq 0$ and $B,D=0$, we believe that this mechanism is still valid in the region $A,B,D\neq 0$, and it explains the emergence of SC orders even in the absence of long-range AFM order, as long as the system is still dominated by short-range AFM spin fluctuations.

It is well known that in the conventional BCS theory, the existence of zero-energy bound fermion state in the vortex of superconductor depends on the nontrivial topological class of the bulk. 
According to the topological classification, 
the \(d+id\) superconductivity discovered on the honeycomb lattice breaks the time reversal symmetry, but has the $SU(2)$ spin-rotation symmetries. Thus it belongs to the class $C$ of ten Altland-Zirnbauer classes \cite{Chiu,Sato}.
Althugh the class $C$ has 
no zero-energy states in the vortex from the perspective of noninteracting topological classification, we argue that the unconventional superconductivity in the $t$-$J$ model on a honeycomb lattice emerges from the spin-charge separation is a different case. The bosonic spinons carrying the spin degrees of freedom condense in the AFM region. The $p+ip$ superconductivity of the spinless holons belongs to the class $D$ and 
there will be a zero-energy bound state in the vortex, similar to the Majorana zero mode in conventional $p+ip$ superconductor \cite{Kopnin,Volovik,Read}, which serves as a solid evidence for the spin-charge separation scenario. Experimentally, the doped spin $1/2$ honeycomb lattice
antiferromagnet $\mathrm{InV_{1/3}Cu_{2/3}O_3}$ \cite{InVCuO} would be an
appealing candidate to examine the emergence of topological Majorana zero mode in its vortex core.

\section*{Acknowledgment}

This work is supported by General Research Fund Grant No. 14302021 and NSFC/RGC Joint Research Scheme No. N-CUHK427/18 from Research Grants Council, and Direct Grant No. 4053416 from the Chinese University of Hong Kong.  WQC is supported by the National Key R\&D Program of China (Grants No. 2022YFA1403700),  NSFC (Grants No. 11861161001), the Science, Technology and Innovation Commission of Shenzhen Municipality (No. ZDSYS20190902092905285), Guangdong Basic and Applied Basic Research Foundation under Grant No. 2020B1515120100, and Center for Computational Science and Engineering at Southern University of Science and Technology.

\appendix

\section{Details of the Grassmann tensor numerical calculation}\label{app:tensor}

\begin{figure*}[t]
    \def\svgwidth{2.0\columnwidth}
    \includegraphics[width=5.8cm]{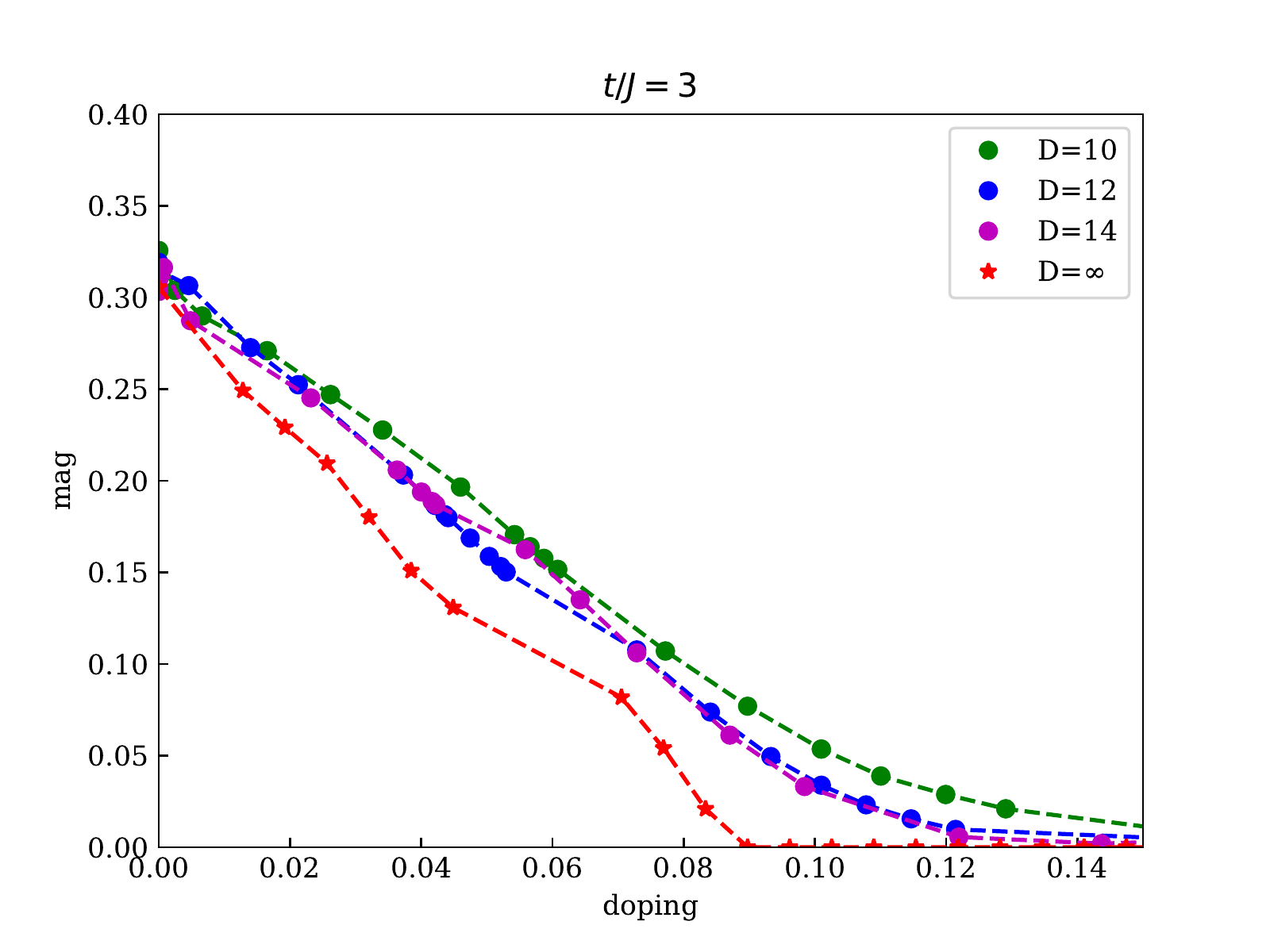}
    \includegraphics[width=5.8cm]{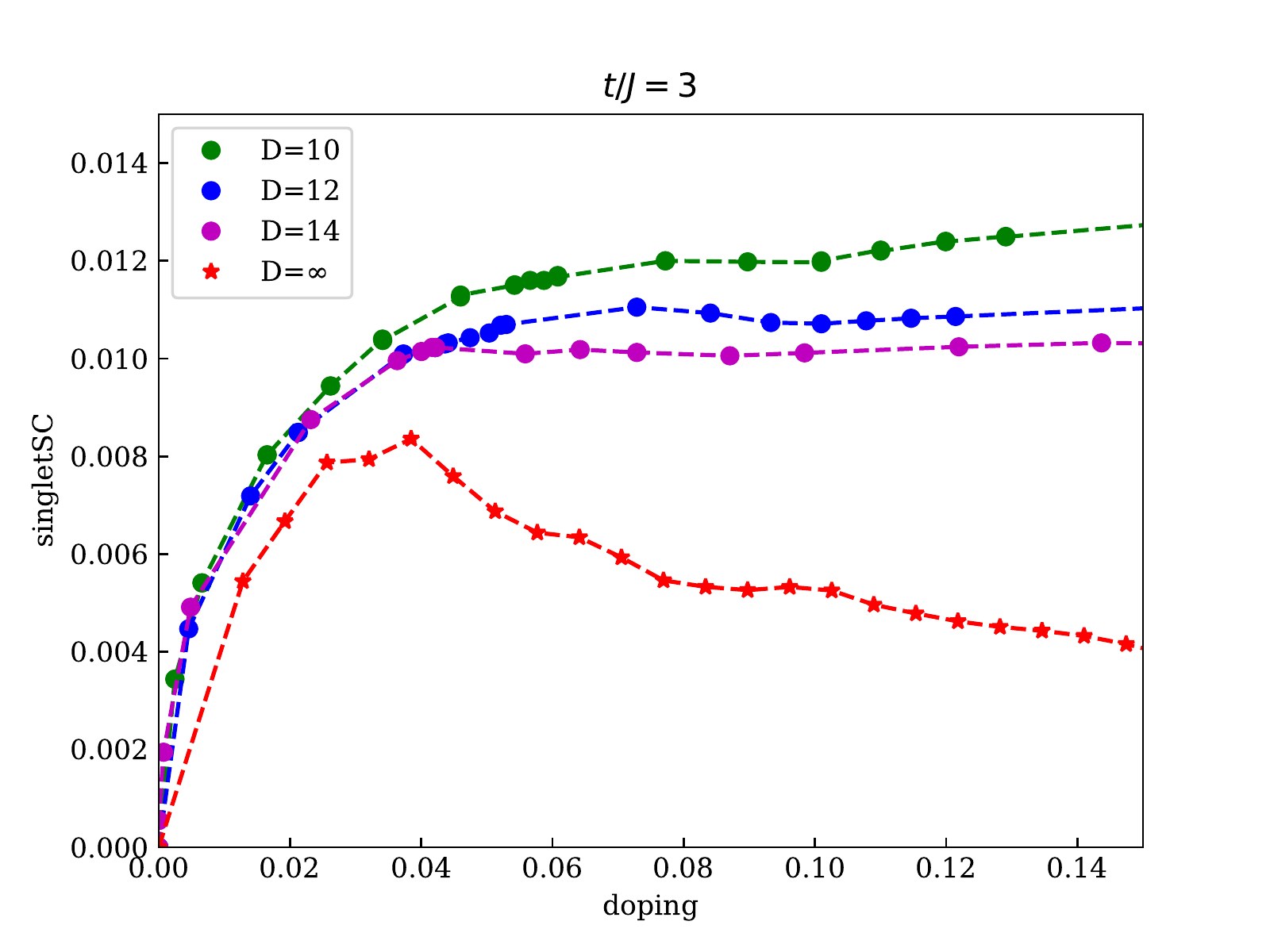}
    \includegraphics[width=5.8cm]{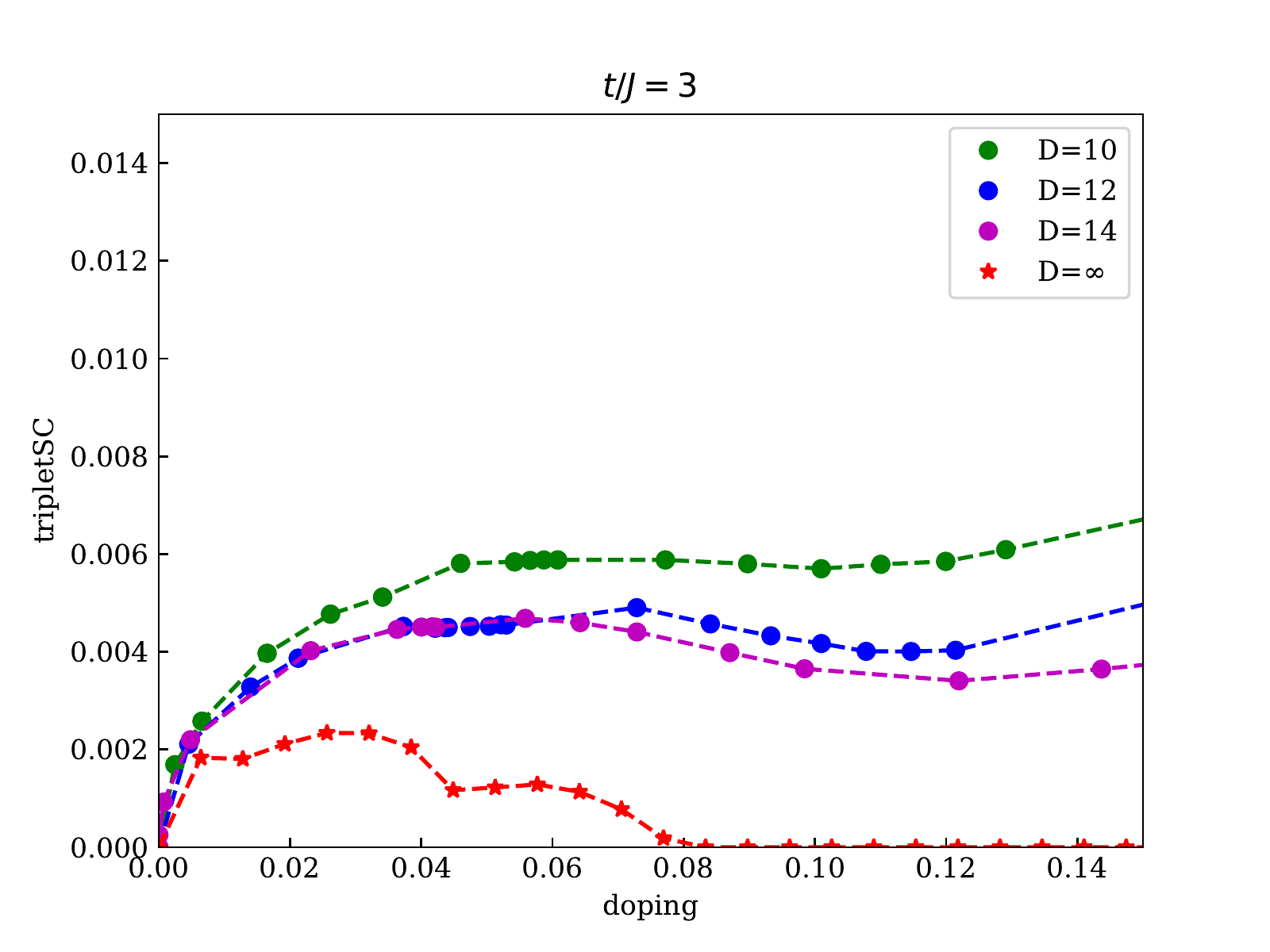}
    \\
    \includegraphics[width=5.8cm]{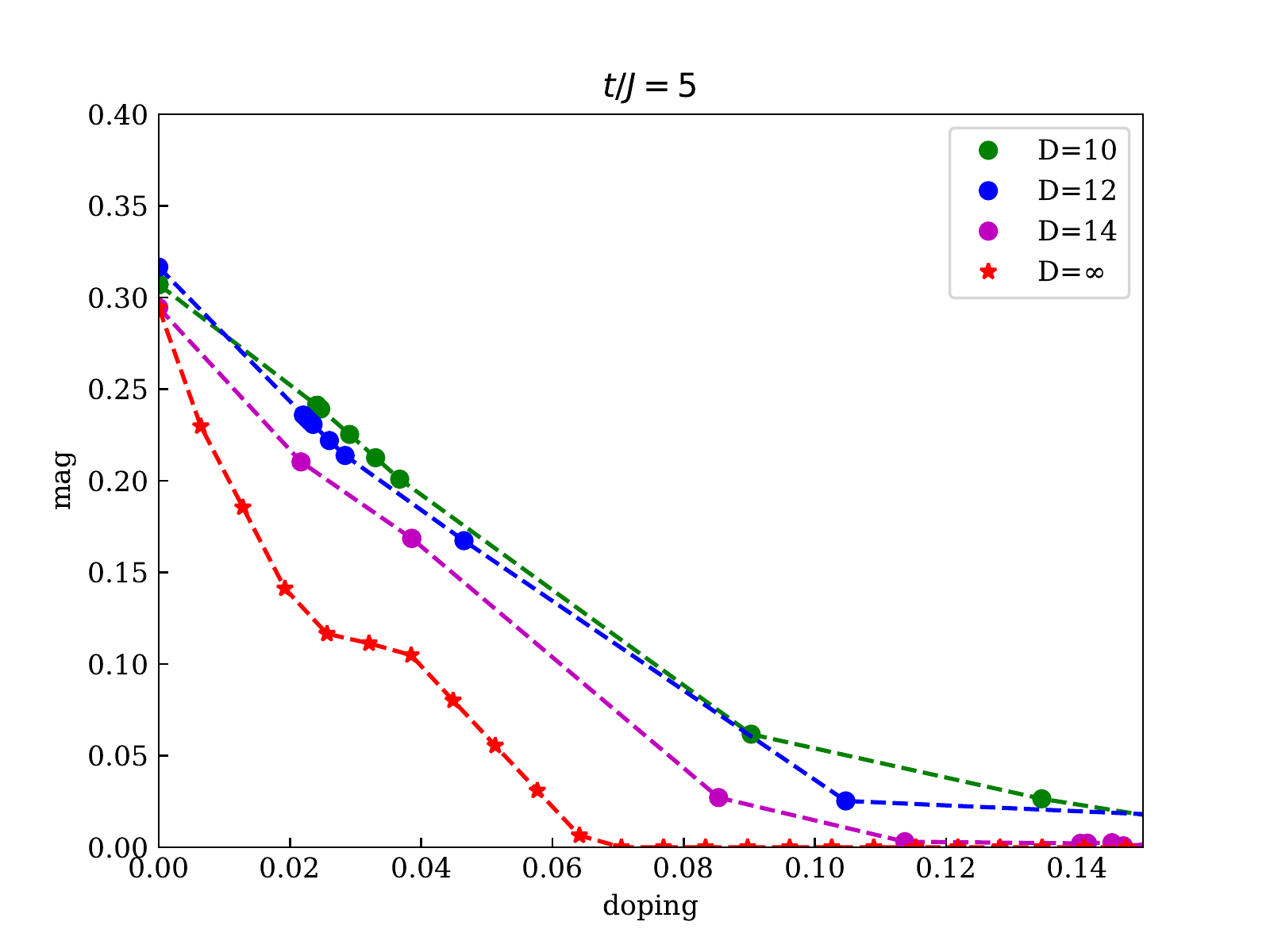}
    \includegraphics[width=5.8cm]{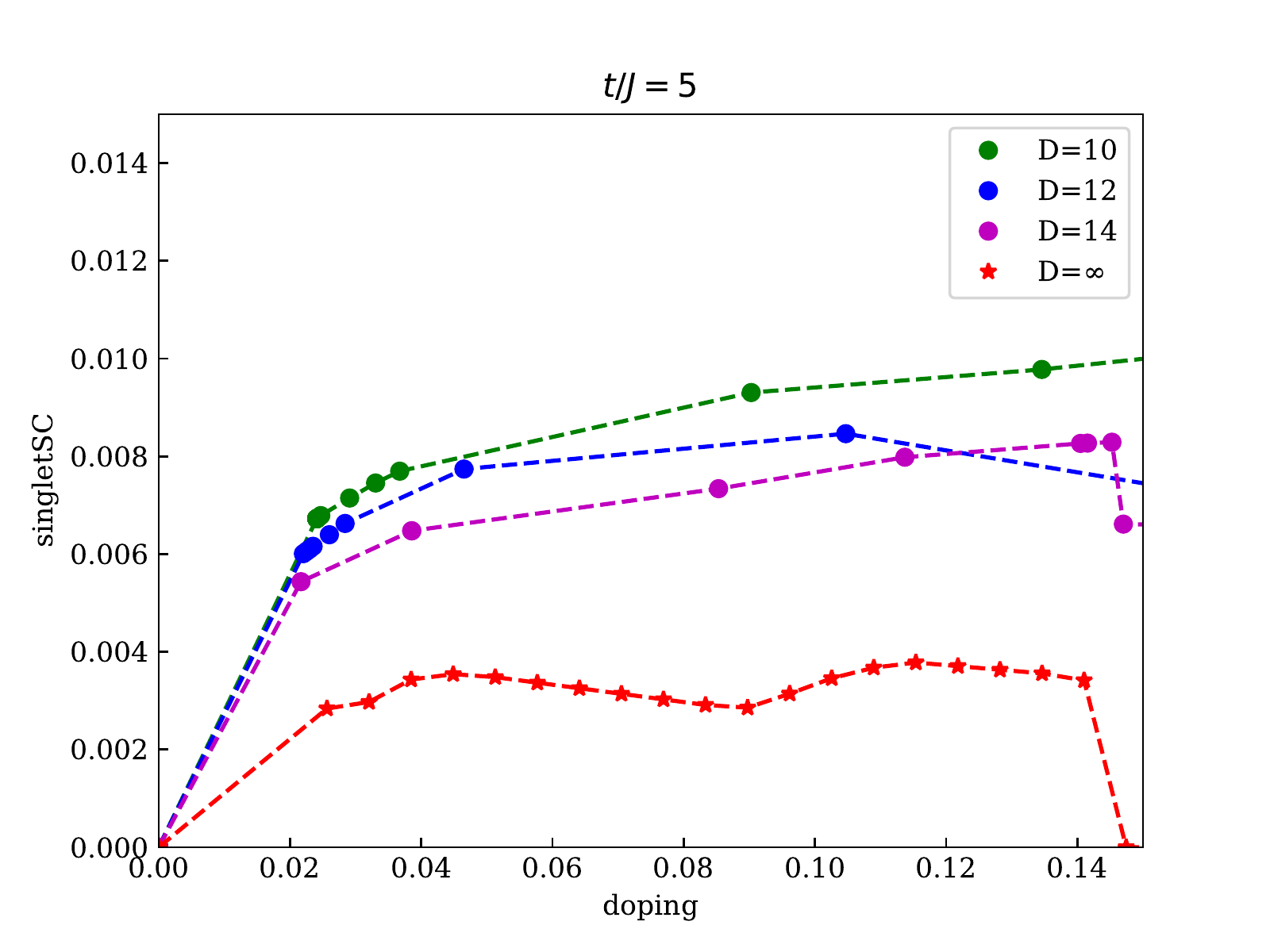}
    \includegraphics[width=5.8cm]{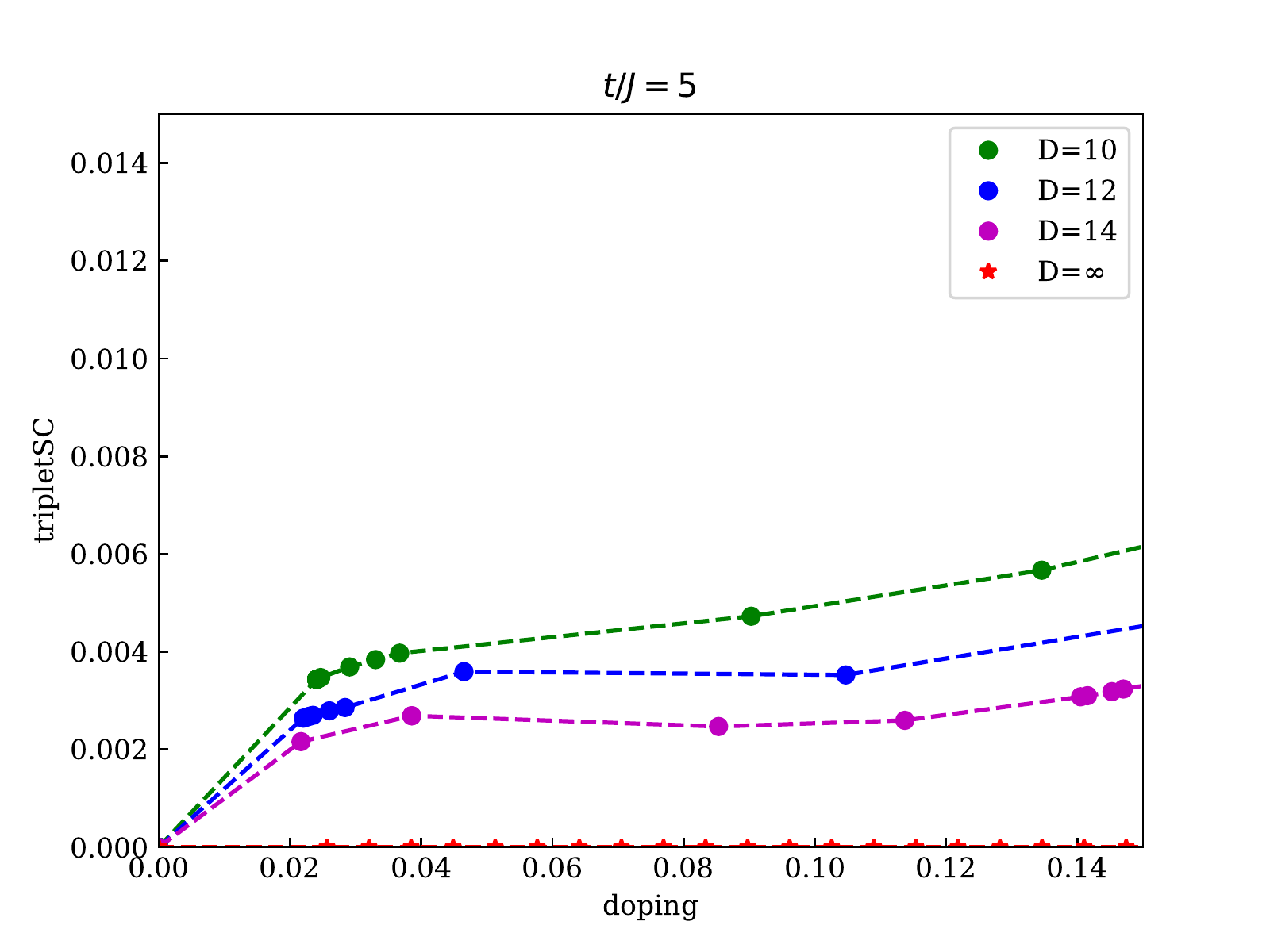}
    \\
    \includegraphics[width=5.8cm]{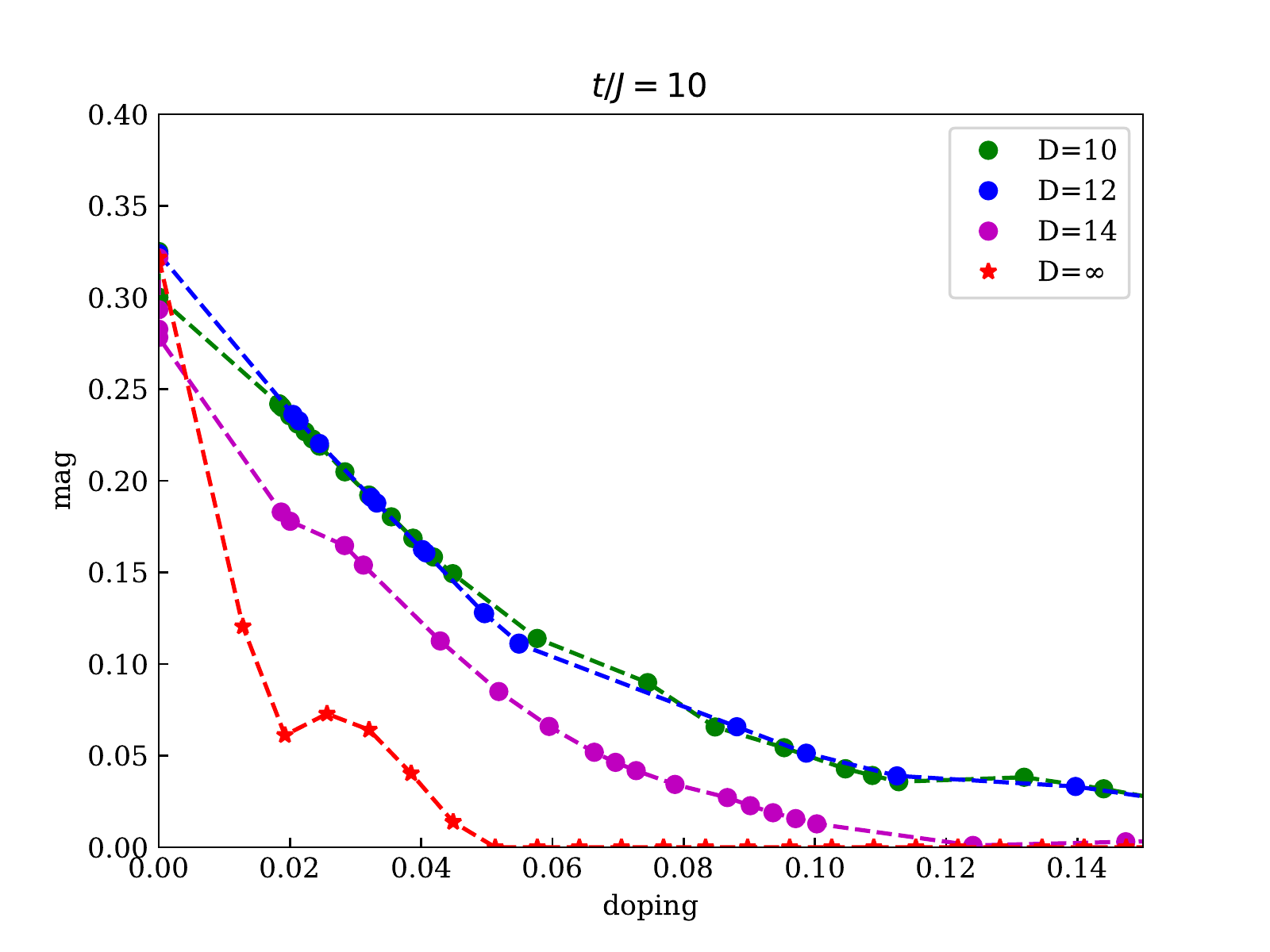}
    \includegraphics[width=5.8cm]{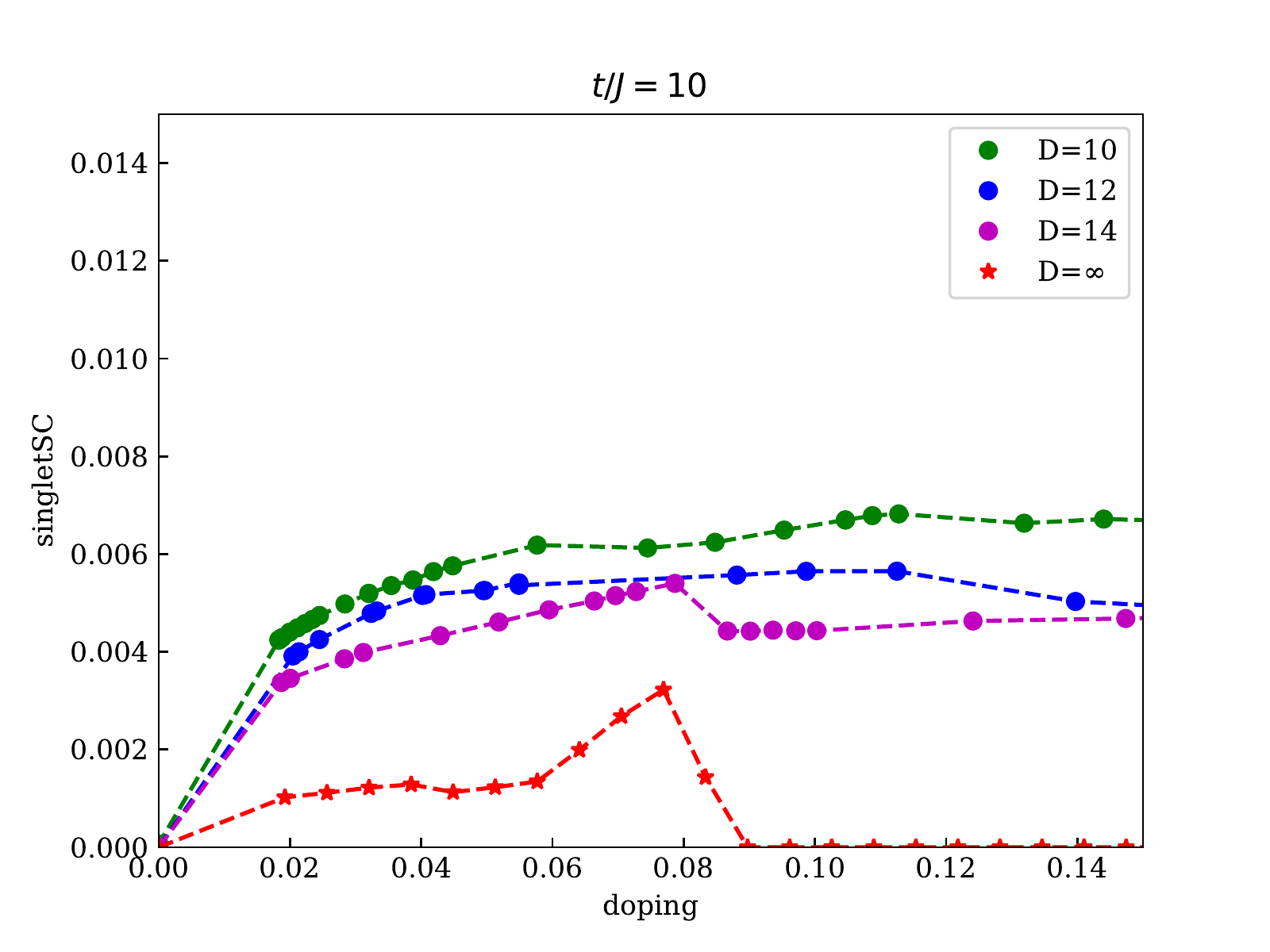}
    \includegraphics[width=5.8cm]{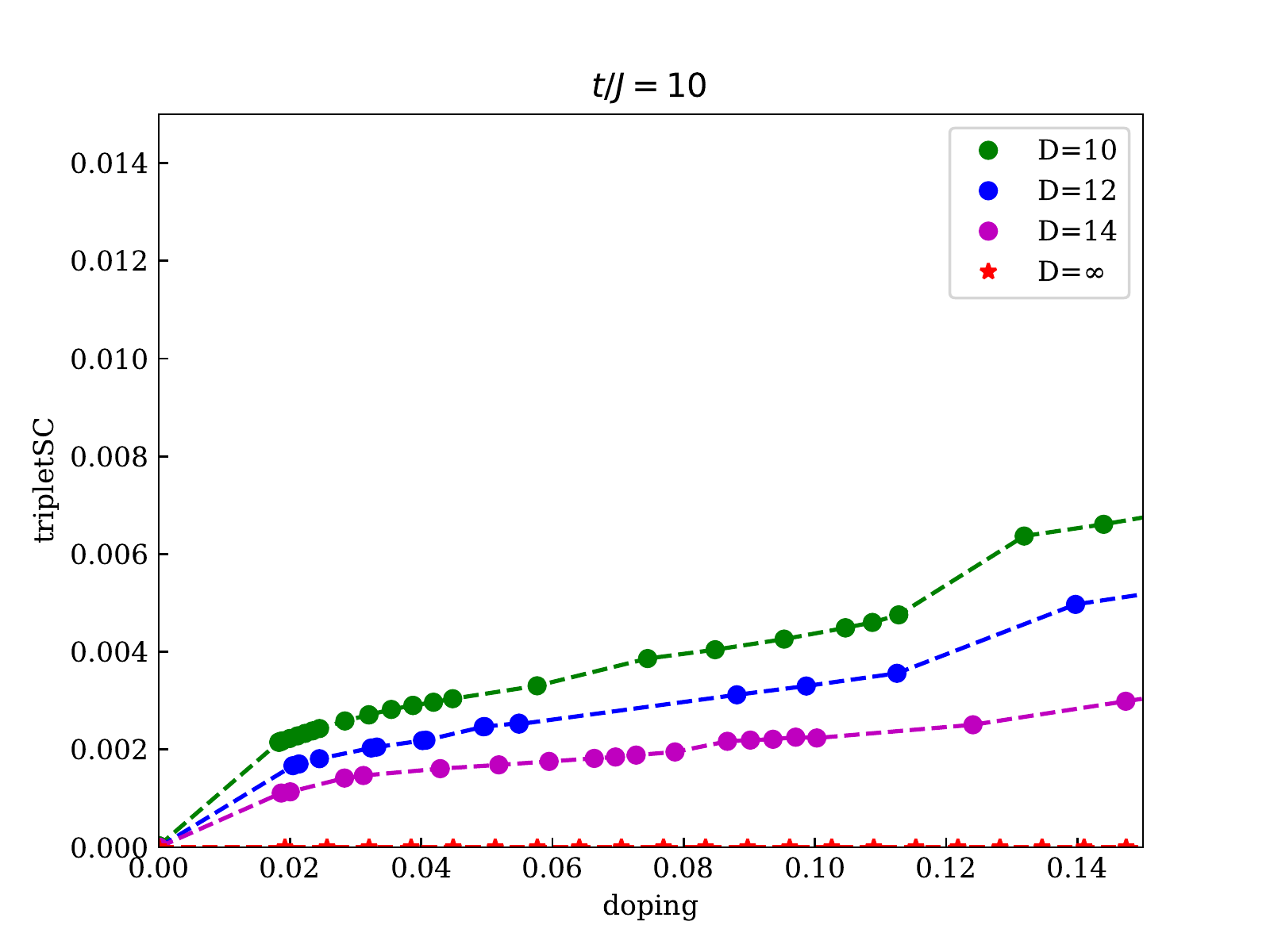}
    \caption{Staggered magnetization, amplitudes of singlet and triplet SC order parameters versus doping at $t/J=3, 5, 10$.}
    \label{fig:grassmann-results-app1}
\end{figure*}

\begin{figure*}[t]
    \def\svgwidth{2.0\columnwidth}
    \includegraphics[width=5.8cm]{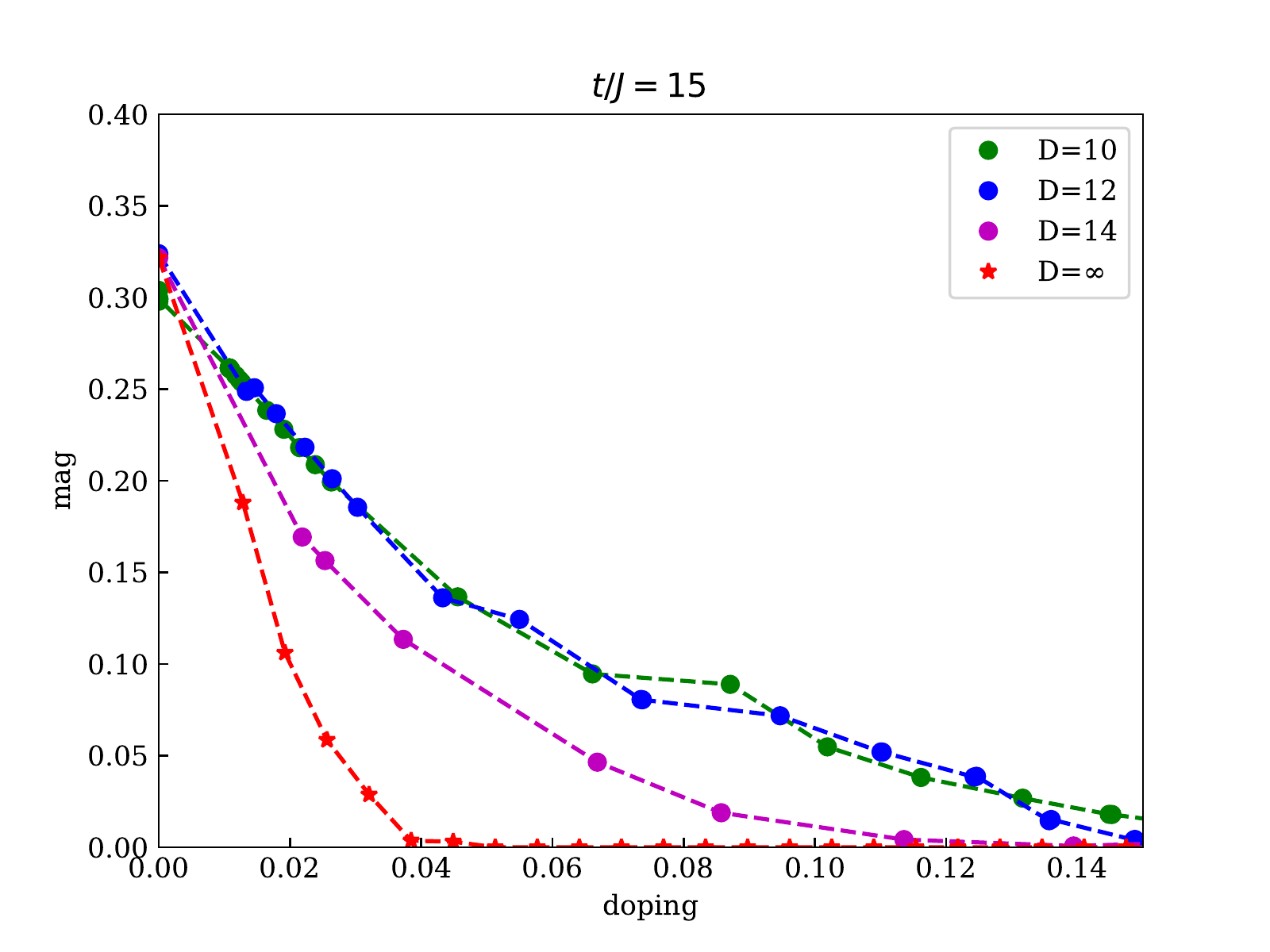}
    \includegraphics[width=5.8cm]{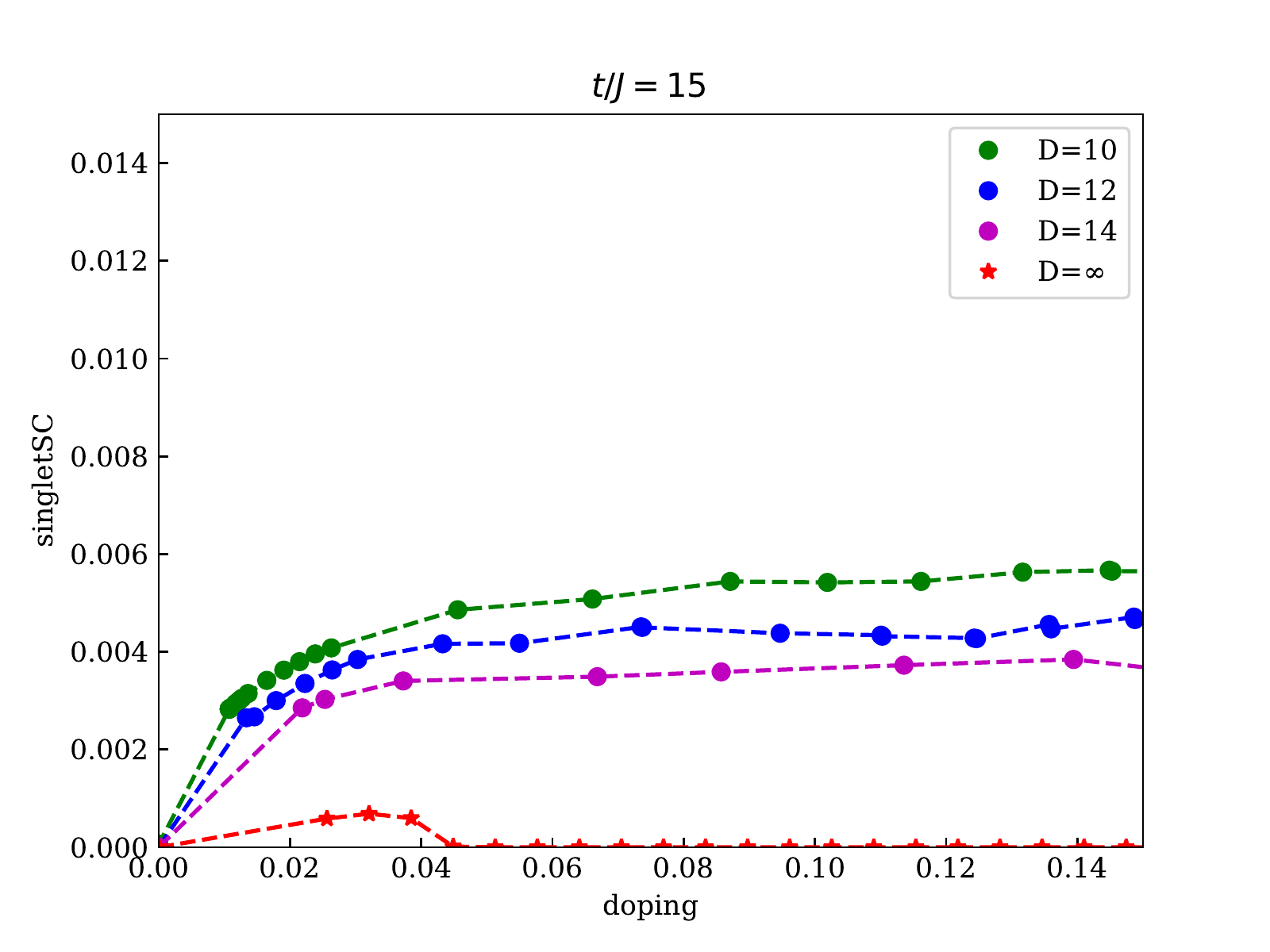}
    \includegraphics[width=5.8cm]{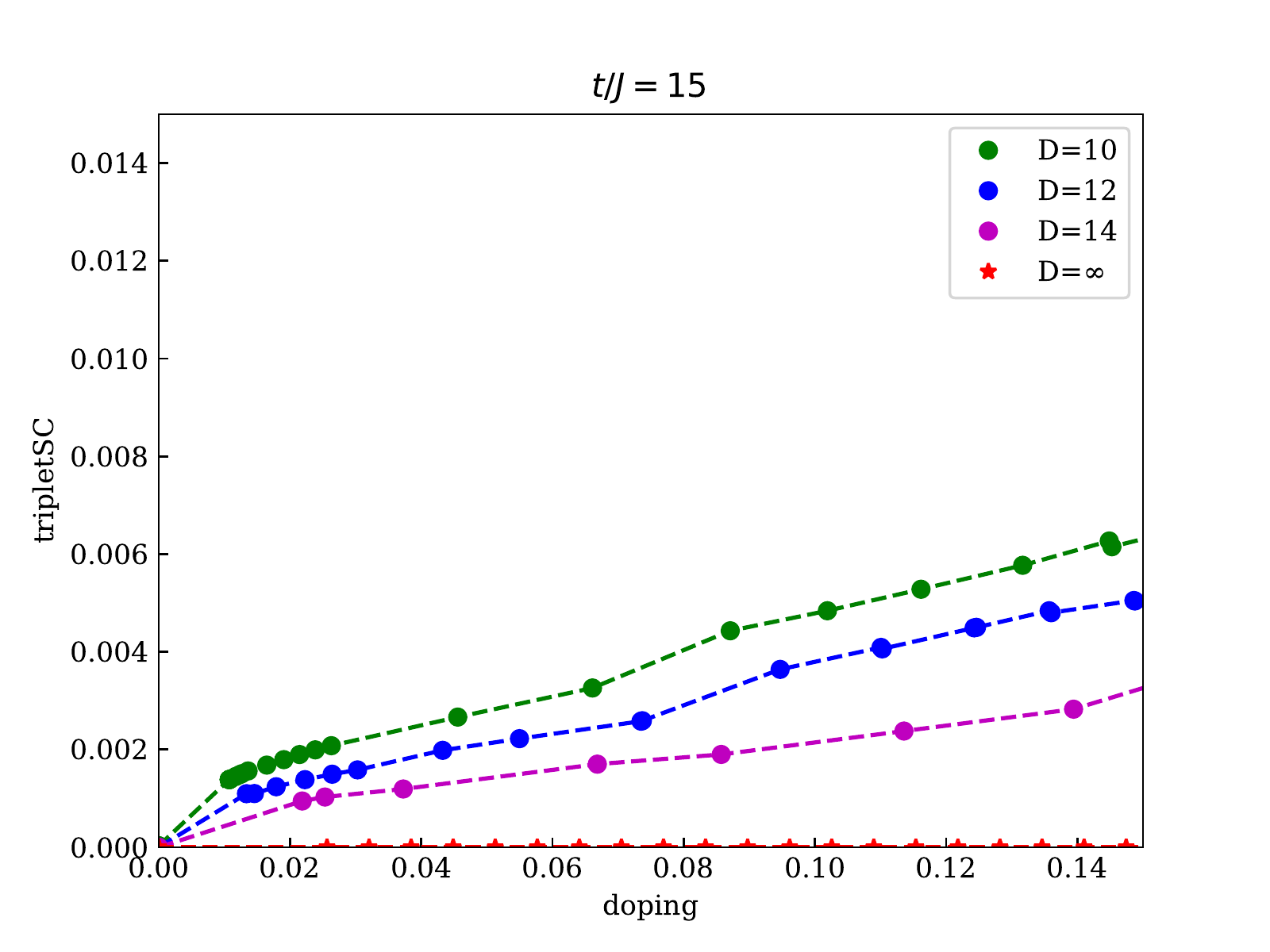}
    \\
    \includegraphics[width=5.8cm]{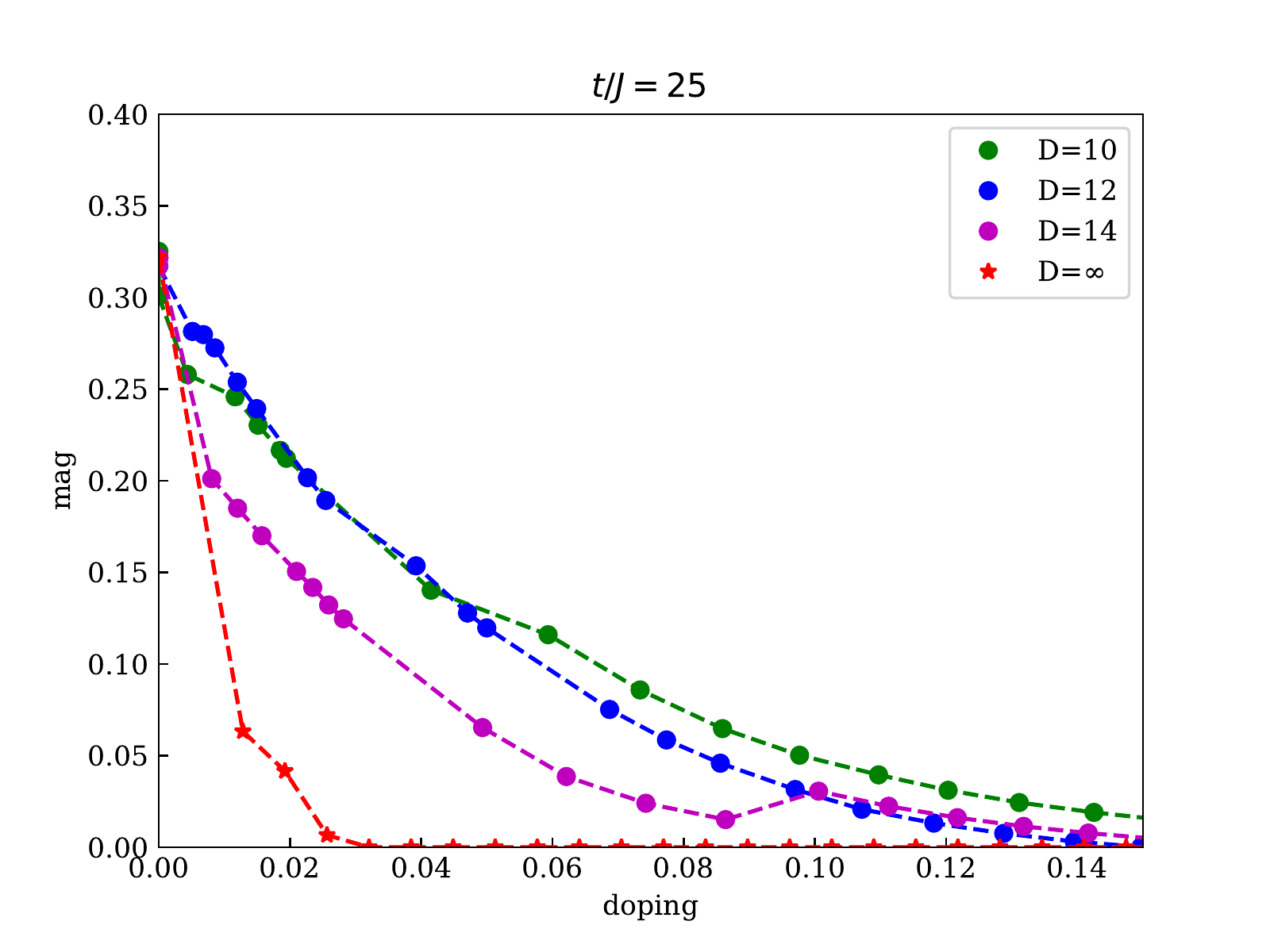}
    \includegraphics[width=5.8cm]{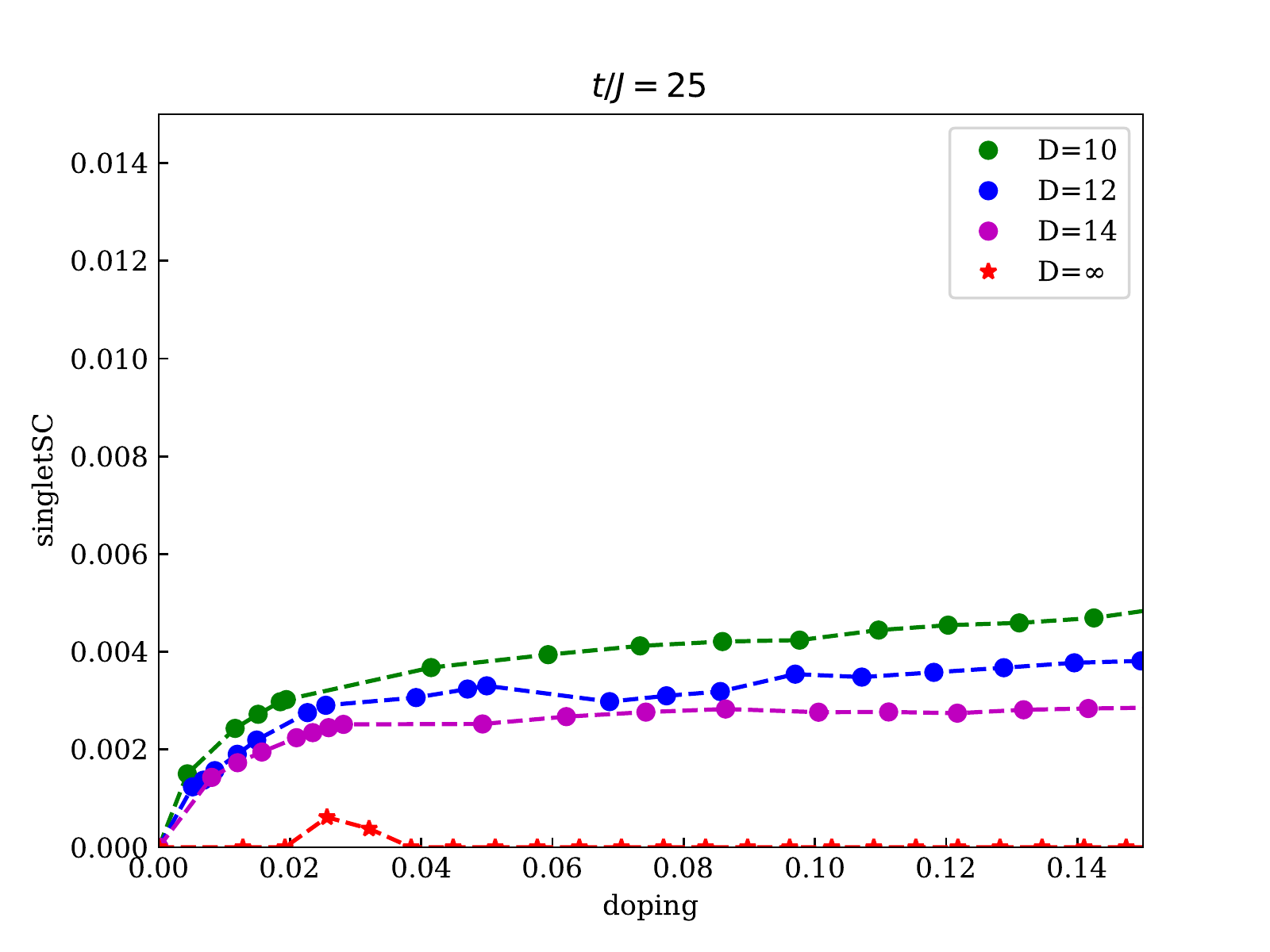}
    \includegraphics[width=5.8cm]{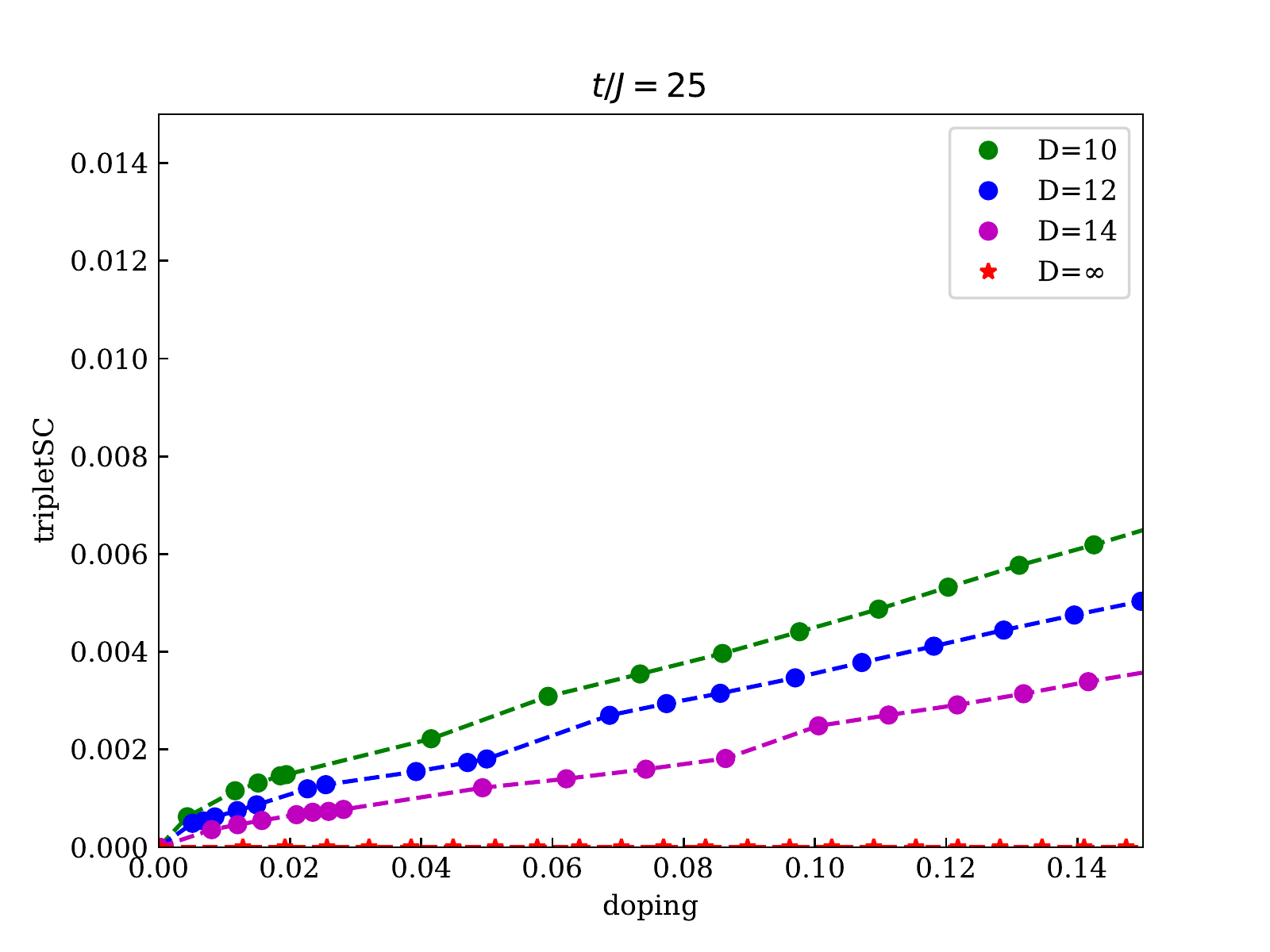}
    \\
    \includegraphics[width=5.8cm]{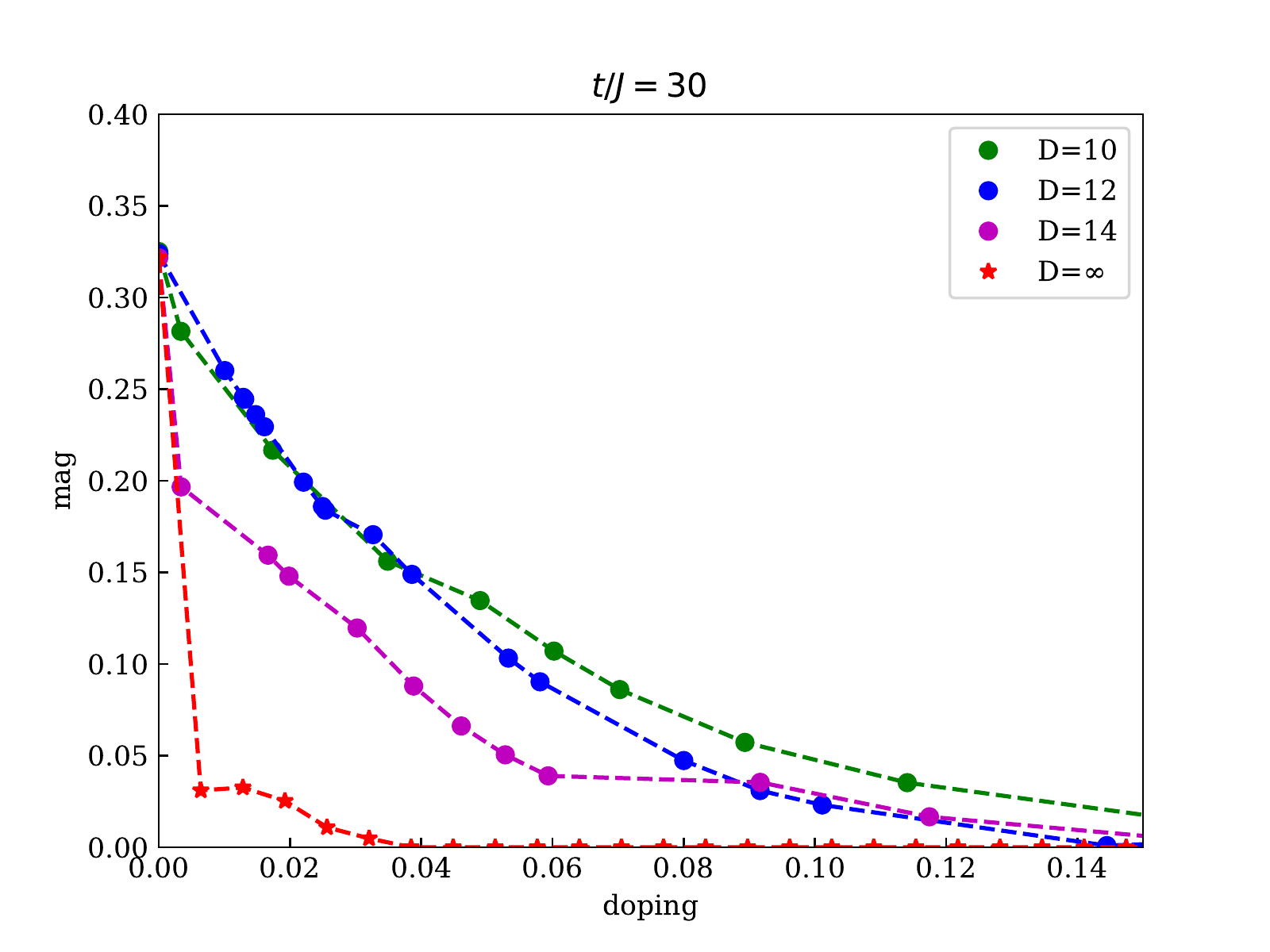}
    \includegraphics[width=5.8cm]{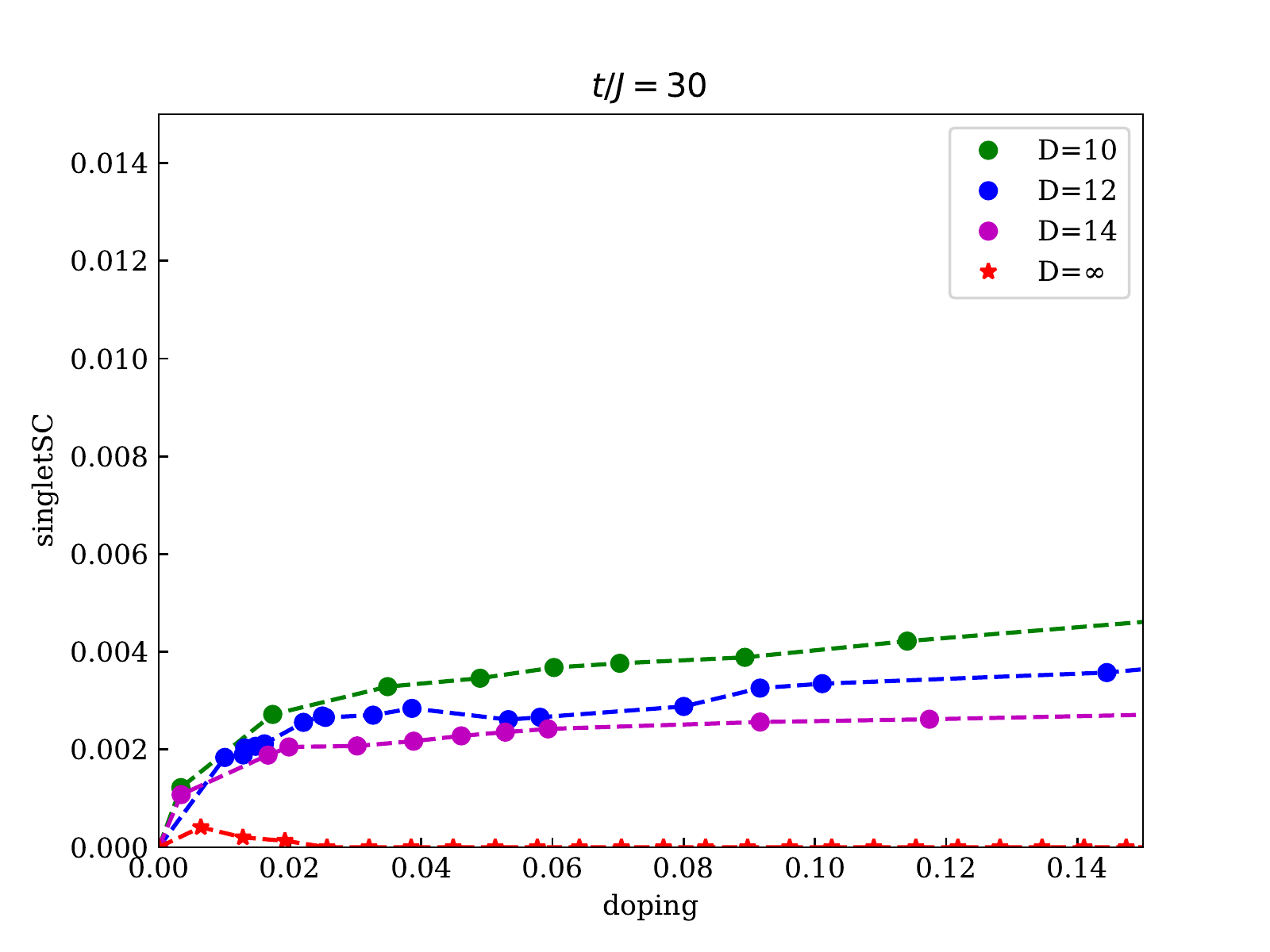}
    \includegraphics[width=5.8cm]{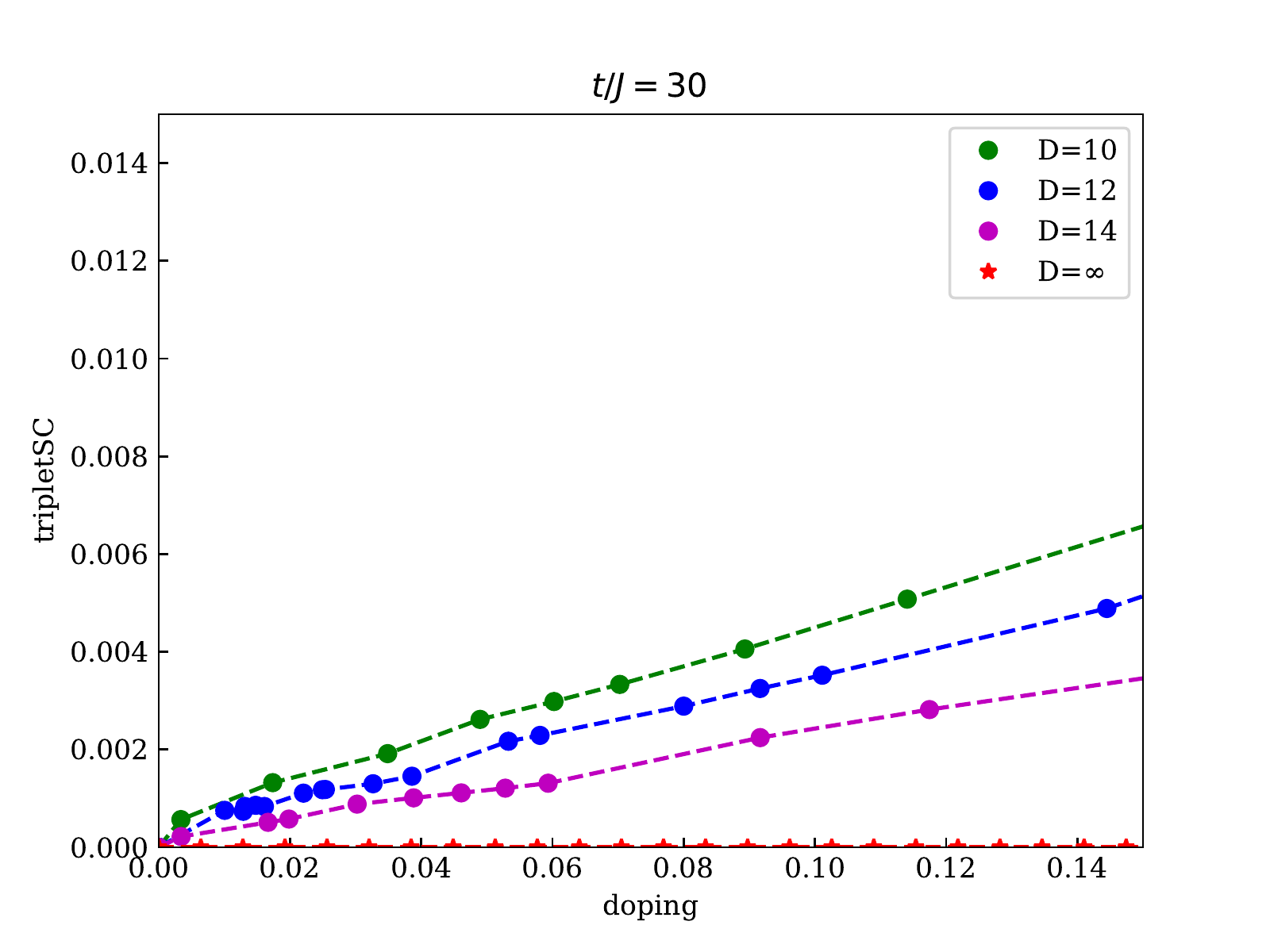}
    \caption{Staggered magnetization, amplitudes of singlet and triplet SC order parameters versus doping at $t/J=15, 25, 30$.}
    \label{fig:grassmann-results-app2}
\end{figure*}




We use state-of-the-art Grassmann tensor product numerical method to obtain data in Sec.~\ref{sec:tensor}. The standard form of GTPS is used as our variational wave function. Translation-invariant ansatz is assumed and is specified by just two different Grassmann tensors $\mathbf{T}_A$ and $\mathbf{T}_B$ on sub-lattices A and B of each unit cell:
\begin{align}
    \Psi({m_i}, {m_j})
    = \text{tTr} \int
    \Pi_{\langle ij\rangle}
    \mathbf{g}_{aa'}\Pi_{i\in A}
    \mathbf{T}^{m_i}_{A;abc}
    \Pi_{j\in B} T^{m_j}_{B;a'b'c'}.
\end{align}
with
\begin{eqnarray}
{\textbf{T}}^{m_{i}}_{A;abc}&=& {T}^{m_{i}}_{A;abc}
\theta_\alpha^{P^f(a)} \theta_\beta^{P^f(b)}
\theta_\gamma^{P^f(c)},  \nonumber\\
 {\textbf{T}}^{m_{j}}_{B,a^\prime b^\prime c^\prime}&=&
{T}^{m_{j}}_{B;a^\prime b^\prime c^\prime}
\theta_{\alpha^\prime}^{P^f(a^\prime)}
\theta_{\beta^\prime}^{P^f(b^\prime)}
\theta_{\gamma^\prime}^{P^f(c^\prime)},    \nonumber\\
\textbf{g}_{aa^\prime}&=& \delta_{aa^\prime}{\dd
\theta}_\alpha^{P^f(a)} {\dd
\theta}_{\alpha^\prime}^{P^f(a^\prime)}.
\end{eqnarray}
More details about this ansatz can be found in the appendix of Ref.~\citenum{Gu_pwave}. The variational ground state is obtained from the simple update imaginary time evolution algorithm \cite{Gu-2013}. The desired doping is achieved by adjusting chemical potential. Then we measure the physical quantities by the renormalization group algorithms (GTERG/wGTERG) in Refs.~\citenum{Gu-2010, Gu-2013}. The total system size is up to $2\times 3^6$ sites and all calculations are performed with periodic boundary conditions. We use three different virtual bond dimensions \(D=10,12,14\) of the GTPS. Different dopings are obtained by tuning the chemical potential. 

The data of $D=\infty$ is obtained by the following approximate extrapolation method. For each fixed $D$, we get a series of data $(\delta_i^D, O_i^D)$, where $\delta_i^D$ is the doping and $O_i^D$ is the physical quantity that we are interested in. We choose some equally spaced doping points $\tilde \delta_i^D$. For each doping point $\tilde \delta_i$ and each $D$, we use the linear interpolation to get the approximated $\tilde O_i^D$ corresponding to $\tilde \delta_i$ using two nearby data points we have, ($\delta_{i,L}^D$,$O_{i,L}^D$) and ($\delta_{i,R}^D$,$O_{i,R}^D$),
\begin{equation}
    \frac{\tilde O_i^D-O_{i,L}^D}{\tilde \delta_i-\delta_{i,L}^D}=\frac{O_{i,R}^D-O_{i,L}^D}{\delta_{i,R}^D-\delta_{i,L}^D}.
\end{equation}
We then use least squares linear fit using data points $(D^{-1}, \tilde O_i^D)$ and then find $\tilde O_i^\infty$ by setting $D^{-1}=0$. The linear fit $O=k_i D^{-1}+b_i$ minimizes the squared error
\begin{equation}
    E_i=\sum_D |k_i D^{-1} + b_i - \tilde O_i^D|^2.
\end{equation}
To reduce fluctuation, we use only points with good fitting (R-squared value greater than 0.7). We also set negative $O_i^\infty$ as 0 due to physical consideration. $O_i^\infty$ at zero doping is set to be 0 for superconducting order parameter and $O_i^{14}$ of smallest doping we have. 
Here we provide figures of other values of \(t/J\) (Figs.~\ref{fig:grassmann-results-app1}, \ref{fig:grassmann-results-app2}) below in addition to the \(t/J=20\) results in the main text (Fig.~\ref{fig:grassmann-results-main}).

\section{Details of the mean-field theory}\label{app:mft}

\subsection{Mean-field Hamiltonian of the \textit{t}-\textit{J} model}\label{app:mf-decouple}

In this appendix, we derive the mean-field Hamiltonian Eqs.~(\ref{eq:mf-ham}) to (\ref{eq:mf-ham-number}). The mean-field decoupling is done by ignoring higher order fluctuations around the mean-field average \cite{bruus-book}: for any two operators \(A\) and \(B\),
\begin{equation}
    AB \approx A \expect{B} + \expect{A} B
    - \expect{A} \expect{B}.
\end{equation}
The \(t\)-terms in Eq.~(\ref{eq:tJ-slave-fermion}) are decoupled to
\begin{align}
    H_t &\equiv 2t \sum_{\expect{ij}} 
    (h^\dagger_i h_j \hat{B}^\dagger_{ij}  + h.c.)
    \nonumber \\
    &\approx 2t \sum_{\expect{ij}} (
        D_{ij} \hat{B}^\dagger_{ij}
        + h^\dagger_i h_j B^*_{ij} + h.c.
    ) \nonumber \\ &\quad
    - 2t \sum_{\expect{ij}} (
        D_{ij} B^*_{ij} + h.c.
    ).
\end{align}
By Wick's theorem, the decoupling of \(J\)-terms in Eq.~(\ref{eq:tJ-slave-fermion})
\begin{equation*}
    H_J 
    = -\frac{J}{2} \sum_{\expect{ij}} \sum_{\sigma, \sigma'}
    \sigma \sigma' 
    b^\dagger_{j,-\sigma} b^\dagger_{i\sigma} 
    b_{i\sigma'} b_{j,-\sigma'} 
\end{equation*}
involves contribution from 3 channels:
\begin{widetext}
\begin{align*}
    H_J &\approx H_J^{(A)} + H_J^{(B)} + H_J^{(\delta)},
    \\
    H_J^{(A)}
    &= -\frac{J}{2} \sum_{\expect{ij}} \sum_{\sigma, \sigma'}
    \sigma \sigma' \Big[
        b^\dagger_{j,-\sigma} b^\dagger_{i\sigma} 
        \expect{b_{i\sigma'} b_{j,-\sigma'}}
        + \expect{b^\dagger_{j,-\sigma} b^\dagger_{i\sigma}}
        b_{i\sigma'} b_{j,-\sigma'} 
        - \expect{b^\dagger_{j,-\sigma} b^\dagger_{i\sigma}}
        \expect{b_{i\sigma'} b_{j,-\sigma'}}
    \Big],
    \\
    H_J^{(B)}
    &= -\frac{J}{2} \sum_{\expect{ij}} \sum_{\sigma, \sigma'}
    \sigma \sigma' \Big[
        b^\dagger_{j,-\sigma} b_{i\sigma'}
        \expect{b^\dagger_{i\sigma} b_{j,-\sigma'}}
        + \expect{b^\dagger_{j,-\sigma} b_{i\sigma'}}
        b^\dagger_{i\sigma} b_{j,-\sigma'}
        - \expect{b^\dagger_{j,-\sigma} b_{i\sigma'}}
        \expect{b^\dagger_{i\sigma} b_{j,-\sigma'}}
    \Big],
    \\
    H_J^{(n)}
    &= -\frac{J}{2} \sum_{\expect{ij}} \sum_{\sigma, \sigma'}
    \sigma \sigma' \Big[
        b^\dagger_{j,-\sigma} b_{j,-\sigma'}
        \expect{b^\dagger_{i\sigma} b_{i\sigma'}}
        + \expect{b^\dagger_{j,-\sigma} b_{j,-\sigma'}}
        b^\dagger_{i\sigma} b_{i\sigma'}
        - \expect{b^\dagger_{j,-\sigma} b_{j,-\sigma'}}
        \expect{b^\dagger_{i\sigma} b_{i\sigma'}}
    \Big].
\end{align*}
With the mean-field ansatz Eqs.~(\ref{eq:abd}) and (\ref{eq:abd2}), we calculate the \(A\)-channel terms:
\begin{align}
    H_J^{(A)}
    &= -2J \sum_{\expect{ij}} (
        A^*_{ij} \hat{A}_{ij} + h.c.
        - |A_{ij}|^2
    ),
\end{align}
the \(B\)-channel terms:
\begin{align}
    H_J^{(B)}
    &= -\frac{J}{2} \sum_{\expect{ij}} \sum_{\sigma,\sigma'}
    \sigma \sigma' \Big[
        B^*_{ij} 
        b^\dagger_{i\sigma} b_{j,-\sigma'} 
        + h.c.
        - |B_{ij}|^2 
    \Big] \delta_{\sigma,-\sigma'}
    \nonumber \\
    &= \frac{J}{2} \sum_{\expect{ij}} \Big[
        B^*_{ij} 
        \sum_\sigma b^\dagger_{i\sigma} b_{j\sigma} 
        + h.c.
        - 2 |B_{ij}|^2 
    \Big]
    = J \sum_{\expect{ij}} (
        B^*_{ij} \hat{B}_{ij}
        + h.c.
        - |B_{ij}|^2 
    ),
\end{align}
and the \(n\)-channel terms:
\begin{align}
    H_J^{(n)}
    &= -\frac{J}{2} \sum_{\expect{ij}} 
    \sum_{\sigma,\sigma'}
    \sigma \sigma' \Big[
        \frac{1-\delta}{2} (
            b^\dagger_{j,-\sigma} b_{j,-\sigma}
            + b^\dagger_{i\sigma} b_{i\sigma}
        )
        - \frac{(1-\delta)^2}{4}
    \Big] \delta_{\sigma \sigma'}
    \nonumber \\
    &= -\frac{J}{2} \sum_{\expect{ij}} \Big[
        \frac{1-\delta}{2} (
            \hat{n}^b_i + \hat{n}^b_j
        )
        - \frac{(1-\delta)^2}{2}
    \Big]
    = -\frac{1}{4} \alpha J (1-\delta) \sum_i \hat{n}^b_i
    + \frac{1}{4} \alpha N J (1-\delta)^2.
\end{align}
\end{widetext}
Here \(\alpha = 3\) is the coordination number. Finally, the mean-field Hamiltonian is
\begin{align}
    H_{\text{MF}}
    &= H_t + H_J + H_\delta,
    \\
    H_\delta &= 
    {\textstyle \sum_i} \left[
        \lambda_i (
            \hat{n}^b_i - 1 + \delta
        )
        - \mu_i (\hat{n}^h_i - \delta)
    \right].
\end{align}
Here \(H_\delta\) is added to impose the no-double-occupancy constraint Eq.~(\ref{eq:doping}), in which
\begin{equation*}
    \hat{n}^b_i = 
    {\textstyle \sum_\sigma} 
    b^\dagger_{i\sigma} b_{i\sigma}
    , \quad
    \hat{n}^h_i = h^\dagger_i h_i.
\end{equation*}
Finally, we redefine \(\lambda_i\) by a constant shift
\begin{align*}
    \lambda_i - \frac{1}{4} \alpha J (1-\delta)
    \to \lambda_i,
\end{align*}
and separate \(H_{\text{MF}}\) to a holon part \(H_h\), a spinon part \(H_b\) and a number term \(H_0\), leading to Eqs.~(\ref{eq:mf-ham}) to (\ref{eq:mf-ham-number}) in the main text. 

\subsection{Spinon condensation at zero temperature}\label{app:spinon-condense}

At zero temperature, Bose condensation of spinons will occur at \(k = 0\), as \(\lambda\) decreases to the critical value \(\lambda_c\) determined from \(
    \min_{k} E^b_{k-} 
    = E^b_{0-} = 0
\), yielding
\begin{equation}
    \lambda_c = \sqrt{p^2 + q^2} \ \alpha.
\end{equation}
We need to modify the mean-field equations Eqs.~(\ref{eq:mfeq-lambda}) to (\ref{eq:mfeq-b}) by including the density of condensed spinons at \(k = 0\):
\begin{equation}
    n_0 \equiv \frac{1}{N} \sum_\sigma
    \expect{b^{s\dagger}_{0\sigma} b^s_{0\sigma}}
    \quad (s = A, B),
\end{equation}
which is the same on the two sub-lattices. In the \(B,D = 0\) (AFM) phase, \(\lim_{\beta \to \infty} \chi_0 \to 0\), and \(n_0\) is equal to the limit
\begin{equation}
    n_0 = \lim_{\beta \to \infty} \frac{1}{N}
    \frac{\lambda}{\chi_0} [1 + 2n_b(\chi_0)].
\end{equation}
In phases with \(B,D \ne 0\), \(\chi_0 \ne 0\). Then \(n_b(E^b_{0-})\) becomes a macroscopic number \(N_0 \gg 1\), which can be related to \(n_0\) by
\begin{equation}
    n_0 = \frac{1}{N} 
    \frac{\lambda}{\chi_0} (1 + N_0)
    \approx \frac{\lambda}{\chi_0}
    \frac{N_0}{N}.
\end{equation}
Then the modified self-consistency equations at \(T = 0\) are
\begin{align}
    1 - \delta
    &= n_0 + \frac{1}{N} \sum_{k \ne 0} \Big[
        \frac{\lambda}{\chi_k} - 1
    \Big],
    \label{eq:mfeq-lambda-t0}
    \\
    A &= \frac{q\alpha}{2\lambda} n_0
    + \frac{q}{2\alpha N} \sum_{k \ne 0}
    \frac{|\gamma_k|^2}{\chi_k} ,
    \label{eq:mfeq-a-t0}
    \\
    B &= -\frac{p\alpha}{2\lambda} n_0
    (1 - \delta_{\chi_0}),
    \label{eq:mfeq-b-t0}
\end{align}
where \(\delta_x = 1\) if \(x = 0\) and \(0\) otherwise. In particular, at half-filling (\(\delta = 0\)), we get \(A = 0.605\) and \(n_0 = 0.484\) as \(N \to \infty\). In the FM phase (\(A = 0\)), we get maximum condensation \(n_0 = 1 - \delta\). 

\section{Effective holon interaction due to spinons}\label{app:eff-holon}

This appendix derives Eq.~(\ref{eq:eff-holon-int-v}), the effective holon interaction by exchanging two spinons. At very small doping, we take the approximation \(B, D \approx 0\). Then the spinon part of the Hamiltonian Eq.~(\ref{eq:mf-spinon-mat}) is simplified to
\begin{equation}
    H^b_k = \begin{bmatrix}
        \lambda & & & -\Delta_k\\
        & \lambda & \Delta^*_k & \\
        & \Delta_k & \lambda & \\
        -\Delta^*_k & & & \lambda
    \end{bmatrix},
\end{equation}
where \(\Delta_k = J_b \gamma_k\) (with \(J_b = J A\)). We then Fourier transform the holons and spinons to momentum-frequency representation:
\begin{align}
    h^s_{k}(\tau)
    &= \frac{1}{\sqrt{\beta}} \sum_{k,\omega} 
    e^{i \omega \tau} h^s_{k \omega},
    \\
    b^s_{k \sigma}(\tau)
    &= \frac{1}{\sqrt{\beta}} \sum_{k,\nu} 
    e^{i \nu \tau} b^s_{k \nu \sigma},
\end{align}
where \(\omega, \nu\) sum over discrete fermion and boson Matsubara frequencies, respectively. The spinon part of the action is
\begin{equation}
    S_b 
    = \sum_{k,k'} \sum_{\nu',\nu'} 
    \left[
        \bar{b}_{k\nu\uparrow},
        b_{-k,-\nu\downarrow}
    \right]
    (G_0)^{-1}_{k\nu, k'\nu'}
    \begin{bmatrix}
        b_{k'\nu'\uparrow} \\[0.2em]
        \bar{b}_{-k',-\nu'\downarrow}
    \end{bmatrix},
\end{equation}
where we introduced the shorthand notation
\begin{equation*}
    b_{k\nu\sigma} \equiv \begin{bmatrix}
        b^A_{k\nu\sigma} \\[0.3em]
        b^B_{k\nu\sigma}
    \end{bmatrix},
\end{equation*}
and the matrix \(G_0\) is (with $\chi_k = \sqrt{\lambda^2 - |\Delta_k|^2}$):
\begin{align}
    & (G_0)_{k\nu, k'\nu'}
    = (\mathcal{G}_0)_{k\nu} \delta_{k-k'} \delta_{\nu-\nu'},
    \\
    & (\mathcal{G}_0)_{k\nu} 
    \equiv \frac{1}{\nu^2 + \chi_k^2} 
    \begin{bmatrix}
        \lambda - i\nu & & & \Delta_k \\
        & \lambda - i\nu & -\Delta^*_k & \\
        & -\Delta_k & \lambda + i\nu & \\
        \Delta^*_k & & & \lambda + i\nu \\
    \end{bmatrix}.
\end{align}
Action terms corresponding to spinon-holon interaction Eq.~(\ref{eq:spinon-holon-int-ham}) is Fourier transformed to (with \(\gamma_k = \sum_\eta e^{ik\cdot \eta}\)):
\begin{align}
    & S_\text{int} = \sum_{k,k'} \sum_{\nu',\nu'} 
    \left[
        \bar{b}_{k\nu\uparrow},
        b_{-k,-\nu\downarrow}
    \right]
    T_{k\nu, k'\nu'}
    \begin{bmatrix}
        b_{k'\nu'\uparrow} \\[0.2em]
        \bar{b}_{-k',-\nu'\downarrow}
    \end{bmatrix},
    \\
    & T_{k\nu, k'\nu'}
    = \begin{bmatrix}
        & t^{BA}_1 & & \\
        t^{AB}_1 & & & \\
        & & & t^{AB}_2 \\
        & & t^{BA}_2 &
    \end{bmatrix}_{k\nu, k'\nu'}.
\end{align}
The matrix elements of \(T\) are
\begin{align}
    t^{AB}_{1, k\nu, k'\nu'}
    &= \frac{t}{\beta N} \sum_q {\sum_{\omega,\omega'}}'
    \bar{h}^A_{k'+q,\omega} h^B_{k+q,\omega'}
    \gamma_q ,
    \label{eq:def-t1ab}
    \\
    t^{AB}_{2, k\nu, k'\nu'}
    &= \frac{t}{\beta N} \sum_q {\sum_{\omega,\omega'}}'
    \bar{h}^A_{k'+q,\omega} h^B_{k+q,\omega'}
    \gamma_{q+k+k'} ,
    \label{eq:def-t2ab}
    \\
    t^{BA}_{1, k\nu, k'\nu'}
    &= \frac{t}{\beta N} \sum_q {\sum_{\omega,\omega'}}'
    \bar{h}^B_{k'+q,\omega'} h^A_{k+q,\omega} 
    \gamma^*_q ,
    \label{eq:def-t1ba}
    \\
    t^{BA}_{2, k\nu, k'\nu'}
    &= \frac{t}{\beta N} \sum_q {\sum_{\omega,\omega'}}'
    \bar{h}^B_{k'+q,\omega'} h^A_{k+q,\omega} 
    \gamma^*_{q+k+k'},
    \label{eq:def-t2ba}
\end{align}
where
\begin{equation}
    {\sum_{\omega,\omega'}}'
    \equiv \sum_{\omega,\omega'} 
    \delta_{(\nu+\omega) - (\nu'+\omega')}.
\end{equation}

To obtain an effective theory for the holons, we integrate out the spinons. Collecting terms involving spinons in the action, we get
\begin{equation*}
    S_b + S_{\text{int}} = \sum_{k,k'} \sum_{\nu',\nu'} 
\left[
    \bar{b}_{k\nu\uparrow},
    b_{-k,-\nu\downarrow}
\right]
G^{-1}_{k\nu, k'\nu'}
\begin{bmatrix}
    b_{k'\nu'\uparrow} \\[0.2em]
    \bar{b}_{-k',-\nu'\downarrow}
\end{bmatrix},
\end{equation*}
where the matrix \(G^{-1}\) is given by
\begin{equation}
    G^{-1}_{k\nu, k'\nu'}
    = (G_0^{-1} + T)_{k\nu, k'\nu'}.
\end{equation}
The effective action for holons
\(S^\text{eff}_h[\bar{h},h]\) is defined by
\begin{align*}
    Z &= \int D\bar{h} \, Dh \, 
    e^{-S^{\text{eff}}_h[\bar{h},h]},
    \\
    e^{-S^{\text{eff}}_h}
    &\equiv \int D\bar{b} \, Db \, e^{-S}
    = e^{-S_h} \int D\bar{b} \, Db \, 
    e^{-(S_b + S_{\text{int}})}.
\end{align*}
Performing Gaussian integration over spinons, we get
\begin{align}
    &\int D\bar{b} \, Db \, e^{-(S_b + S_{\text{int}})}
    \nonumber \\
    &\propto (\det G^{-1})^{-1}
    = \exp(- \ln \det G^{-1}).
\end{align}
Thus the effective holon action is given by
\begin{equation}
    S^{\text{eff}}_h = S_h + \ln \det G^{-1}.
\end{equation}
The log-determinant can be expanded as a power series of \(t\); using \(\ln \det A = \tr \ln A\) and \(\ln(1+x) = \sum_{n=1}^{\infty} (-1)^{n+1} x^n / n\), we obtain
\begin{align}
    \ln \det G^{-1}
    &= \tr \ln (G_0^{-1} + T)
    \nonumber
    \\
    &= \tr [
        \ln G_0^{-1} + \ln (1 + G_0 T)
    ]
    \nonumber
    \\
    &= \tr\ln G_0^{-1}
    + \sum_{n=1}^\infty \frac{(-1)^{n+1}}{n}
    \tr (G_0 T)^n.
\end{align}
The term \(\tr\ln G_0^{-1}\) is a constant independent of \(h\) and can be omitted. Then the effective holon action is
\begin{align}
    S^{\text{eff}}_h[\bar{h},h]
    &= S_h^0 + \sum_{n=1}^\infty S_h^n, 
    \\
    S_h^n 
    &\equiv \frac{(-1)^{n+1}}{n} \tr (G_0 T)^n.
\end{align}
We keep up to second order terms. The first order term \(\tr (G_0 T) = 0\). The second order term is
\begin{widetext}
\begin{align*}
    S_h^2 
    &\equiv -\frac{1}{2} \tr (G_0 T)^2
    = -\frac{1}{2} \sum_{k,\nu} \sum_{k_1,\nu_1}
    \sum_{k_2,\nu_2} \sum_{k',\nu'} \tr [
        (G_0)_{k\nu,k_1\nu_1} T_{k_1\nu_1,k_2\nu_2}
        (G_0)_{k_2\nu_2,k'\nu'} T_{k'\nu',k\nu}
    ]
    \\
    &= \frac{1}{2} \sum_{k,\nu} \sum_{k',\nu'} 
    \frac{1}{(\nu^2 + \chi_k^2)(\nu'^2 + \chi_{k'}^2)}
    \Big[
        \Delta_k \Delta_{k'} (
            t^{AB}_1 t'^{BA}_2
            + t^{BA}_2 t'^{AB}_1
        ) 
        + \Delta^*_{k} \Delta^*_{k'} (
            t^{BA}_1 t'^{AB}_2
            + t^{AB}_2 t'^{BA}_1
        ) \\ &\qquad
        - (i\nu - \lambda)(i\nu' - \lambda) (
            t^{AB}_1 t'^{BA}_1
            + t^{BA}_1 t'^{AB}_1
        )
        - (i\nu + \lambda)(i\nu + \lambda') (
            t^{AB}_2 t'^{BA}_2
            + t^{BA}_2 t'^{AB}_2
        )
    \Big].
\end{align*}
Here we simply write \(t^{ss'}_i = t^{ss'}_{i,k\nu,k'\nu'}\), \(t'^{ss'}_i = t^{ss'}_{i,k'\nu',k\nu}\) (where \(s,s' = A,B\) and \(i = 1,2\)). Note that each term is invariant under the exchange of \((k,\nu) \leftrightarrow (k',\nu')\). Then
\begin{align*}
    S_h^2 
    &= \sum_{k,\nu} \sum_{k',\nu'} 
    \frac{1}{(\nu^2 + \chi_k^2)(\nu'^2 + \chi_{k'}^2)}
    \Big[
        \Delta_k \Delta_{k'} t^{BA}_2 t'^{AB}_1
        + \Delta^*_k \Delta^*_{k'} t^{BA}_1 t'^{AB}_2
        \\ &\qquad
        - (i\nu - \lambda)(i\nu' - \lambda) t^{BA}_1 t'^{AB}_1
        - (i\nu + \lambda)(i\nu' + \lambda) t^{BA}_2 t'^{AB}_2
    \Big].
\end{align*}
Evaluating summation over \(\nu'\) first after substituting in the definition of \(t_i^{ss'}\) (Eqs.~(\ref{eq:def-t1ab}) to (\ref{eq:def-t2ba})), we obtain:
\begin{align}
    S_h^2 
    &= \bigg(\frac{t}{\beta N}\bigg)^2
    \sum_{k,\nu} \sum_{k',\nu'} 
    \sum_{q_1,q_2} \sum_{\omega_1,\omega'_1}
    \sum_{\omega_2,\omega'_2}
    \frac{
        \delta_{(\nu+\omega_1) - (\nu'+\omega'_1)}
        \delta_{(\nu'+\omega_2) - (\nu+\omega'_2)}
    }{(\nu^2 + \chi_k^2)(\nu'^2 + \chi_{k'}^2)}
    \bar{h}^B_{k'+q_1,\omega'_1} h^A_{k+q_1,\omega_1} 
    \bar{h}^A_{k+q_2,\omega_2} h^B_{k'+q_2,\omega'_2}
    \nonumber \\ &\quad \times
    \Big[
        \Delta_k \Delta_{k'} 
        \gamma^*_{q_1+k+k'} \gamma_{q_2}
        + \Delta^*_k \Delta^*_{k'} 
        \gamma^*_{q_1} \gamma_{q_2+k+k'}
        - (i\nu - \lambda)(i\nu' - \lambda) 
        \gamma^*_{q_1} \gamma_{q_2}
        - (i\nu + \lambda)(i\nu' + \lambda) 
        \gamma^*_{q_1+k+k'} \gamma_{q_2+k+k'}
    \Big]
    \nonumber \\
    &= \bigg(\frac{t}{\beta N}\bigg)^2
    \sum_{k,k',\nu} 
    \sum_{q_1,q_2} {\sum_{\{\omega\}}}'
    \frac{1}{
        (\nu^2 + \chi_k^2)
        [(\nu+\omega_1-\omega'_1)^2 + \chi_{k'}^2]
    } \bar{h}^B_{k'+q_1,\omega'_1} h^A_{k+q_1,\omega_1} 
    \bar{h}^A_{k+q_2,\omega_2} h^B_{k'+q_2,\omega'_2}
    \nonumber \\ &\quad \times
    \Big[
        \Delta_k \Delta_{k'} 
        \gamma^*_{q_1+k+k'} \gamma_{q_2}
        + \Delta^*_k \Delta^*_{k'} 
        \gamma^*_{q_1} \gamma_{q_2+k+k'}
        \nonumber \\ &\qquad
        - (i\nu - \lambda)[i(\nu+\omega_1-\omega'_1) - \lambda] 
        \gamma^*_{q_1} \gamma_{q_2}
        - (i\nu + \lambda)[i(\nu+\omega_1-\omega'_1) + \lambda] 
        \gamma^*_{q_1+k+k'} \gamma_{q_2+k+k'}
    \Big],
    \label{eq-app:s2h-exact}
\end{align}
where
\begin{equation*}
    {\sum_{\{\omega\}}}' \equiv 
    \sum_{\omega_1,\omega'_1}
    \sum_{\omega_2,\omega'_2}
    \delta_{(\omega_1+\omega_2) - (\omega'_1+\omega'_2)}
\end{equation*}
We keep only the \emph{instantaneous} holon interaction by taking the \emph{static limit}, i.e. setting \(\omega_1 = \omega'_1\) in the coefficient of \(\bar{h}^B h^A \bar{h}^A h^B\). Then Eq.~\ref{eq-app:s2h-exact} reduces to
\begin{align}
    S_h^2 
    &= \bigg(\frac{t}{\beta N}\bigg)^2
    \sum_{k,k',\nu} 
    \sum_{q_1,q_2} {\sum_{\{\omega\}}}'
    \frac{
        \bar{h}^B_{k'+q_1,\omega'_1} h^A_{k+q_1,\omega_1} 
        \bar{h}^A_{k+q_2,\omega_2} h^B_{k'+q_2,\omega'_2}
    }{
        (\nu^2 + \chi_k^2)
        (\nu^2 + \chi_{k'}^2)
    }
    \nonumber \\ & \times
    \Big[
        \Delta_k \Delta_{k'} 
        \gamma^*_{q_1+k+k'} \gamma_{q_2}
        + \Delta^*_k \Delta^*_{k'} 
        \gamma^*_{q_1} \gamma_{q_2+k+k'}
        - (i\nu - \lambda)^2
        \gamma^*_{q_1} \gamma_{q_2}
        - (i\nu + \lambda)^2
        \gamma^*_{q_1+k+k'} \gamma_{q_2+k+k'}
    \Big].
\end{align}
\end{widetext}
In the zero-temperature limit, the summation over \(\nu\) can be replaced by an integral:
\begin{align*}
    \frac{1}{\beta} \sum_\nu 
    &\to \int \frac{d\nu}{2\pi},
    \\
    \int \frac{d\nu}{2\pi} \frac{1}{
        (\nu^2 + \chi_k^2)
        (\nu^2 + \chi_{k'}^2)
    } &= \frac{1}{
        2 \chi_k \chi_{k'}
        (\chi_k + \chi_{k'})
    },
    \\
    \int \frac{d\nu}{2\pi} \frac{(i\nu \pm \lambda)^2}{
        (\nu^2 + \chi_k^2)
        (\nu^2 + \chi_{k'}^2)
    } &= \frac{\lambda^2 - \chi_k \chi_{k'}}{
        2 \chi_k \chi_{k'} (\chi_k + \chi_{k'})
    }.
\end{align*}
We can then inverse Fourier transform \(\omega\) to \(\tau\):
\begin{align}
    & \frac{1}{\beta} {\sum_{\{\omega\}}}'
    \bar{h}^B_{k'+q_1,\omega'_1} \bar{h}^A_{k+q_2,\omega_2} 
    h^A_{k+q_1,\omega_1} h^B_{k'+q_2,\omega'_2}
    \nonumber \\ &
    = \int d\tau \,
    \bar{h}^B_{k'+q_1,\tau} \bar{h}^A_{k+q_2,\tau} 
    h^A_{k+q_1,\tau} h^B_{k'+q_2,\tau}.
\end{align}
Therefore (the order of \(h\) is changed)
\begin{align}
    & S_h^2 
    = -\frac{t^2}{2 N^2}
    \sum_{k,k'} \sum_{q_1,q_2}
    \frac{
        \bar{h}^B_{k'+q_1} \bar{h}^A_{k+q_2} 
        h^A_{k+q_1} h^B_{k'+q_2}
    }{
        \chi_k \chi_{k'} (\chi_k + \chi_{k'})
    }
    \nonumber \\ &\ \ \times
    \Big[
        \Delta_k \Delta_{k'} 
        \gamma^*_{q_1+k+k'} \gamma_{q_2}
        + \Delta^*_k \Delta^*_{k'} 
        \gamma^*_{q_1} \gamma_{q_2+k+k'}
        \nonumber \\ &\quad
        + (\chi_k \chi_{k'} - \lambda^2)(
            \gamma^*_{q_1} \gamma_{q_2}
            + \gamma^*_{q_1+k+k'} \gamma_{q_2+k+k'}
        )
    \Big].
\end{align}
Redefining summation variables:
\begin{equation*}
\left. \begin{aligned}
    k_1 &= k + q_1 \\ 
    k_2 &= k' + q_2 \\
    q &= k' - k \\ 
    p &= -k'
\end{aligned} \right\} 
\ \Rightarrow \ \left\{ \begin{aligned}
    k &= -p-q \\
    k' &= -p \\
    q_1 &= k_1+p+q \\
    q_2 &= k_2+p
\end{aligned} \right.,
\end{equation*}
we obtain
\begin{align}
    & S_h^2 
    = -\frac{t^2}{2 N^2}
    \sum_{k_1,k_2} \sum_{p,q}
    \frac{
        \bar{h}^B_{k_1+q} \bar{h}^A_{k_2-q} 
        h^A_{k_1} h^B_{k_2} 
    }{
        \chi_{p+q} \chi_p (\chi_{p+q} + \chi_p)
    } \nonumber \\ &\ \times
    \Big[
        \Delta^*_{p+q} \Delta^*_p 
        \gamma^*_{k_1-p} \gamma_{k_2+p}
        + \Delta_{p+q} \Delta_p 
        \gamma^*_{k_1+p+q} \gamma_{k_2-p-q}
        \nonumber \\ &\quad
        + (\chi_{p+q} \chi_p - \lambda^2)(
            \gamma^*_{k_1+p+q} \gamma_{k_2+p}
            + \gamma^*_{k_1-p} \gamma_{k_2-p-q}
        )
    \Big].
\end{align}
Here we used
\begin{equation*}
    \chi_{-k} = \chi_k, \quad
    \gamma_{-k} = \gamma^*_k, \quad
    \Delta_{-k} = \Delta^*_k.
\end{equation*}
From this effective action, we read off terms in the Hamiltonian that represent interaction between holons (rename \(k_1, k_2\) to \(k, k'\)):
\begin{align}
    &H^{\text{eff}}_{\text{int}} 
    = \frac{1}{N} \sum_{k,k',q} V_{kk'q}
    h^{B\dagger}_{k+q} h^{A\dagger}_{k'-q} 
    h^A_{k} h^B_{k'},
    \\
    &V_{kk'q} 
    \equiv -\frac{t^2}{2 N}
    \sum_p \frac{1}{
        \chi_{p+q} \chi_p (\chi_{p+q} + \chi_p)
    } \nonumber \\ &\ \ \times
    \Big[
        \Delta^*_{p+q} \Delta^*_p 
        \gamma^*_{k-p} \gamma_{k'+p}
        + \Delta_{p+q} \Delta_p 
        \gamma^*_{k+p+q} \gamma_{k'-p-q}
        \nonumber\\ &\quad
        + (\chi_{p+q} \chi_p - \lambda^2)(
            \gamma^*_{k+p+q} \gamma_{k'+p}
            + \gamma^*_{k-p} \gamma_{k'-p-q}
        )
    \Big].
\end{align}
This effective holon interaction is due to the exchange of two spinons of momenta \(p\) and \(p+q\). Let us call the holon momenta as
\begin{equation*}
    k + q = k_1, \quad
    k = k'_1, \quad
    k' = k'_2, \quad
    k'_2+k'_1-k_1 = k_2.
\end{equation*}
Then we get an alternative expression of \(H^\text{eff}_\text{int}\):
\begin{align}
    & H^\text{eff}_\text{int} 
    = \frac{1}{N}
    \sum_{\{k,k'\}} V_{k_1 k_2 k'_1 k'_2}
    h^{B\dagger}_{k_1} h^{A\dagger}_{k_2} 
    h^A_{k'_1} h^B_{k'_2},
    \\
    & V_{k_1 k_2 k'_1 k'_2}
    \equiv -\frac{t^2}{2 N}
    \sum_p \frac{\delta_{(k_1+k_2)-(k'_1+k'_2)}}{
        \chi_p \chi_{p+k_1-k'_1} (\chi_p + \chi_{p+k_1-k'_1})
    } \nonumber \\ & \times
    \Big[
        \Delta^*_p \Delta^*_{p+k_1-k'_1}
        \gamma^*_{k'_1-p} \gamma_{k'_2+p}
        + \Delta_p \Delta_{p+k_1-k'_1}
        \gamma^*_{k_1+p} \gamma_{k_2-p}
        \nonumber \\ &\ 
        + (\chi_p \chi_{p+k_1-k'_1} - \lambda^2)(
            \gamma^*_{k_1+p} \gamma_{k'_2+p}
            + \gamma^*_{k'_1-p} \gamma_{k_2-p}
        )
    \Big].
    \label{eq-app:eff-holon-int-v}
\end{align}
Define \(p' \equiv p+k_1-k'_1\). We then rewrite the expression by keeping $k'_1, k_2$ (the momenta of $h^A$) only:
\begin{align*}
    & V_{k_1 k_2 k'_1 k'_2}
    \equiv -\frac{t^2}{2 N} \sum_p \bigg\{ 
        \frac{\delta_{(k_1+k_2)-(k'_1+k'_2)}}{
            \chi_p \chi_{p'} (\chi_p + \chi_{p'})
        } \\ & \times
        \Big[
            \Delta^*_p \Delta^*_{p'}
            \gamma^*_{k'_1-p} \gamma_{k_2+p'}
            + \Delta_p \Delta_{p'}
            \gamma^*_{k'_1+p'} \gamma_{k_2-p}
            \\ &\ 
            + (\chi_p \chi_{p'} - \lambda^2)(
                \gamma^*_{k'_1+p'} \gamma_{k_2+p'}
                + \gamma^*_{k'_1-p} \gamma_{k_2-p}
            )
        \Big]
    \bigg\}.
\end{align*}
Finally, with
\begin{equation*}
    {\sum_{\{k\}}}'
    \equiv \sum_{k_1,k_2} \sum_{k'_1,k'_2}
    \delta_{(k_1+k_2)-(k'_1+k'_2)}
\end{equation*}
we get (substituting in \(\Delta_k = J_b \gamma_k\))
\begin{align}
    & 
    H^\text{eff}_\text{int} = \frac{1}{N}
    {\sum_{\{k\}}}' V_{k_1 k_2 k'_1 k'_2}
    h^{B\dagger}_{k_1} h^{A\dagger}_{k_2} 
    h^A_{k'_1} h^B_{k'_2}.
    \\
    & V_{k_1 k_2 k'_1 k'_2}
    = -\frac{t^2}{2 N} \sum_p \bigg\{ 
        \frac{1}{
            \chi_p \chi_{p'} (\chi_p + \chi_{p'})
        } \nonumber \\ &\ \ \times
        \Big[
            J_b^2 (
                \gamma^*_p \gamma^*_{p'}
                \gamma^*_{k'_1-p} \gamma_{k_2+p'}
                + \gamma_p \gamma_{p'}
                \gamma^*_{k'_1+p} \gamma_{k_2-p}
            ) \nonumber\\ &\quad
            + (\chi_p \chi_{p'} - \lambda^2)(
                \gamma^*_{k'_1+p} \gamma_{k_2+p'}
                + \gamma^*_{k'_1-p} \gamma_{k_2-p}
            )
        \Big]
    \bigg\},
\end{align}
which is Eq.~(\ref{eq:eff-holon-int-v}). 

\section{Continuum limit of the \emph{t-J} model Lagrangian}\label{app:continuum}

This section derives the Lagrangian Eqs.~(\ref{eq:l-holon-con}) to (\ref{eq:l-0-con}) in the continuum limit. The following identities on the honeycomb lattice are useful:
\begin{equation*}
\begin{aligned}
    \textstyle
    \sum_{\delta} \delta \cdot \mathbf{A} &= 0,
    \\
    \textstyle
    \sum_{\delta} (\delta \cdot \nabla)^2 f
    &= \textstyle
    (\alpha/2) a^2 \nabla^2 f,
    \\
    \textstyle
    \sum_{\delta} (\delta \cdot \mathbf{A}) 
    (\delta \cdot \mathbf{B})
    &= \textstyle
    (\alpha/2) a^2 \mathbf{A} \cdot \mathbf{B},
    \\
    \textstyle
    \sum_{\delta} (\delta \cdot \mathbf{A})^2
    &= \textstyle
    (\alpha/2) a^2 \mathbf{A}^2.
\end{aligned}
\end{equation*}
Here \(a\) is the nearest neighbor distance, \(\delta\) sums over nearest neighbors of a site on sub-lattice \(A\) (see Fig.~\ref{fig:honeycomb}), \(f,g\) are any scalar fields and \(\mathbf{A},\mathbf{B}\) are any vector fields. The continuum limit of \(L_h\) (Eq.~(\ref{eq:l-holon})) and \(L_0\) (Eq.~(\ref{eq:l0})) are
\begin{align}
    L_h &= \int \frac{d^2x}{2\Omega} \sum_s \Big[
        \bar{h}^s (\partial_\tau + i s A_\tau) h^s
        + \lambda \bar{h}^s h^s
    \Big],
    \\
    L_0 &= \int \frac{d^2x}{2\Omega} \bigg[
        - \sum_s (\lambda + i s A_\tau)
        + \frac{\alpha J}{2} \Delta^2
    \bigg]
    \nonumber \\
    &= \int \frac{d^2x}{2\Omega} \bigg[
        - 2\lambda
        + \frac{\alpha J}{2}\Delta^2
    \bigg].
\end{align}
The continuum limit of the spinon-holon interaction \(L_t\) (Eq.~(\ref{eq:l-spinon-holon})) is
\begin{align*}
    & L_t
    \to t \int \frac{d^2x}{2\Omega}
    \sum_{\delta,\sigma} 
    \\ &\quad \Big[
        \bar{h}^B(x+\delta) h^A(x)
        \bar{b}^A_\sigma(x) b^B_\sigma(x+\delta)
        + h.c.
    \Big]
    \\
    &\approx t \int \frac{d^2x}{2\Omega}
    \sum_{\delta,\sigma} \Big\{
        [
            (1 + \delta\cdot \nabla 
            + \tfrac{1}{2} (\delta\cdot \nabla)^2)
            \bar{h}^B 
        ] h^A 
        \\ &\quad \times \bar{b}^A_\sigma [
            (1 + \delta\cdot \nabla 
            + \tfrac{1}{2} (\delta\cdot \nabla)^2)
            b^B_\sigma
        ]
        + h.c.
    \Big\}.
\end{align*}
We expand this in powers of the nearest neighbor distance \(a\). The first order terms in $a$ turn out to vanish. The second order terms are not important in describing the spinon-holon interaction. We then only keep the zeroth order terms:
\begin{align}
    L_t & \approx t \int \frac{d^2x}{2\Omega}
    \sum_{\delta,\sigma} (
        \bar{h}^B h^A \bar{b}^A_\sigma b^B_\sigma
        + h.c.
    )
    \nonumber \\
    &= \alpha t \int \frac{d^2x}{2\Omega}
    \sum_\sigma (
        \bar{h}^B h^A \bar{b}^A_\sigma b^B_\sigma
        + h.c.
    ).
\end{align}
Finally we calculate the continuum limit of \(L_\text{heis}\) (Eq.~(\ref{eq:l-heis})). Let us separate it to two parts:
\begin{align}
    L_\text{heis} &= L_b^0 + L_J,
    \\
    L_b^0 &= \sum_{s,\delta} \sum_{i \in s} \Big[
        \bar{b}_{i\sigma} 
        (\partial_\tau + i s A_\tau(i)) b_{i\sigma}
        + \lambda \bar{b}_{i\sigma} b_{i\sigma}
    \Big],
    \\ 
    L_J &=
    - \frac{J \Delta}{2} 
    \sum_{\substack{i\in A \\ \delta,\sigma}}
    \sigma \Big[
        e^{i \delta \cdot \mathbf{A}(i,i+\delta)}
        \bar{b}_{i\sigma} \bar{b}_{i+\delta,-\sigma}
        + h.c.
    \Big].
\end{align}
The continuum limit of \(L_b^0\) is
\begin{equation}
L_b^0 \to \int \frac{d^2x}{2\Omega}
\sum_{s,\sigma} \Big[
    \bar{b}^s_\sigma (\partial_\tau + i s A_\tau) b^s_\sigma
    + \lambda \bar{b}^s_\sigma b^s_\sigma 
\Big].
\end{equation}
Next, 
\begin{align}
    & L_J \to 
    -\frac{J\Delta}{2} \int \frac{d^2x}{2\Omega}
    \sum_{\delta,\sigma} \sigma 
    \nonumber \\ &\times \Big[
        e^{i \delta \cdot \mathbf{A}(x)}
        \bar{b}^A_\sigma (x-\tfrac{\delta}{2}) 
        \bar{b}^B_{-\sigma}(x+\tfrac{\delta}{2})
        + h.c.
    \Big]
    \nonumber \\
    &\approx -\frac{J\Delta}{2}
    \int \frac{d^2x}{2\Omega}
    \sum_{\delta,\sigma} \sigma \bigg\{
        [
            1 + i \delta \cdot \mathbf{A}
            + \tfrac{1}{2} (i \delta \cdot \mathbf{A})^2
        ] 
        \nonumber \\ &\quad \times 
        \Big[
            (
                1 - \tfrac{\delta}{2} \cdot \nabla
                + \tfrac{1}{2} (\tfrac{\delta}{2} \cdot \nabla)^2
            ) \bar{b}^A_\sigma 
            \nonumber \\ &\qquad \cdot 
            (
                1 + \tfrac{\delta}{2} \cdot \nabla
                + \tfrac{1}{2} (\tfrac{\delta}{2} \cdot \nabla)^2
            ) \bar{b}^B_{-\sigma}
        \Big] 
        + h.c.
    \bigg\}.
\end{align}
We keep terms up to second order in \(a\). The zeroth order terms in \(a\) are
\begin{align}
    L_J^0 
    &= -4Q \int \frac{d^2x}{2\Omega}
    \sum_\sigma \sigma (
        \bar{b}^A_\sigma  \bar{b}^B_{-\sigma}
        + h.c.
    ).
\end{align}
Here $Q \equiv \alpha J\Delta/8$. The first order terms \(L_J^1 = 0\). Using the identity
\begin{align}
    &\int d^2x \, \Big[
        (\nabla^2 \bar{b}^A_\sigma)
        \bar{b}^B_{-\sigma}
        + \bar{b}^A_\sigma  
        (\nabla^2 \bar{b}^B_{-\sigma})
    \Big]
    \nonumber \\
    &= \int d^2x \, \Big[
        \nabla^2(
            \bar{b}^A_\sigma  
            \bar{b}^B_{-\sigma}
        ) - 2 \nabla \bar{b}^A_\sigma 
        \cdot \nabla \bar{b}^B_{-\sigma}
    \Big]
    \nonumber \\
    &= -2 \int d^2x \, \nabla \bar{b}^A_\sigma 
    \cdot \nabla \bar{b}^B_{-\sigma},
\end{align}
we calculate the second order terms in \(a\):
\begin{widetext}
\begin{align}
    & L_J^2
    = -\frac{J \Delta}{2}  \int \frac{d^2x}{2\Omega}
    \sum_{\delta,\sigma} \sigma \bigg\{
        \Big[
            \tfrac{1}{2} [
                (-\tfrac{\delta}{2} \cdot \nabla)^2
                \bar{b}^A_\sigma] 
            \bar{b}^B_{-\sigma}
            + \bar{b}^A_\sigma \tfrac{1}{2} [
                (\tfrac{\delta}{2} \cdot \nabla)^2 
                \bar{b}^B_{-\sigma}
            ]
            + (-\tfrac{\delta}{2} \cdot \nabla) \bar{b}^A_\sigma 
            (\tfrac{\delta}{2} \cdot \nabla) \bar{b}^B_{-\sigma}
        \Big]
        \nonumber \\ 
        &\qquad \qquad \qquad \qquad \qquad \quad
        + (i\delta \cdot \mathbf{A}) \Big[
            [
                (-\tfrac{\delta}{2} \cdot \nabla)
                \bar{b}^A_\sigma
            ] \bar{b}^B_{-\sigma}
            + \bar{b}^A_\sigma [
                (\tfrac{\delta}{2} \cdot \nabla) 
                \bar{b}^B_{-\sigma}
            ]
        \Big]
        + \tfrac{1}{2}(i\delta \cdot \mathbf{A})^2
        \bar{b}^A_\sigma  \bar{b}^B_{-\sigma}
    \bigg\} + h.c.
    \nonumber \\
    &= Q a^2 \int \frac{d^2x}{2\Omega}
    \sum_\sigma \sigma \Big\{
        \tfrac{-1}{4} \Big[
            (\nabla^2 \bar{b}^A_\sigma)
            \bar{b}^B_{-\sigma}
            + \bar{b}^A_\sigma  
            (\nabla^2 \bar{b}^B_{-\sigma})
        \Big]
        + \tfrac{1}{2} \nabla \bar{b}^A_\sigma 
        \cdot \nabla \bar{b}^B_{-\sigma}
        + i \mathbf{A}\cdot \Big[
            (\nabla \bar{b}^A_\sigma)
            \bar{b}^B_{-\sigma}
            - \bar{b}^A_\sigma  
            (\nabla \bar{b}^B_{-\sigma})
        \Big]
        + \mathbf{A}^2 
        \bar{b}^A_\sigma  \bar{b}^B_{-\sigma}
    \Big\}
    + h.c.
    \nonumber \\
    &= Q a^2 \int \frac{d^2x}{2\Omega}
    \sum_\sigma \sigma \Big\{
        \nabla \bar{b}^A_\sigma 
        \cdot \nabla \bar{b}^B_{-\sigma}
        + i \mathbf{A}\cdot \Big[
            (\nabla \bar{b}^A_\sigma)
            \bar{b}^B_{-\sigma}
            - \bar{b}^A_\sigma  
            (\nabla \bar{b}^B_{-\sigma})
        \Big]
        + \mathbf{A}^2 
        \bar{b}^A_\sigma  \bar{b}^B_{-\sigma}
    \Big\} + h.c.
    \nonumber \\
    &= Q a^2 \int \frac{d^2x}{2\Omega} 
    \sum_\sigma \sigma \Big[
        (\nabla - i\mathbf{A}) \bar{b}^A_\sigma  \cdot 
        (\nabla + i\mathbf{A}) \bar{b}^B_{-\sigma}
        + h.c.
    \Big].
\end{align}
\end{widetext}

\bibliography{tJhoneycomb}
\end{document}